\documentclass[12pt,onecolumn]{article}
\pdfoutput=1
\usepackage{hyperref}
\hypersetup{colorlinks=true,linkcolor=blue,citecolor=blue}
\usepackage{amsmath,amssymb,amsfonts,amsthm}
\usepackage{caption,subcaption}
\usepackage[pdftex]{graphicx}
\usepackage{mathtools}
\usepackage{young}
\usepackage{comment}
\usepackage{mathrsfs}
\usepackage{titlesec}
\usepackage[normalem]{ulem}
\usepackage[affil-it]{authblk}
\usepackage{listings}
\usepackage{soul}
\usepackage{multirow}

\def\be{\begin{equation}}
\def\ee{\end{equation}}

\def\arg{\operatorname{arg}}

\def\Tr{{\rm Tr}\,}

\def\CP{{\mathcal P}}
\def\ICP{{\mathbb{CP}}}
\def\CW{{\mathcal W}}
\def\tCW{{\widetilde\CW}}
\def\CB{{\mathcal B}}

\def\det{\mathrm{det}}

\def\eff{{\mathrm{eff}}}
\def\tZ{{\widetilde{Z}}}
\def\tm{{\widetilde{m}}}

\def\CK{{\mathcal K}}
\def\CN{{\mathcal N}}

\def\CS{{\mathcal S}}
\def\CR{{\mathcal R}}
\def\IR{{\mathbb{R}}}
\def\IZ{{\mathbb{Z}}}

\def\IC{{\mathbb{C}}}

\def\e{{\rm e}}

\def\scale{{\mu}}
\def\d{{\rm d}}

\newcommand{\SD}[3]{{\mathbb{S}}_{#3}{[#1,#2]}}

\titleclass{\subsubsubsection}{straight}[\subsection]

\newcounter{subsubsubsection}[subsubsection]
\renewcommand\thesubsubsubsection{\thesubsubsection.\arabic{subsubsubsection}}

\titleformat{\subsubsubsection}
  {\normalfont\normalsize\bfseries}{\thesubsubsubsection}{1em}{}
\titlespacing*{\subsubsubsection}
{0pt}{3.25ex plus 1ex minus .2ex}{1.5ex plus .2ex}

\makeatletter
\renewcommand\paragraph{\@startsection{paragraph}{5}{\z@}%
  {3.25ex \@plus1ex \@minus.2ex}%
  {-1em}%
  {\normalfont\normalsize\bfseries}}
\renewcommand\subparagraph{\@startsection{subparagraph}{6}{\parindent}%
  {3.25ex \@plus1ex \@minus .2ex}%
  {-1em}%
  {\normalfont\normalsize\bfseries}}
\def\toclevel@subsubsubsection{4}
\def\toclevel@paragraph{5}
\def\toclevel@paragraph{6}
\def\l@subsubsubsection{\@dottedtocline{4}{7em}{4em}}
\def\l@paragraph{\@dottedtocline{5}{10em}{5em}}
\def\l@subparagraph{\@dottedtocline{6}{14em}{6em}}
\makeatother

\newcommand{\weight}{\nu}

\newcommand{\gA}{\mathrm{A}}
\newcommand{\gD}{\mathrm{D}}
\newcommand{\gE}{\mathrm{E}}

\newcommand{\cS}{\mathcal{S}}

\newcommand{\fg}{\mathfrak{g}}

\newcommand{\SU}{\mathrm{SU}}
\newcommand{\SO}{\mathrm{SO}}

\newcommand{\UU}{\mathrm{U}}

\newcommand{\UE}{\gE}

\newcommand{\dd}{\mathrm{d}}

\usepackage[usenames,dvipsnames]{xcolor}

\definecolor{darkergreen}{rgb}{0,0.667,0}
\definecolor{gold}{rgb}{1.,0.843137,0.}
\definecolor{deepskyblue}{rgb}{0.,0.74902,1.}
\definecolor{mediumorchid}{rgb}{0.729412,0.333333,0.827451}

\usepackage{tikz}
\usetikzlibrary{positioning}
\usetikzlibrary{decorations.pathmorphing}
\usetikzlibrary{decorations.markings}
\usetikzlibrary{shapes.misc}
\usetikzlibrary{intersections}

\usetikzlibrary{arrows}
\tikzset{>={angle 60}}

\tikzstyle{every picture}=[scale=1,baseline=(current bounding box.south)]

\tikzstyle{X}=[cross out, draw, scale = 0.75, thick]
\tikzstyle{Z}=[draw, circle, minimum size=1em, scale=1, inner sep=2pt]
\tikzstyle{W}= [circle, draw, minimum size=1em]
\tikzstyle{D}= [circle, minimum size=1em]

\makeatletter
\def\@maketitle{%
  \newpage
  \null
  \vskip 3.5em%
  \begin{center}%
  \let \footnote \thanks
    {\huge \@title \par}%
    \vskip 3.0em%
    {\large
      \lineskip 1.5em%
      \begin{tabular}[t]{c}%
        \@author
      \end{tabular}\par}%
    \vskip 1em%
  \end{center}%
  \par
  \vskip 1.5em}
\makeatother

\usepackage{listings}

\numberwithin{equation}{section}

\begin{document}

\allowdisplaybreaks

\title{
ADE Spectral Networks and\\ Decoupling Limits of Surface Defects
}

\author[1]{{Pietro Longhi}\thanks{pietro.longhi@physics.uu.se}}
\author[2]{{Chan Y. Park}\thanks{chan@physics.rutgers.edu}}

\affil[1]{{\normalsize Department of Physics and Astronomy, Uppsala university, Sweden}}
\affil[2]{\normalsize{NHETC and Department of Physics and Astronomy, Rutgers University, NJ, USA}}

\maketitle

\begin{abstract}
We study vacua and BPS spectra of canonical surface defects of class $\mathcal{S}$ theories in different decoupling limits using $\mathrm{ADE}$ spectral networks. 
In some regions of the IR moduli spaces of these 2d-4d systems, the mixing between 2d and 4d BPS states is suppressed, and the spectrum of 2d-4d BPS states becomes that of a 2d $\CN=(2,2)$ theory. 
For some decoupling limits, we identify the 2d theories describing the surface defects with nonlinear sigma models and coset models that have been previously studied. 
We also study certain cases where the decoupling limit of a surface defect exhibits a set of vacua and a BPS spectrum that appear to be entirely new. A detailed analysis of these spectra and their wall-crossing behavior is performed.
\end{abstract}

\setcounter{tocdepth}{2}
\tableofcontents

\section{Overview}

Four dimensional $\CN=2$ theories admit various types of half-BPS surface defects, in this paper we focus on 2d $\mathcal{N} = (2,2)$ surface defects of class $\mathcal{S}$ theories \cite{Gukov:2006jk, Gaiotto:2009fs, Gaiotto:2011tf, Gaiotto:2013sma}. 
In suitable decoupling limits, the low energy dynamics of the surface defects are described by purely two dimensional theories. 
Focusing on two of these limits, we employ surface defects as tools to study nonperturbative aspects of certain 2d $\CN=(2,2)$ models.
Despite a long and celebrated history, the study of these models continues to be a seemingly inextinguishable source of surprises and interesting developments \cite{Hori:2006dk, Hori:2011pd, Hori:2013gga, Benini:2013nda,  Benini:2014mia, Gaiotto:2015aoa, Aharony:2016jki, Gomis:2016sab}.

The BPS spectra of a certain type of surface defects, known as \emph{canonical defects}  in theories of class $\CS$, can be computed with geometric techniques, involving a construction known as spectral networks \cite{Gaiotto:2012rg}. 
Previous studies focused on surface defects of 4d $\mathcal{N}=2$ class $\CS$ theories of $\gA_n$-type, which arise from the worldvolume theory of M2 branes ending on M5 branes \cite{Gaiotto:2011tf, Longhi:2012mj, Hori:2013ewa}. 
However much less is known about surface defects associated with other simply-laced Lie algebras. For example, an M2-M5 brane description is not available for canonical surface defects of $\gE$-type theories.
It is nevertheless natural to attempt to extend the field theoretic interpretation of spectral networks to these settings, i.e.\ to interpret network data in terms of soliton counting in presence of surface defects.
In \cite{Longhi:2016rjt} we advanced an argument in favor of this hypothesis, showing that it passes a nontrivial consistency check, relating 2d-4d BPS spectra to 4d BPS spectra through the 2d-4d wall-crossing phenomenon of \cite{Gaiotto:2011tf}.

This paper is a companion to \cite{Longhi:2016rjt}, where we focused on the formal definition of ADE spectral networks. In this paper we study aspects of their field theoretic interpretation.
By adopting suitable decoupling limits, in which the low energy dynamics of a surface defect is well-described by a 2d $\CN = (2,2)$ effective theory, we study predictions of spectral networks about the BPS spectra of purely 2d models. 
On the one hand, this provides nontrivial checks of the field theoretic interpretation of spectral networks when compared to known results. On the other hand, we show how spectral networks can be used as a powerful tool to compute new 2d BPS spectra for theories which have not been studied previously. 
Decoupling limits of surface defects have been previously discussed in \cite{Gaiotto:2011tf} and \cite{Hori:2013ewa}, within broader contexts.
The present paper aims to extend these previous analyses to $\gA\gD\gE$ theories, and to employ such limits to support the field theoretic interpretation of $\gA\gD\gE$  spectral networks.
On the practical level, an obstacle that this program faces is the sheer degree of complexity of a generic spectral network.
A convenient approach for overcoming these difficulties is to tune the moduli of a theory (for example the twisted masses of a 2d $\mathcal{N} = (2,2)$ model) in such a way that the Seiberg-Witten curve acquires an enhanced discrete symmetry.
In this case the topology of the corresponding spectral network assumes a simplified, although degenerate form.  
Early examples of such degenerate spectral networks can be found in \cite{Gaiotto:2012rg}, as well as in applications to the study of 2d models in \cite{Hori:2013ewa}. More recently a further development on this theme appeared in \cite{Hollands:2016kgm}, where it has been put to effective use for the study of the $\gE_6$ Minahan-Nemeschansky theory.
Using spectral networks, we focus on two particular low-energy decoupling limits of surface defects: a \emph{semiclassical} and a \emph{superconformal} limit.%
\footnote{Renormalization and decoupling do not commute in general, as recently observed in \cite{Aharony:2016jki} for 2d $\mathcal{N}=(2,2)$ models. This phenomenon occurs for 2d-4d theories of surface defects. For definiteness, we first consider the low energy 2d effective theory and then study a decoupling limit of that.}

The semiclassical limit corresponds to a region of the 4d Coulomb branch where BPS states of the 4d theory become very heavy. In this limit we are able to identify a sub-sector of the 2d-4d BPS spectrum of solitons supported by the defect, whose mixing with the 4d spectrum is suppressed. It is this subset of BPS states which can be identified with the BPS spectrum of a 2d $\mathcal{N} = (2,2)$ model.
In Section \ref{sec:spectral_networks_from_field_theory} we carry out a detailed analysis of canonical defects of 4d $\mathcal{N} = 2$ pure gauge theories with gauge group $G = \SU(3)$, $\SO(8)$ in the semiclassical limit.
In the case of $G = \SU(3)$ we recover the $\ICP^2$ sigma model, and we are able to match exactly the spectrum and chamber structure with field theoretic results obtained by \cite{Bolokhov:2012dv,Dorey:2012dq}.
The decoupling limit of the $G = \SO(8)$ defect does not have a known field theoretic description, to the best of our knowledge. In this case we provide a detailed prediction of the 2d spectrum and chamber structure of this theory. 
To compute the spectrum of the $\SO(8)$ theory we develop novel ``soliton traffic rules''. 
The underlying principle from which these rules follow is \emph{twisted homotopy invariance} of the formal parallel transport, a defining property of spectral networks \cite{Gaiotto:2012rg}. 
The new traffic rules greatly simplify the task of computing 2d-4d soliton spectra, enabling us to study in detail both the spectrum and the chamber structure of 2d wall crossing.
It would be interesting to check our results with a first-principles field theoretic computation.

The superconformal limit corresponds to tuning the couplings of the surface defect to a region where a subset of the 2d-4d states become arbitrarily light. 
As in the semiclassical limit, the large mass hierarchy between this sector and the masses of 4d BPS states suppresses the effects of 2d-4d mixing, and we are able to identify the light states with the spectrum of a 2d $\CN=(2,2)$ model.
Due to the nature of canonical defects, the masses of their 2d-4d BPS states depend on a choice of point $z\in C$ on the UV curve of the class $\CS$ theory. 
Superconformal points correspond to choosing $z$ near a branch point of the Hitchin spectral curve, whose ramification structure and location on the UV curve are determined by a choice of 4d Coulomb vacuum. 
In Section \ref{sec:superconformal_limit} we analyze superconformal limits of surface defects in 4d $\mathcal{N} = 2$ pure gauge theories of $\gA\gD\gE$ types, associated with minuscule representations. We find exact agreement between the spectrum of light 2d-4d solitons and the BPS spectrum of Landau-Ginzburg models describing most-relevant deformations of Kazama-Suzuki coset models. 
A superconformal limit of a canonical defect is \emph{a priori} distinct from the semiclassical limit described previously. Nonetheless, semiclassical limits of defect typically admit a further specialization to a superconformal limit: in these instances the relation between the two can be understood in the context of the GLSM-LG correspondence \cite{Witten:1993yc}.

The paper is organized as follows.
Section \ref{sec:background}  collects useful background on the field theoretic description of surface defects of 4d $\mathcal{N}=2$ theories, and contains a general description of the suppression of 2d-4d mixing of BPS spectra.
In Section \ref{sec:decoupling_limits_SYM}  we describe the two decoupling limits on which we focus: the semiclassical and superconformal limits.
Section \ref{sec:semiclassical} contains detailed analyses of semiclassical limits of defects for 4d $\mathcal{N} = 2$ pure gauge theories with gauge group $G = \SU(3)$, $\SO(8)$. In Section \ref{sec:superconformal_limit} we study the superconformal limit of the defects of 4d $\mathcal{N} = 2$ pure gauge theories with gauge group $G = \SU(N)$, $\SO(2N)$, $\gE_6$, $\gE_7$, and discuss evidence for their identification with Landau-Ginzburg models.

\section{Decoupling limits of 2d-4d BPS spectra of surface defects} \label{sec:background}
In this paper we will be mostly interested in regimes where surface defects are described by purely 2d theories, and where 2d-4d BPS spectra are well-approximated by 2d BPS spectra. 
BPS spectra of surface defects differ from those of purely 2d theories due to the \emph{2d-4d wall crossing phenomenon} of \cite{Gaiotto:2011tf}, which describes the mixing of 2d-4d BPS spectra and 4d BPS spectra.
In order to suppress the effects of 2d-4d wall crossing, we will tune the moduli of a coupled 2d-4d system to make 4d BPS states very heavy, and will restrict our attention to those 2d-4d BPS states that are much lighter than any 4d BPS state.

\subsection{Some background on coupled 2d-4d systems}
A prototypical example of 2d-4d system is a 2d $\mathcal{N} = (2,2)$ $U(1)$ gauged linear sigma model (GLSM) with a global symmetry $G$, coupled to the vector multiplet of a 4d $\CN=2$ gauge theory with gauge group $G$.
A large family of surface defects can be defined in this way, by varying the choice of 2d gauge and chiral matter content as well as the choice of how 2d chiral fields couple to the 4d gauge fields \cite{Gaiotto:2011tf, Gaiotto:2013sma}.
When the 4d theory is in a Coulomb branch vacuum, the vevs of 4d vector multiplet scalars play the role of twisted masses for the 2d chiral multiplets, lifting the latter.\footnote{2d chiral fields may gain twisted masses generated by three different sources: the 4d vector multiplet scalars, purely 2d twisted mass parameters $m$, and the 2d vector multiplet scalar $\sigma$. We will not consider purely 2d twisted masses.}
Integrating out their degrees of freedom yields an effective action for the 2d U(1) vector multiplet scalar $\sigma$: a 2d chiral multiplet in representation $\CR$ of $G$  contributes to the 2d effective twisted superpotential a term
\be
	2\pi i\tCW_\CR = -\Tr_\CR\,(\sigma+\Phi) \log[(\sigma+\Phi)/e]\,,
\ee
where $\Phi$ is the scalar field in the 4d vector multiplet. 
The vacuum expectation value of $\Phi$ is parameterized by Coulomb branch moduli $u_i$, corresponding to the Casimirs of $\mathfrak{g} = \mathrm{Lie}(G)$.
The 2d superpotential also includes an FI-$\theta$ term, which reads
\be
	2\pi i \tCW_{\text{FI},\theta} = t \sigma\,.
\ee

To capture the interaction between 2d and 4d degrees of freedom, it was proposed in \cite{Gaiotto:2013sma} to evaluate $\tCW_\CR$ by replacing $\Tr_\CR\Phi^k$ with suitable functions of the Coulomb moduli, as we will review in a moment. 
The twisted chiral ring equations for the surface defect can then be derived by extremizing the 2d effective twisted superpotential 
\begin{align}
	\tCW_{\eff} = \tCW_\CR+\tCW_{\text{FI},\theta} 
\end{align}
with respect to $\sigma$. 
A key observation of \cite{Gaiotto:2013sma} is that these equations take the following form
\be\label{eq:chiral-ring-resolvent}
	t  = \int d\sigma \left\langle\Tr_\CR\,\frac{1}{\sigma+\Phi}\right\rangle\,,
\ee
and for certain 4d theories the integrand may be identified as the resolvent of the corresponding matrix model \cite{Dijkgraaf:2002dh}.

In this paper we shall be interested in certain decoupling limits of 2d-4d systems, and for our purposes it will be sufficient to take the 4d theory to be a pure gauge theory with simply-laced gauge group $G$.
When $G = \SU(N)$ or $\SO(2N)$ and $\CR$ is the first fundamental representation, 
the resolvent can be computed explicitly (see for example  \cite{Cachazo:2002ry, Feng:2002gb, Nekrasov:2003rj})
\be\label{eq:AD-resolvents}
\begin{split}
	\gA_{N-1}\,:&\qquad \frac{P'_N(\sigma)}{\sqrt{P^2_N(\sigma) - 4 \Lambda^{2N}}} \\
	\gD_N\,:&\qquad \frac{P'_{2N}(\sigma) - \frac{2}{\sigma}P_{2N}(\sigma)}{\sqrt{P^2_{2N}(\sigma) - 4 \sigma^4 \Lambda^{4N-4}}}-\frac{2}{\sigma}
\end{split}
\ee
where $P_N$ and $P_{2N}$ are polynomials of degrees $N$ and $2N$, and the latter is an even function of $\sigma$.
The coefficients of these polynomials correspond to Coulomb branch moduli, and the chiral ring equations can be cast into the standard forms 
\cite{Martinec:1995by}
\be\label{eq:AD-chiral-rings}
\begin{split}
	\gA_{N-1}:&\qquad \left( \Lambda^{N} z + 2u_N + \frac{\Lambda^{N}}{z} \right)  = P_N(\sigma)\,,\qquad z = {e^t}, \\
	\gD_{N}:&\qquad  \left(\Lambda^{2N-2} z + 2u_{2N-2} + \frac{\Lambda^{2N-2}}{z}\right) \sigma^2 = P_{2N}(\sigma)\,,\qquad z = {\sigma^2 \, e^t},
\end{split}
\ee
where the Coulomb moduli of highest degree $u_{h^\vee}$ have been separated from $P_k(\sigma)$ for later convenience.
For generic $t$, there is a finite number of massive vacua $\sigma_i(t)$: respectively $N$ and $2N$.

For $\gE$-type gauge theories there does not seem to be a known matrix model description, to the best of our knowledge. Nevertheless the right hand side of (\ref{eq:chiral-ring-resolvent}) should have a meaning in the 4d gauge theory, and it may be possible to compute it by other techniques, for example from direct analysis of the generalized Konishi anomaly \cite{Cachazo:2002ry}.
As a working hypothesis, we will assume that the chiral ring equation of an $\gE$-type theory coincides with its Seiberg-Witten curve written in the form of a spectral curve \cite{Martinec:1995by,Longhi:2016rjt}. As we have just reviewed, this is indeed the case for  $\gA$- and $\gD$-type theories.%
\footnote{
One test of this assumption is that the 2d-4d BPS spectrum obtained via spectral networks from the spectral curve equation reproduces correctly the 4d BPS spectrum of the gauge theory \cite{Longhi:2016rjt}.
}
The spectral curves for $\gE_6 $ and $\gE_7$ pure gauge theories are
\be\label{eq:E-chiral-rings}
\begin{split}
	\gE_{6}:& \qquad \sigma^3 \left( z\Lambda^{12} + 2u_{12}  + \frac{\Lambda^{12}}{z}\right)^2  + P_{15}(\sigma) \, \left(  z\Lambda^{12} + 2u_{12}  + \frac{\Lambda^{12}}{z} \right)  + P_{27}(\sigma) = 0\\
	\gE_{7}:&\qquad \sigma^2 \Bigg[\left( z\Lambda^{18} + 2u_{18}  + \frac{\Lambda^{18}}{z} \right)^3  + P_{18}(\sigma) \, \left( z\Lambda^{18} + 2u_{18}  + \frac{\Lambda^{18}}{z} \right)^2   \\
	&\qquad \qquad\qquad \qquad\qquad \qquad\qquad  + P_{36}(\sigma) \, \left( z\Lambda^{18} + 2u_{18}  + \frac{\Lambda^{18}}{z} \right)  + P_{54}(\sigma) \Bigg] = 0\,,
\end{split}
\ee
where $P_N(\sigma)$ is a homogeneous polynomial of maximum degree $N$, counting the weight of $u_i$, the $i$-th Casimir of $\Phi$, as $i$ and the weight of $\sigma$ as 1.

4d $\mathcal{N} = 2$ pure gauge theories of ADE types are theories of class $\CS$.
As such, they admit a distinguished type of surface defects, termed \emph{canonical surface defects}. 
A detailed discussion for the canonical defect of 4d $\mathcal{N} = 2$ pure SU(2) gauge theory can be found in \cite[\S 8.3]{Gaiotto:2011tf}.  
More generally, canonical defects of $\gA_n$-type class $\CS$ theories have been studied in the language of M2 branes ending on M5 branes in \cite{Gaiotto:2009fs, Gaiotto:2011tf, Hori:2013ewa}. 
A defining property of canonical defects is that they exhibit a parameter space of UV couplings which coincides with $C$, the UV curve characterizing the 4d class $\CS$ theory.
Recall that, in the case of 4d $\CN=2$ pure gauge theory, $C$ has the topology of a cylinder.
For generic $z\in C$ and generic vacuum $u$ in the 4d Coulomb branch, the defect has a finite number of massive vacua. 
Each vacuum corresponds to a preimage of $z$ on the spectral curve $\Sigma$ of the class $\CS$ theory, which is naturally presented as a ramified covering of $C$ \cite{Gaiotto:2009hg}. 
Recall in fact that the spectral curve lives inside $T^*C$, and is defined by the equation
\be
    \det_\CR(\lambda-\varphi(z)) = 0\,.\label{eq:spectral_curve_definition}
\ee
The 2d massive vacua at $z\in C$ are determined by the canonical projection map $\pi:T^*C\to C$.
The number and positions of the 2d vacua are further determined by a choice of representation $\CR$ \cite{Longhi:2016rjt}. 
When $\CR$ is the vector representation\footnote{%
Here we call \textbf{27} of $\gE_6$ and \textbf{56} of $\gE_7$ their vector representations.
} of $\mathfrak{g}$, the spectral curve coincides with the Seiberg-Witten curve of the 4d theory. 
By virtue of (\ref{eq:AD-chiral-rings}) the spectral curve also coincides with the chiral ring of the a 2d $\UU(1)$ GLSM with chiral matter in the $\CR$ representation coupled to the 4d vector multiplet. In other words, the canonical defects corresponding to $\CR$ admit a description in terms of such a gauged linear sigma model, at least for $G = \SU(N)$ or $\SO(2N)$.

This identification means that the tools available for studying 2d-4d BPS spectra of canonical defects can be used to study the BPS spectra of these GLSMs in suitable limits.
In particular spectral networks are a powerful tool for studying degeneracies of such 2d-4d BPS spectra. Below we shall illustrate how they adapt naturally to the study 2d BPS spectra.
Spectral networks were introduced in \cite{Gaiotto:2012rg} for class $\CS$ theories of A-type, and later extended in \cite{Longhi:2016rjt} to D- and E-type theories, with spectral curves in minuscule representations. 
A choice of representation for a spectral curve was associated in \cite{Longhi:2016rjt} with a type of surface defects. 
The main motivation behind this proposal was the physical interpretation of spectral networks in terms of the 2d-4d wall-crossing phenomenon\footnote{More precisely, its \emph{framed} version.} of \cite{Gaiotto:2011tf}, which captures the intricate interplay between 2d and 4d BPS spectra. 
Surface defects ``probe" the BPS spectrum of the 4d theory, and the latter is entirely encoded into the former. 
As a consequence, different surface defects for the same 4d theory must all encode the same 4d BPS spectrum. In this sense 2d-4d wall crossing imposes consistency constraints on the BPS spectra of different surface defects. 
A nontrivial check of the construction of \cite{Longhi:2016rjt} was indeed to prove that the 2d-4d soliton data encoded by spectral networks in different  representations of $\mathfrak{g}$ complies with these constraints.

\subsection{Suppressing 2d-4d mixing of BPS spectra}\label{subsec:suppressing-2d-4d-mixing}

Here we briefly review 2d and 2d-4d BPS solitons and define the decoupling limit in which the mixing of 2d-4d and 4d BPS spectra is suppressed.
The 2d effective action for the 2d $U(1)$ vector multiplet is (see for example \cite{Witten:1993yc})
\be
	\frac{1}{4} \int d^4\theta K_\eff(\Sigma, \Sigma^\dagger) + \left( \int d^2\tilde\theta \, \tCW_{\eff}(\Sigma)\, + \, c.c.\right)
\ee
where $\Sigma$ is the twisted chiral multiplet whose lowest component is the $\UU(1)$ field strength scalar.
The ground state energy density is 
\be\label{eq:potential}
	U_{\eff} = \frac{1}{2} e_{\eff}^2 \left|  \widehat{\partial_\sigma \tCW_{\eff}} \right|^2
\ee
where $(2 e^2_{\eff})^{-1} = \partial_\sigma\partial_{\bar\sigma}K_\eff$, and the hat denotes the choice $\hat\theta^2 = \min_{n\in\IZ}\{(\theta-2\pi n)^2\}$ for
\be
	t = i r + \frac{\hat\theta}{2\pi}\,.
\ee
It follows from (\ref{eq:potential}) that vacua of the 2d theory coincide with critical points of the twisted superpotential.
At fixed $t$, there is a discrete and finite set of vacua $\sigma_i(t)$ labeled by $i=1,\dots, d$, where $d = \dim \CR$. 

In general there may be abelian global symmetries surviving in the low energy 2d theory. Let $S_k$ be the corresponding charges, and $\tm_k$ the corresponding twisted masses. The central charge of a 2d-4d soliton interpolating between vacua $\sigma_i$ and $\sigma_j$ and carrying flavor charges $S_k$ is \cite{Hanany:1997vm, Gaiotto:2011tf}
\be\label{eq:central-charge}
	\tZ_{ij,\vec S} = 4 (\tCW_\eff(\sigma_j) - \tCW_\eff(\sigma_i)) + \sum_k \tm_k S_k\,,
\ee
and 2d-4d BPS states saturate the Bogomolny bound $M \geq |\tZ|$.
We will take the 4d bulk theory to be a pure gauge theory with gauge group $G$. Then the global abelian symmetries of the 2d low energy theory will coincide with the Cartan torus of $G$, the residual 4d gauge symmetry at generic points on the 4d Coulomb branch. 
2d flavor charges $\vec S$ correspond to 4d electromagnetic charges, understood in a suitable choice of electromagnetic duality frame, while the twisted masses $\tm_i$ are identified with the corresponding electric and magnetic periods of the Seiberg-Witten differential.

The BPS soliton equation for a 2d $\mathcal{N} = (2,2)$ theory is \cite{Cecotti:1992rm}
\be\label{eq:2d-BPS-eqn-W-plane}
	\partial_{x_1} \sigma = \alpha\, \overline{\partial_\sigma\tCW}_{\eff}
\ee
where $\partial_{x_1}$ is the derivative along the spatial direction of the 2d worldsheet of the defect. $\alpha$ is a phase fixed by the topological sector of the soliton, it is
\be
	\alpha = \frac{\tCW_{\eff}(\sigma_j) -\tCW_{\eff}(\sigma_i)  }{|\tCW_{\eff}(\sigma_j) -\tCW_{\eff}(\sigma_i)  |}
\ee
for a soliton interpolating between vacua $\sigma_i$ and $\sigma_j$.

To disentangle the effects of the coupling to 4d dynamics we consider a mass filtration of the 2d-4d BPS spectrum, retaining only those states that are much lighter than the lightest 4d BPS soliton
\be\label{eq:2d-BPS-limit}
	 \big|\tCW_\eff(\sigma_j) - \tCW_\eff(\sigma_i) \big|   \ll M_0 \,.
\ee
2d-4d wall crossing can occur whenever the phases of central charges of 4d BPS states and 2d-4d BPS states align $\arg Z_{\gamma} = \arg \tZ_{ij,\vec S}$.
At fixed $u\in\CB$, this happens at certain special loci $z\in C$ (recall that $\tCW$ depends on $\sigma_i(t)$, and that $t$ is related to $z$ by a change of variable).\footnote{%
In the language of spectral networks, these loci correspond to \emph{two-way streets}, appearing at \emph{$\CK$-wall jumps} of a network. Note that such streets end on branch points, where the mass of a 2d-4d state goes to zero, for this reason such loci are generically open curves on $C$.
} Across these loci, the spectrum of 2d-4d states can jump due to the appearance or disappearance of states with shifted flavor charges $\vec S+\gamma$, carrying mass $|\tZ_{ij,\vec S+\gamma}|\simeq |\tZ_{ij,\vec S}|+|Z_{\gamma}| \geq M_0$.
In the limit (\ref{eq:2d-BPS-limit}) the effects of 2d-4d wall-crossing become negligible, if we restrict attention to light 2d-4d BPS states. In this sense the light 2d-4d BPS spectrum of the defect resembles that of a purely 2d field theory.

\subsection{Spectral networks from a field-theoretic perspective}\label{sec:spectral_networks_from_field_theory}
The spectrum of 2d-4d BPS states in the presence of a surface defect is naturally computed by spectral networks. 
Therefore in the limit (\ref{eq:2d-BPS-limit}) spectral networks should compute the BPS spectrum of a 2d theory describing the defect.
An example of such a limit is when $t$ is chosen such that a number of vacua of the 2d theory collide together, and arbitrarily light BPS solitons appear. 
In this regime we can see how spectral networks arise naturally by studying the BPS equation. This viewpoint has been previously emphasized in \cite{Gaiotto:2011tf}, here we review and further elaborate on it.
The emergence of spectral networks within a purely field theoretic framework will be especially relevant for studying $\gE$-type theories, for which a brane interpretation of surface defects and 2d-4d solitons is currently unavailable. 
The use of spectral networks for computing 2d BPS spectra in these cases will be mainly motivated by their field theoretic origin.

To illustrate the point we consider the $\ICP^1$ model coupled to 4d $\mathcal{N}=2$ pure $\SU(2)$ gauge theory \cite{Gaiotto:2011tf, Gaiotto:2013sma}. Setting the dynamical scale of the 4d theory $\Lambda=1$ for simplicity, the vacua are determined by
\be
	\sigma =\pm \sqrt{u+2\cosh(t)},
\ee
where $u$ is the Coulomb branch parameter of the 4d theory. When $u=0$,\footnote{Recall that $u=0$ is a generic point in the strong coupling regime, not a singular point of the 4d theory. The conclusions do not depend on the specific choice of $u$.} the critical points are at $t_\pm = \pm i \pi/2$. 
Introducing a parametrization $t = \log(i+z/2)$ near $t_+$, the 2d superpotential can be obtained by integrating the chiral ring equation $\partial_\sigma\tCW$. 
Note that critical points are at $\sigma=0$. Expanding around small $z$ and small $\sigma$ gives
\be
	\tCW_\eff = \frac{\sigma^3}{3} - z\sigma +\dots
\ee
up to an additive constant and a positive rescaling.
Near $z=0$ the two vacua of the superpotential are 
\be
	\sigma_\pm = \mp\sqrt z\,,\qquad \tCW_\pm = \pm\frac{2}{3}z^{3/2}\,.
\ee
Looking back at (\ref{eq:2d-BPS-eqn-W-plane}), it is easy to see that 2d BPS equations imply that a 2d soliton configuration must project to a straight line in the $\tCW$ plane \cite{Cecotti:1992rm}. This observation was used to compute the number of solitons by studying certain wave fronts in the $\sigma$-plane.
In the case at hand, the wavefronts can be constructed explicitly by expanding around the vacuum at $\sigma_+ = -\sqrt{z}$
\be
	\tCW_\eff(-\sqrt{z} + y) = \tCW_{+} - \sqrt{z}\, y^2 + \dots
\ee
The BPS equation in the neighborhood of this vacuum is 
\be
	\partial_{x_1} \sigma = %
	\partial_{x_1} y = %
	\alpha \, \overline{\partial_y {\tCW}} %
	= -2 \alpha \sqrt{ \bar z}\, \bar{y}\,,
\ee
where $\alpha = \Delta \tCW / |\Delta \tCW |$. 
Noting that $\alpha \sqrt{\bar z} \sim z / |\sqrt z|$, we choose to fix $z$ to be real and positive for simplicity. 
Defining $f(z) = -2 z / |\sqrt{z}|$, the BPS equation becomes $\partial_{x_1} y = f(z) \bar y$, whose solution is
\be\label{eq:soliton-thimble}
	y = c \exp(f(z) x_1),
\ee
where $c$ is an undetermined real constant. Either sign choice for $c$ gives a component of the wavefront,
which consists of two points near $\sigma_+$, in fact the wavefront is a zero-dimensional sphere. 
For example, asking for the loci where $|\tCW(y) - \tCW_+| =  \epsilon^2$ intersects the wavefront, we find $y = \pm\epsilon / z^{1/4}$. A similar construction gives the wavefront near $\sigma_-$. There is precisely one intersection of these two wavefronts, and it occurs along the straight segment connecting $\sigma_+$ to $\sigma_-$ on the real axis because we chose $z$ to be real and positive. 
The intersection of the wavefronts captures the presence of one soliton interpolating the two vacua.

A simple change of perspective shows how spectral networks emerge. 
To follow the growth of a wavefront, one may consider the intersection of (\ref{eq:soliton-thimble}) with the locus $|\tCW_\eff^{-1}(\tCW_+ \pm \epsilon)|$ as we grow $\epsilon$. 
On the other hand one may as well choose to vary $z$ instead of $\epsilon$, e.g. taking $z\to 0$ along the positive real axis in such a way that $\tCW_\pm$ move towards each other, along the segment that connects them. 
This is equivalent to ``zooming in'' into the soliton on the worldsheet of the surface defect, because the new soliton configuration that one finds interpolates between two new critical values $\tCW'_\pm$, and these are just intermediate values for the field configuration of the original soliton.
From the viewpoint of (\ref{eq:soliton-thimble}) the situation is similar: we are not changing the differential equation itself, but only the boundary conditions, and the new boundary conditions are chosen along the flow of the previous equation, therefore the new soliton profile $\sigma(x_1)$ overlaps with that of the previous soliton. 
In other words, the wavefronts will travel along the same straight line between $\sigma_\pm$ as they did before, and the only difference is that the distance between $\sigma_\pm$ is now shorter.
Such a trajectory $z(\tau)$ is characterized by the constraint
\be
	\frac{\d }{\d \tau}\Delta\tCW\, \sim - \Delta\tCW  \qquad \Leftrightarrow \qquad %
	{\dot z} \, \frac{\partial}{\partial z}\Delta\tCW\,   \sim - \alpha\,.
\ee
Identifying the Seiberg-Witten differential with $\lambda = x \, \d z/z$, where  $x_\pm = \partial_{\log z}\tCW(\sigma_\pm)$ (see e.g. \cite{Gaiotto:2009fs}), the above constraint reads then
\be
	(\partial_\tau, \lambda_+ - \lambda_-) \in e^{i\arg(\alpha)} \IR^{-}
\ee
which is precisely the $\CS$-wall equation describing the geometry of a spectral network \cite{Gaiotto:2012rg}.
This simple argument can be generalized to various types of branch points for spectral curves for ADE class $\CS$ theories, recovering the generalized $\CS$-wall equations describing ADE spectral networks \cite{Longhi:2016rjt}.

Intersections of $\CS$-walls and the propagation of soliton content of spectral networks are determined by flatness equations for a formal parallel transport, which are known  to coincide with the Cecotti-Vafa wall crossing formula for 2d BPS solitons \cite{Gaiotto:2012rg}. 
Therefore the emergence of spectral networks from a 2d field theory analysis is expected to extend globally, away from branch points.

\section{Decoupling limits for defects of pure gauge theories}\label{sec:decoupling_limits_SYM} 
For the rest of the paper we will restrict our focus to surface defects of pure gauge theories. We describe two interesting decoupling limits, that will be studied in greater detail in the rest of the paper.

\begin{figure}[ht]
\begin{center}
\includegraphics[width=0.30\textwidth]{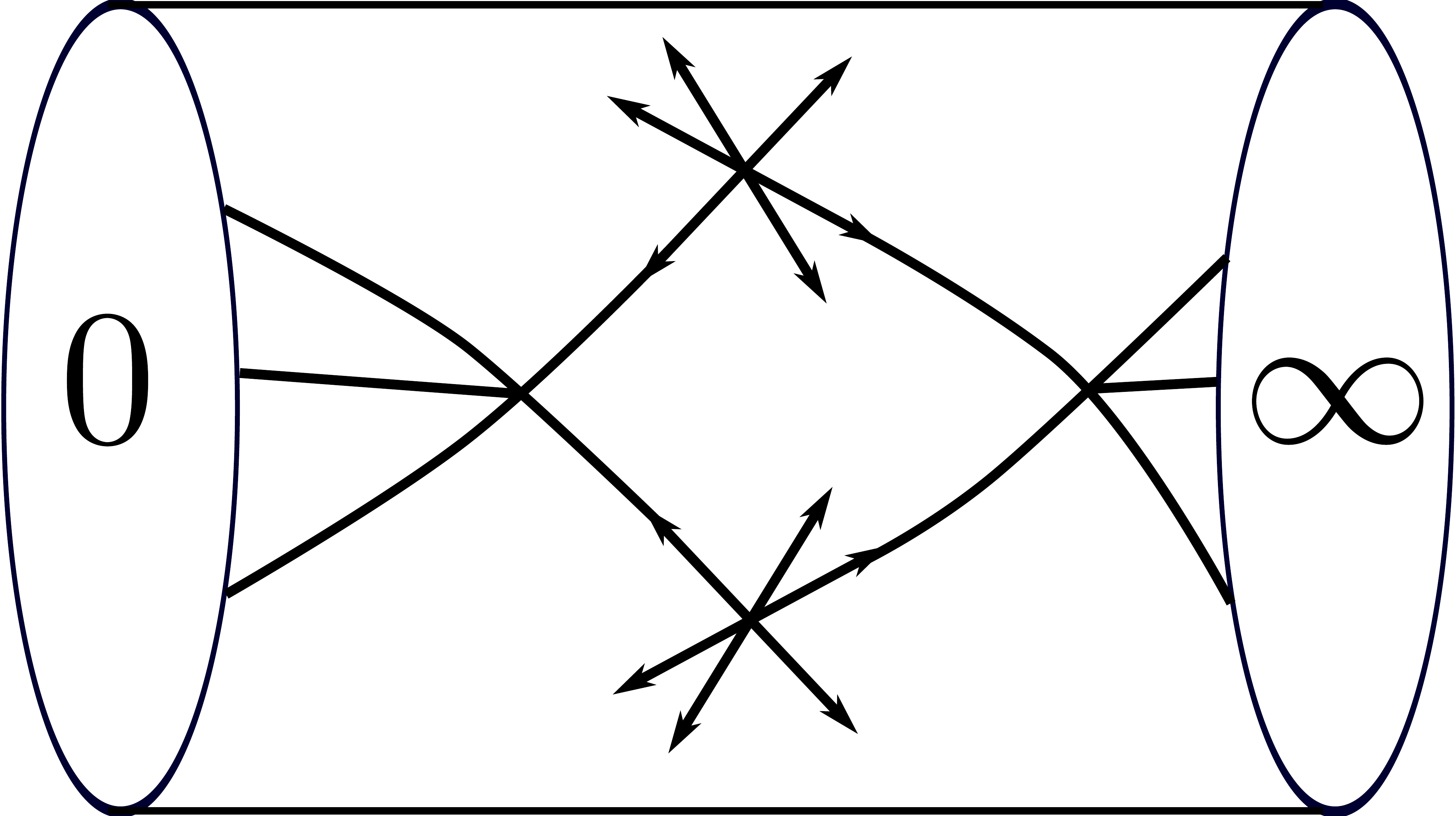}
\caption{A cartoon of the spectral network of pure gauge theory with $\gA\gD\gE$ gauge group, at $u_i=0$.}
\label{fig:strong-coupling-SYM}
\end{center}
\end{figure}

Consider a canonical surface defect for 4d $\mathcal{N} = 2$ pure gauge theory of type $\mathfrak{g}$. 
The space of 2d vacua coincides with the spectral curve of the Class $\CS$ Hitchin system, whose explicit form has been recalled in (\ref{eq:AD-chiral-rings}), (\ref{eq:E-chiral-rings}).  
Both the effective 2d low energy dynamics and the 2d-4d BPS spectrum depend on the 4d Coulomb branch vacuum, parameterized by Casimirs $u_i$. 
We will choose to work on a slice of complex-dimension one, by setting $u_{i<h^\vee}=0$, and leaving $u_{h^\vee}$ as the only coordinate along this slice. 
The polynomials $P_k(\sigma)$ in (\ref{eq:AD-chiral-rings}), (\ref{eq:E-chiral-rings}) then acquire a very simple form
\begin{align}
	P_k(\sigma) \sim \sigma^k
\end{align}
up to an overall multiplicative constant, which is a common feature to all choices of $\fg$.
Branch points of the spectral curve's projection map $\Sigma\to C$ coalesce into two groups, resulting in two distinct branch points of order $h^\vee$ at
\begin{align} 
	z=-u_{h^\vee} \Lambda^{-h^\vee}\pm\Lambda^{-h^\vee} \sqrt{u_{h^\vee}^2-\Lambda ^{2 h^\vee}}.
\end{align}
Although some branch points have collided, this is not a singular limit for the spectral curve, as no cycle shrinks to zero for generic $u_{h^\vee}$.\footnote{%
However we do observe the presence of two singularities on the $u_{h^\vee}$-plane, located at $\pm\Lambda^{h^\vee}$.
}
At the boundaries of the UV curve $C$ there are two irregular singularities, located at $z=0$, $\infty$.
Spectral networks also acquire a simplified form on this slice of $\CB$, as sketched in Figure \ref{fig:strong-coupling-SYM}.
The particular case of $u_{h^\vee}=0$ was studied in detail in \cite[\S 5]{Longhi:2016rjt}. 
As noted there, the simple topology of spectral networks is owed to the fact that several $\CS$-walls overlap entirely, in turn this is due to the fact that $\Sigma$ enjoys an enhanced discrete symmetry, see for example Figure 35 of \cite{Longhi:2016rjt}.

Each $\CS$-wall of a spectral network carries combinatorial data, counting 2d-4d BPS states for a surface defect at $z$ on the wall's trajectory. 
At intersections of $\CS$-walls, additional $\CS$-walls may be generated. This phenomenon admits a physical interpretation in terms of 2d wall-crossing: the soliton data carried by the newborn walls counts the bound states of the incoming 2d-4d BPS solitons.
Therefore a spectral network neatly captures the evolution of the 2d-4d BPS spectrum across several regions of the parameter space $C$ of a canonical surface defect \cite{Gaiotto:2012rg}.

There are two distinguished decoupling limits of the type described in Section \ref{subsec:suppressing-2d-4d-mixing} on which we will focus: the semiclassical limit and the superconformal limit, both of which will be introduced shortly.
In either limit the effects of 2d-4d mixing between 2d-4d BPS stats and 4d BPS states become suppressed, and the spectral network describes the BPS spectrum of a 2d $\CN=(2,2)$ model.

\subsection{Semiclassical limit}\label{sec:semiclassical_limits_definition}

The semiclassical limit corresponds to the region at infinity in the $u_{h^\vee}$-plane.
At large $\left|u_{h^\vee}/\Lambda^{h^\vee}\right|$, one of the branch points moves towards the singularity at $z=0$ while the other moves towards $z=\infty$.
It is convenient to introduce a rescaled coordinate $\tilde z = z \, \Lambda^{h^\vee} / 2u_{h^\vee}$, which remains finite as we zoom into the $z\to\infty$ region of $C$. 
In this coordinate, one of the branch points ends up at $\tilde z=-1$ while the other merges with the irregular puncture at $\tilde z=0$, turning it into a regular puncture. 
The spectral curve becomes
\be\label{eq:weak-coupling-curves}
\begin{split}
	\gA_{N-1}:&\qquad {\tilde\sigma}^N - (1+ \tilde z) = 0\,, \\
	\gD_{N}:&\qquad {\tilde\sigma}^2 \left( {\tilde\sigma}^{2N-2} - (1+ \tilde z)\right) = 0\,, \\
	\gE_{6}:& \qquad {\tilde\sigma}^3 \left( 108\, {\tilde\sigma}^{24} + 540\, {\tilde\sigma}^{12} (1+\tilde z) - (1+\tilde z)^2 \right)  = 0\\
	\gE_{7}:&\qquad  {\tilde\sigma}^2 \left(    c_{54} \, \tilde\sigma ^{54}  + c_{36}\, \tilde\sigma^{36}  (1+\tilde z) + c_{18}\,\tilde\sigma^{18} (1+\tilde z)^2 + c_0  (1+\tilde z)^3    \right)   = 0
\end{split}
\ee
where we adopted a rescaling $\tilde\sigma = \sigma / (2 u_{h^\vee})^{1/h^\vee}$, and $c_i$ are numerical coefficients.\footnote{$c_{54}=633053478050784$, $c_{36}=4778352$, $c_{18} = -1808136 $, $c_0 = 8 $.} 
Each spectral curve is characterized by the presence of an irregular singularity at $\tilde{z}=\infty$, a regular singularity at $\tilde{z}=0$, and a branch point of order $h^\vee$ at $\tilde z=-1$. The branch point has a sheet monodromy corresponding to a Coxeter element of the Weyl group of $\fg$ The permutation of sheets is identified with a Coxeter element of the Weyl group of $\fg$ with a choice of trivialization which assigns a weight to each sheet of the curve, as described in \cite{Longhi:2016rjt}.

We call this the \emph{semiclassical limit} because it may be viewed as squeezing the cylinder $C$ infinitely long, and pinching it in the middle. This corresponds to decoupling the 4d gauge degrees of freedom \cite{Gaiotto:2009we}, and the regular puncture at $\tilde z=0$ emerging from such a manipulation carries flavor symmetry $G$.

Let us now explain in what sense this limit satisfies condition (\ref{eq:2d-BPS-limit}).
Before taking the limit, 4d BPS states of the gauge theories come from 1-cycles connecting the two branch points on $C$. The masses of 4d BPS states are the integrals of the Seiberg-Witten differential along 1-cycles connecting the ramification points above them, for more details see \cite[\S 5]{Longhi:2016rjt}. 
In the semiclassical limit one of the branch points moves infinitely far away from the other, ending up into the puncture at $\tilde z=0$. Therefore 4d BPS states become infinitely massive, as required by (\ref{eq:2d-BPS-limit}). 
All mixing between 2d-4d and 4d BPS spectra is thus suppressed, and the spectral network counts purely 2d BPS states in this limit.

For example, the spectral curve for $\fg=\gA_{N-1}$ coincides with the chiral ring equation of the $\ICP^{N-1}$ sigma model deformed by twisted masses \cite{Hanany:1997vm, Dorey:1998yh}.\footnote{More precisely, taking eq. (111) of \cite{Dorey:1998yh} and identifying $\tilde\Lambda^N = e^{2\pi i \tau(\mu)} \equiv w$, gives $\sigma^N - u_2  \sigma^{N-2} - \dots - u_N -w = 0$. Taking $u_{i<N}=0$ and rescaling $\tilde z = w / u_{N}$ recovers the curve in (\ref{eq:weak-coupling-curves}).} 
An exact 2d BPS spectrum of this theory was computed previously in \cite{Dorey:1998yh} for the regime where the effective FI parameter is large and negative, which corresponds to placing the defect at $\tilde z$ near $\infty$.\footnote{Because we identify $\tilde\Lambda^N = e^{2\pi i \tau(\mu)} \equiv \tilde z u_{N}$ and the regime studied in \cite[\S6]{Dorey:1998yh} is for $r \ll 0$ with $\tau = i r+\theta / 2\pi$.} 
As we will see in Section \ref{sec:semiclassical_A2}, spectral networks correctly reproduce the 2d BPS spectrum in this regime. In addition, spectral networks enable us to analyze wall-crossings of the 2d BPS spectrum and find them in all other regions of the complexified FI parameter space $C$. 
Previous studies of the $\ICP^1$ and $\ICP^2$ models using spectral networks can be found in \cite{Gaiotto:2011tf} and \cite[\S6.2.4]{Park:2013osa}, respectively.

We will study 2d BPS spectra and the chamber structure of semiclassical limits of surface defects in Section \ref{sec:semiclassical}.

\subsection{Superconformal limit}\label{sec:superconformal_limit_definition}
A superconformal limit corresponds to taking a surface defect located at $z \in C$ close to one of the branch points of order $h^\vee$. In this limit, condition (\ref{eq:2d-BPS-limit}) for decoupling 4d physics from 2d physics is satisfied if we focus on the spectrum of light 2d solitons whose masses vanish when $z$ coincides with the branch point. 
Spectral curves (\ref{eq:AD-chiral-rings}) and  (\ref{eq:E-chiral-rings}) reduce to
\be\label{eq:curves-scft-limit}
\begin{split}
	\gA_{N-1} : & \qquad \tilde\sigma^{N} - \tilde z = 0, \\
	\gD_{N} :& \qquad  \tilde\sigma^{2N} - \tilde z\, \tilde\sigma^{2} = 0, \\
	\gE_6 : & \qquad \tilde\sigma^{27} - 5 \,\tilde z\, \tilde\sigma^{15} - \frac{1}{108}{\tilde  z}^2 \, \tilde\sigma^{3} = 0,\\
	\gE_{7}: &\qquad  \tilde\sigma^{56} + \frac{2458}{3}  \tilde z \, \tilde\sigma^{38} + \frac{8371}{27} {\tilde z}^{2} \,\tilde\sigma^{20} + \frac{1}{729} {\tilde z}^{3} \, \tilde\sigma^{2} = 0
\end{split}
\ee
with $\lambda = \tilde\sigma \, \dd \tilde z$. 
Each of these curves is a local curve of a spectral curve of a class $\mathcal{S}$ theory around a ramification point with a maximal ramification index, and therefore can be obtained by expanding (\ref{eq:weak-coupling-curves}) and the Seiberg-Witten differential near $\tilde z = -1$.\footnote{
$\tilde z$ used here is not the same as the one used in (\ref{eq:weak-coupling-curves}), it is a local coordinate near $-1$.
}
Note that the spectral curves now have a single branch point of order $h^\vee$ at $\tilde z=0$ and a singularity at $\tilde z=\infty$, which implies that there is no 4d BPS state.
A physical realization of this defect was described in \cite{Tong:2006pa}, 
where the superconformal limit of a type $\gA_{N-1}$ defect is identified with the low energy description of a 2d vortex string in a deformed Argyres-Douglas theory.
This setup was cast into the context of class $\CS$ in \cite{Hori:2013ewa}, and similarly a type $\gD_{N}$ defect can be obtained from a surface defect of a 4d $\mathcal{N}=2$ Argyres-Douglas theory associated with $\mathfrak{g} = \gD_N$ \cite{Longhi:2016rjt}. 
In the following we will elaborate on the description of the limit in conjunction with spectral networks of class $\CS$ theory

\begin{figure}[t]
    \centering
    \includegraphics[width=.5\textwidth]{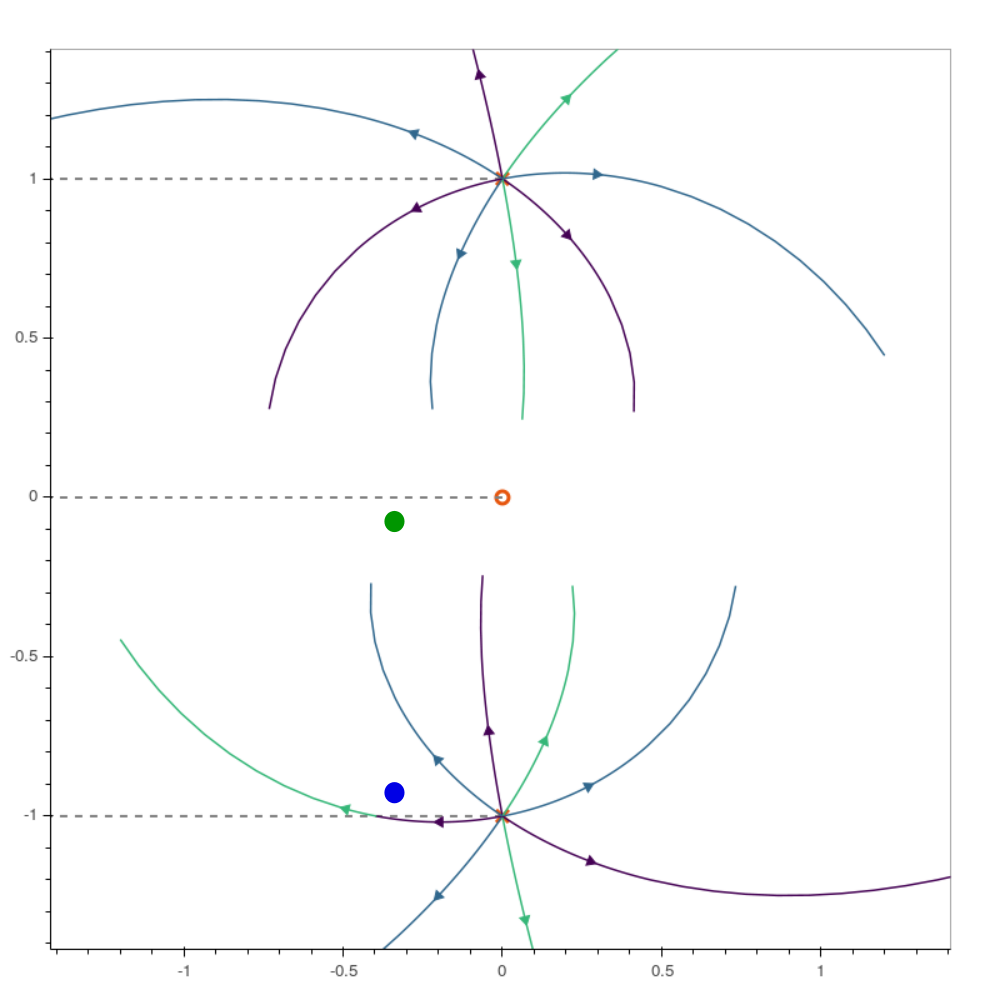}  
    \caption{With a sufficiently low mass cutoff $\scale$, the BPS spectrum of a defect away from a branch point becomes empty. When a defect is near a branch points, however, there will be arbitrarily light BPS states for any $\scale$.}
    \label{fig:su3-mass-filtration}
\end{figure}

For concreteness, consider a canonical surface defect that is coupled to 4d $\mathcal{N} = 2$ pure $\SU(3)$ gauge theory and is located at generic $z\in C$.
Introducing a mass cutoff $\scale$ and probing the 2d-4d BPS spectrum below this scale, we find that for a sufficiently low $\scale$ there is no 2d-4d BPS soliton with mass below $\Lambda$. 
Schematically, the mass of a 2d-4d BPS soliton interpolating between vacua $z_i, z_j$ is
\be
	\frac{1}{\pi} \left|\int_{z_i}^{z_j}\lambda\right|,
\ee
where $\lambda$ is the Seiberg-Witten differential and the integration path is determined by the topological charge of the soliton \cite{Gaiotto:2012rg}.
In particular, note that the path must travel from sheet $i$ to sheet $j$ and therefore must pass through one or more ramification points.
In fact, taking $z$ arbitrarily close to one of the branch points, we always find a fixed number of 2d solitons for arbitrarily low cutoff scale $\scale$.
This is reflected by the fact that $\CS$-walls emanate from branch points, see Figure \ref{fig:su3-mass-filtration} for an illustration.

In fact, bringing $z$ close to a branch point makes the 2d BPS states massless, and this limit corresponds to flowing the 2d theory to a SCFT \cite{Tong:2006pa,Gaiotto:2011tf,Hori:2013ewa}. 
There is a classification of such fixed points based on the spectrum of BPS states that become massless \cite{Cecotti:1992rm}. 
In the present example, the SCFT is $\gA_2$ minimal model.

Superconformal decoupling limits of canonical defects can be defined for any branch point of a generic spectral curve, here we restrict to degree $h^\vee$ branch points of pure gauge theories for illustrative purposes.
Superconformal limits are therefore \emph{a priori} distinct from the semiclassical limit described previously. Nevertheless, in the setup at hand where a superconformal limit can be ``embedded'' into a semiclassical limit, the relation between the two can be understood in the context of the GLSM-LG correspondence \cite{Witten:1993yc}.
We will return to BPS spectra of surface defects in superconformal limits in Section \ref{sec:superconformal_limit}.

\section{Semiclassical limits of canonical defects} \label{sec:semiclassical}

In this section we study two examples of the semiclassical limit of a canonical defect, as defined in Section \ref{sec:semiclassical_limits_definition}.
The first example will be the canonical surface defect of 4d $\mathcal{N}=2$ pure $\SU(3)$ gauge theory, whose semiclassical limit coincides with the $\ICP^2$ sigma model. 
We will analyze the spectrum and chamber structure in detail, and discuss the relation to results obtained via field-theoretic analysis, extending a previous study of the $\ICP^2$ model with spectral networks in \cite{Park:2013osa}.
The second example will be the canonical defect of 4d $\mathcal{N}=2$ pure $\SO(8)$ gauge theory, whose semiclassical limit does not seem to reproduce any previously studied model. We provide a detailed analysis of the BPS spectrum and chamber structure.
In both examples a convenient choice of Coulomb moduli confers an enhanced discrete symmetry to the spectral curve $\Sigma$. As a consequence we study spectral networks that have a simpler topology than generic spectral networks.  
The price to pay for the simpler topology is that the standard soliton equations of spectral networks need to be slightly generalized, as will become clear in the second example we study. 
The novelty arises from the fact that several $\CS$-walls can now overlap entirely, and the 2d wall-crossing occurring at joints of such degenerate walls is somewhat complex.
Other interesting alternatives for simplifying the study of complicated spectral networks appeared for instance in \cite{Hollands:2016kgm, Longhi:2016wtv}. To avoid potential confusion we stress that both are different from the spectral networks we study here.

\subsection{Semiclassical limit of the canonical defect of \texorpdfstring{$\SU(3)$}{SU(3)}  pure gauge theory}\label{sec:semiclassical_A2}
\begin{figure}[h!]
	\centering
	\begin{subfigure}{.32\textwidth}	
		\includegraphics[width=\textwidth]{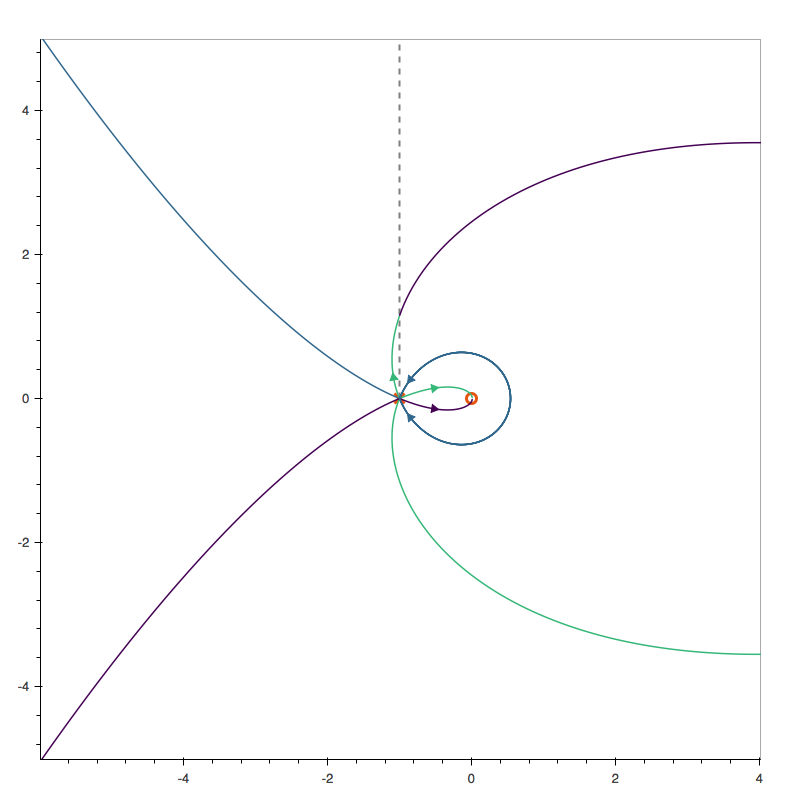}
		\caption{$\vartheta = 0$}
		\label{fig:CP2SN_1}
	\end{subfigure}
	\begin{subfigure}{.32\textwidth}	
		\centering
		\includegraphics[width=\textwidth]{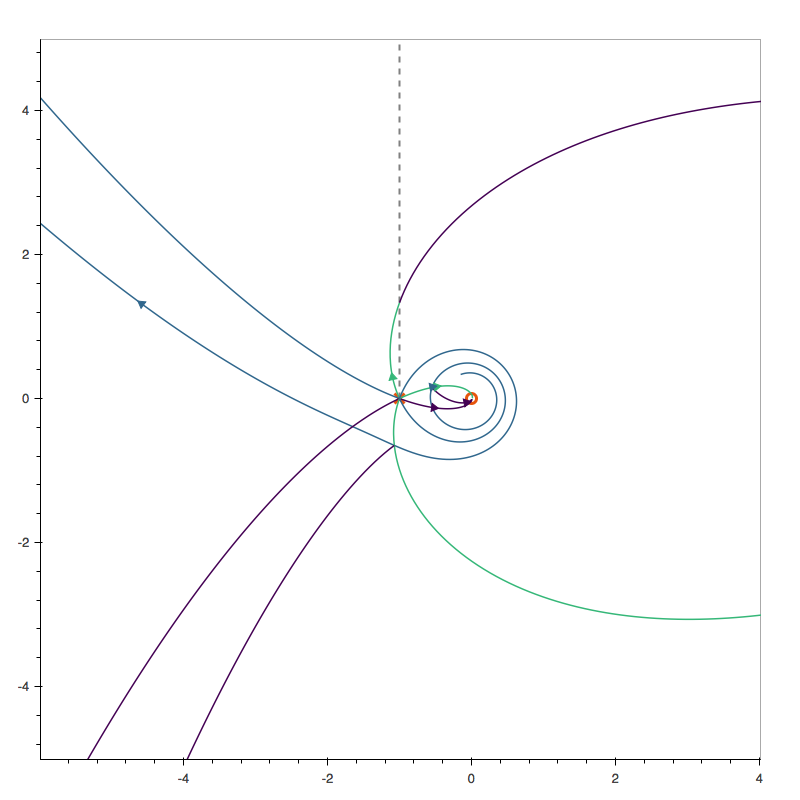}
		\caption{$\vartheta = \pi /60$}
		\label{fig:CP2SN_2}
	\end{subfigure}	
	\begin{subfigure}{.32\textwidth}	
		\includegraphics[width=\textwidth]{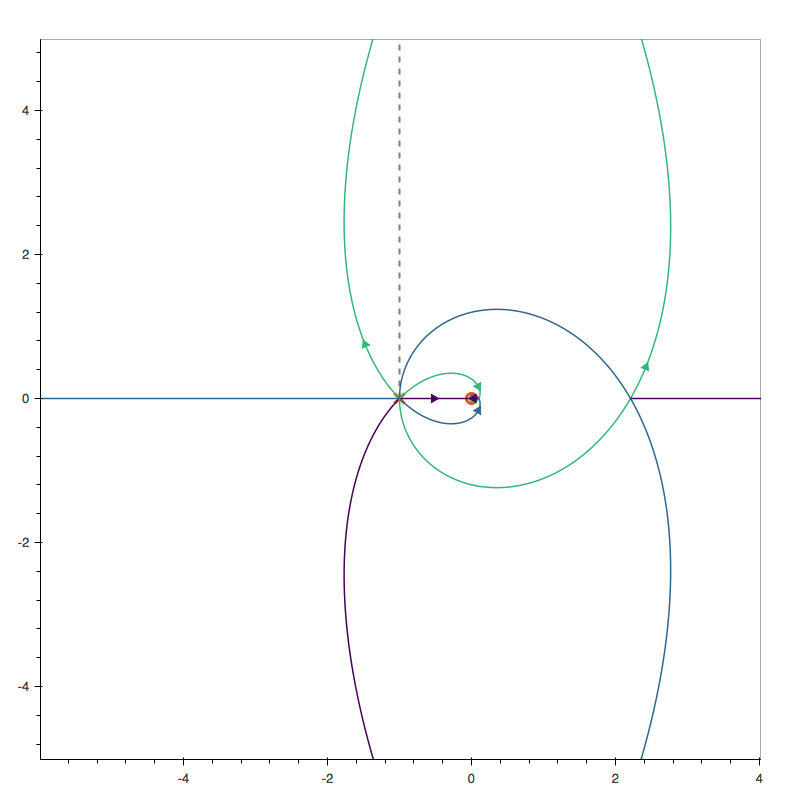}
		\caption{$\vartheta = 10 / 60 \pi$}
		\label{fig:CP2SN_3}
	\end{subfigure}	
	\begin{subfigure}{.32\textwidth}	
		\centering
		\includegraphics[width=\textwidth]{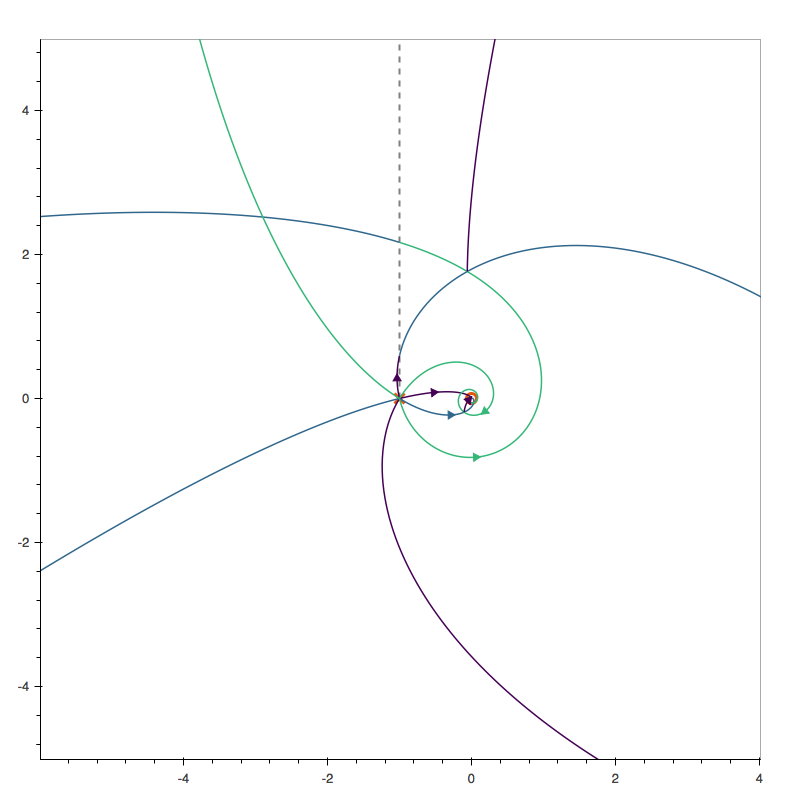}
		\caption{$\vartheta=16/60 \pi$}
		\label{fig:CP2SN_4}
	\end{subfigure}	
	\begin{subfigure}{.32\textwidth}	
		\centering
		\includegraphics[width=\textwidth]{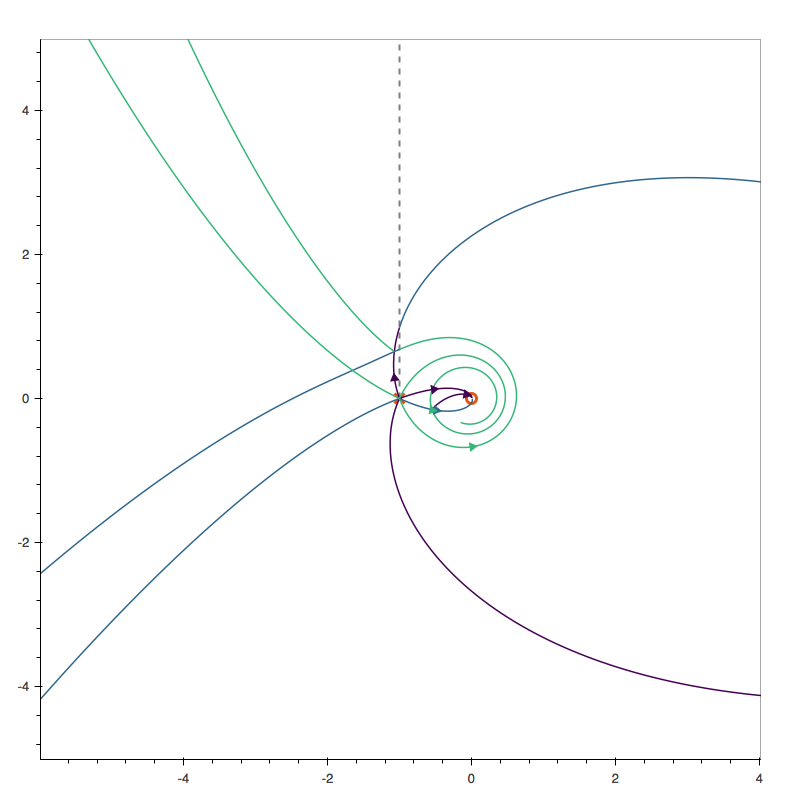}
		\caption{$\vartheta =19/60\pi$}
		\label{fig:CP2SN_5}
	\end{subfigure}		
	\begin{subfigure}{.32\textwidth}	
		\centering
		\includegraphics[width=\textwidth]{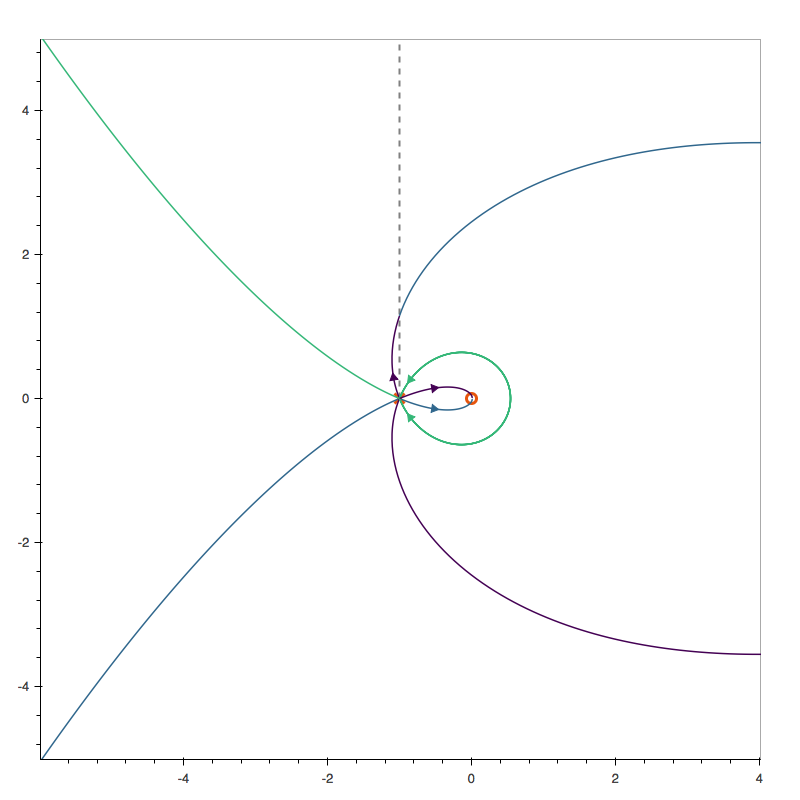}
		\caption{$\vartheta=20/60\pi$}
		\label{fig:CP2SN_6}
	\end{subfigure}	
	\caption{The spectral network for the $\ICP^2$ model. Purple walls carry solitons interpolating between vacua 1 and 2; green walls between 2 and 3; blue walls between 1 and 3.}
	\label{fig:CP2SN}
\end{figure}

The spectral curve equation follows from (\ref{eq:weak-coupling-curves}) with $\lambda = \tilde\sigma \, d\tilde z  / \tilde z$.
\be
	\lambda^3 = \left(  \frac{1}{\tilde z^3} + \frac{1}{\tilde z^2} \right) \, (d\tilde z)^3 \,.
\ee
The corresponding spectral network is shown in Figure \ref{fig:CP2SN} for various phases.

There are three distinguished phases $\vartheta = k \pi / 3\ (k = 0, 1, 2)$. At these phases spectral networks show topological jumps due to the appearance of finite $\CS$-walls encircling the puncture at $\tilde z=0$.
Note that the evolution shown in Figure \ref{fig:CP2SN} covers only a third of the whole phase range. 
At large $|\tilde z|$, for example at $\tilde z = -2$, there are three $\CS$-walls of the same type sweeping through $\tilde z$ and each $\CS$-wall carries a single soliton according to the basic traffic rules of spectral networks \cite{Gaiotto:2012rg, Longhi:2016rjt}.
Overall there are therefore 9 solitons in such region, an SU(3) triplet of (phase-wise) consecutive ones of the same $(ij)$ type, for each $(ij)\in\{(21),(23),(13)\}$.\footnote{In our conventions, an $(ij)$ soliton interpolates between vacuum $i$ at the spatial boundary $x_1 = -\infty$ and vacuum $j$ at the opposite boundary $x_1=+\infty$. Its charge is therefore distinguished from that of the opposite type, a $(ji)$ soliton (see \cite[eq. (3.26)]{Longhi:2016rjt} for further detail on conventions).} 
As explained above, this region of $C$ corresponds to the one analyzed in \cite{Dorey:1998yh}, and we find exact agreement for the 2d BPS spectrum \cite{Park:2013osa}: for the $\ICP^{N-1}$ model there should be $\binom{N}{k}$ solitons interpolating between vacua $i$ and $i+k$.

\begin{figure}[ht]
	\centering
	\includegraphics[width=.4\textwidth]{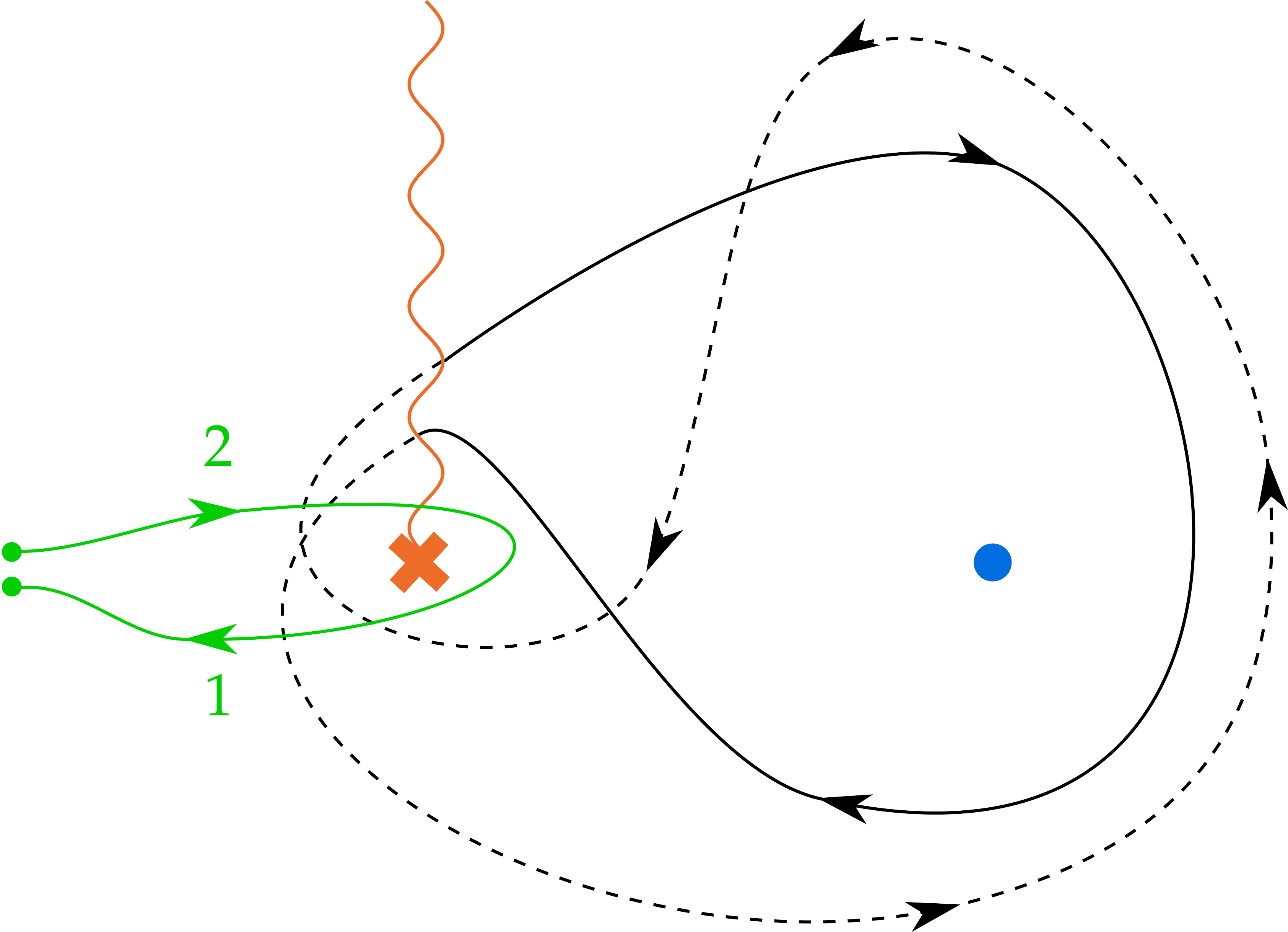}
	\caption{The branch point at $\tilde z=-1$ is marked by a cross, while the blue dot is the puncture at $\tilde z=0$. The counter-clockwise monodromy at the branch point permutes sheets as $1\to 2\to 3\to 1$. We define three cycles $\gamma_i$ for $i=1,2,3$ as depicted by the black path, where the solid and dashed lines run respectively on sheets 1 and 2 for $\gamma_1$, on sheets 2 and 3 for $\gamma_2$ and sheets 3 and 1 for $\gamma_3$. Note that $\gamma_1+\gamma_2+\gamma_3 = 0$. The values of their central charges are $Z_{\gamma_k} = M_0\, e^{\frac{2\pi i}{3} \, k}$ with $M_0>0$. The solid green line denotes a representative for the relative homology class  $a$, of solitons of type $(21)$ supported at $\tilde z$ on the negative real axis; it runs on sheets $2$ and $1$ as indicated.}
	\label{fig:CP2-cycles}
\end{figure}

To characterize these solitons more precisely, let us introduce three homology classes $\gamma_1,\gamma_2,\gamma_3$ corresponding to cycles around the puncture at $\tilde z=0$, see Figure \ref{fig:CP2-cycles}.
Among the 9 solitons we found at $\tilde z=-2$, let us focus on those of type $(21)$. 
Their charges are, in order of increasing central charge phase: 
\be\label{eq:CP2_21_soliton_charges}
	a + \gamma_1\,, \quad a\,,\quad  a -\gamma_3 \,,
\ee
where $a$ denotes the relative homology class in $H_1^{\text{rel}}(\Sigma,\IZ; (\tilde z^{(2)},\tilde z^{(1)}))$ of paths starting from the lift of $\tilde z$ to sheet 2, to its lift to sheet 1. A representative is shown in green in Figure \ref{fig:CP2-cycles}.
Let $c_1, c_2, c_3 \in H_1(\Sigma,\IZ)$ be counter-clockwise cycles around the puncture, running respectively on sheets $1,2,3$. In terms of $c_i$, we have 
\begin{align}
	\gamma_1 = c_2-c_1, \gamma_2 = c_3-c_2, \gamma_3 = c_1-c_3.
\end{align}	
Introducing $a' \equiv a - c_1$, the charges in (\ref{eq:CP2_21_soliton_charges}) assume the suggestive form
\be
	a'+c_2\,,\quad a'+c_1\,,\quad a'+c_3 \,.
\ee
It is now manifest that the three solitons belong to a triplet of $\SU(3)$: if the symmetry is restored by sending $Z_{c_i} \to 0$ the three central charges become equal.

\begin{figure}[ht]
	\centering
	\begin{subfigure}{.3\textwidth}
		\includegraphics[width=\textwidth]{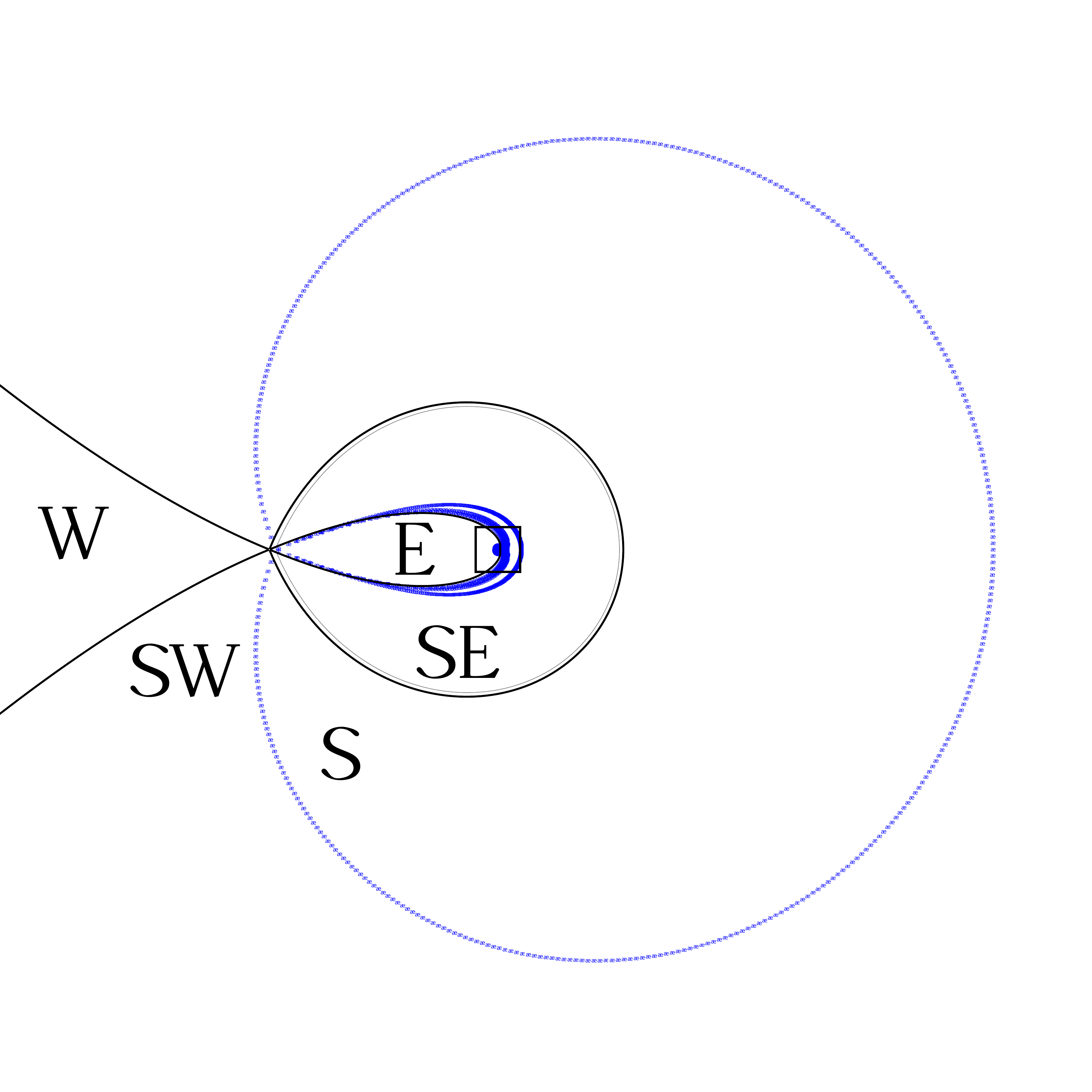}
		\caption{}
		\label{fig:CP2-chambers-merged}
	\end{subfigure}
	\begin{subfigure}{.3\textwidth}
		\includegraphics[width=\textwidth]{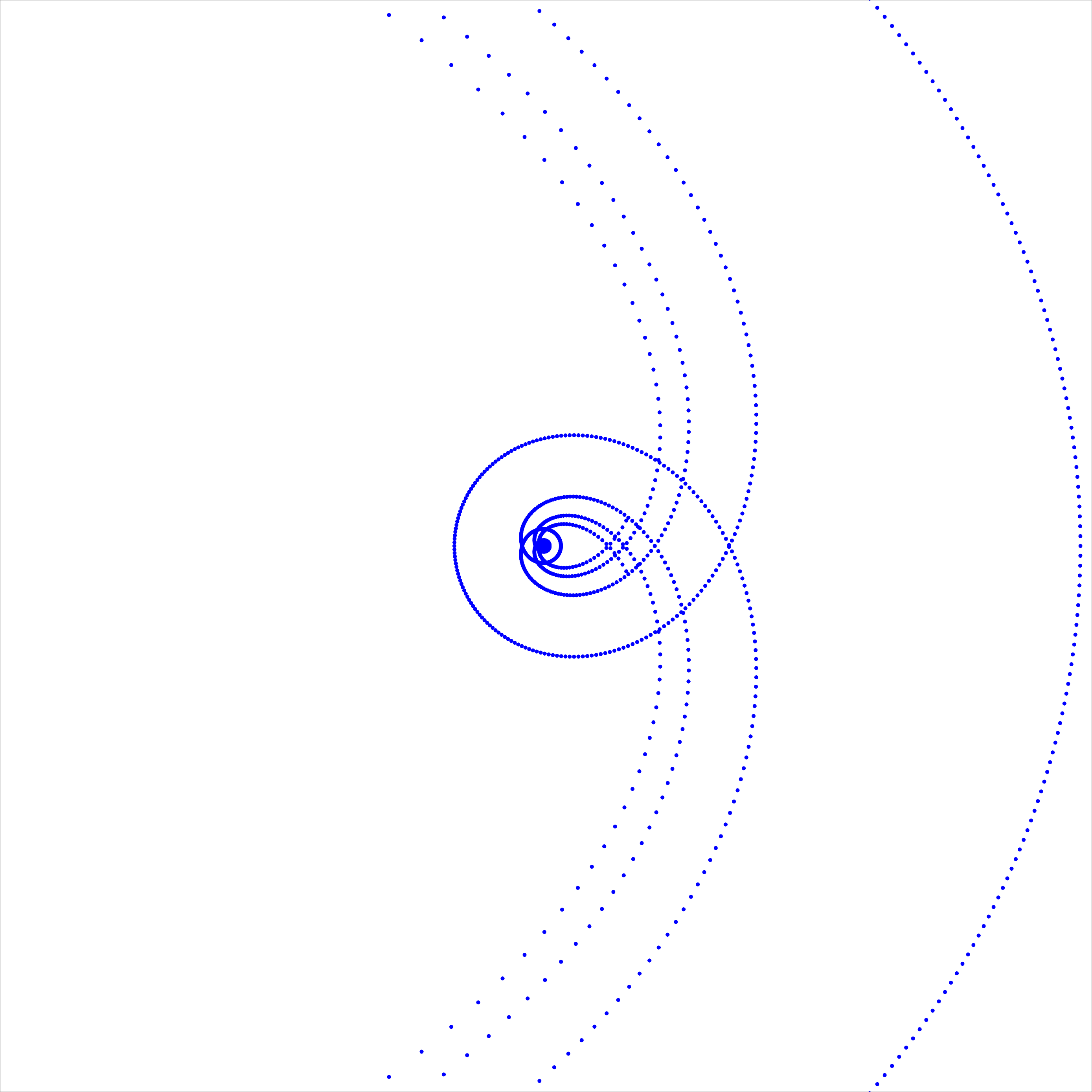}
		\caption{}
		\label{fig:CP2SN_2d_BPS_Wall_02}
	\end{subfigure}
	\caption{2d walls of marginal stability separating different phases of the $\ICP^2$ model. In the right frame a detail of the chamber structure near the puncture is shown. 
	The curve separating regions $S$ and $SE$ is the finite $\CS$-wall which appears at each of the three critical phases, also shown in the first and last frames of Figure \ref{fig:CP2SN}.
	}
	\label{fig:CP2-chambers}
\end{figure}

While the flavor central charges $Z_{\gamma_k}$ do not depend on $\tilde z$, $Z_{a}$ does. 
Therefore varying $\tilde z$ can result in crossing a 2d wall of marginal stability, where the central charges of two solitons acquire the same phase. We can find the 2d walls of marginal stability by studying the evolution of the network throughout the range $\vartheta\in[0,\pi]$. 
2d walls of marginal stability occur where intersecting $\CS$-walls form a joint\footnote{Intersecting $\CS$-walls forms a joint when the sum of roots of the two $\CS$-walls at the point of intersection is again a root.}, across which the 2d BPS spectrum undergoes wall-crossing of the Cecotti-Vafa type.
Additional marginal stability walls are determined by the topological jumps of the network: these correspond to the collection of all $\CS$-walls taken at each of the critical phases. 
The resulting chamber structure is shown in Figure \ref{fig:CP2-chambers}.

To discuss the wall-crossing behavior, we introduce soliton charges 
\begin{align}
    b\in H_1^{\text{rel}}(\Sigma,\IZ; (\tilde z^{(2)},\tilde z^{(3)})), \\
    c\in H_1^{\text{rel}}(\Sigma,\IZ; (\tilde z^{(1)},\tilde z^{(3)})), 
\end{align}
again these are understood with real and negative $\tilde z<-1$. These are the analogues of the soliton charge $a$.
Counting all the $\CS$-walls sweeping through $\tilde z$, and keeping track of their phase ordering, the complete spectrum generator for $\tilde z$ in region W is then easily written down
\be
	\mathbb{S_{\text{W}}} = %
	\CS_{a-\gamma_3}\,\CS_{a}\,\CS_{a+\gamma_1} \,%
	\CS_{b+\gamma_1} \, \CS_{b} \, \CS_{b-\gamma_2}\, %
	\CS_{c-\gamma_2} \, \CS_{c} \, \CS_{c+\gamma_3} \,.
\ee
Varying $\tilde z$ smoothly from region W into SW involves crossing a wall along which 
\be
	\arg(Z_a) = \arg(Z_{-\gamma_3})\,,\qquad%
	\arg(Z_b) = \arg(Z_{\gamma_1})\,,\qquad%
	\arg(Z_c) = \arg(Z_{-\gamma_2})\,,
\ee
this involves a reordering of operators $\CS_{a}$ and $\CS_{a-\gamma_3}$, and so on. However these operators commute, and the overall number of BPS states remains constant, hence
\be
	\mathbb{S_{\text{SW}}} = %
	\CS_{a}\,\CS_{a-\gamma_3}\,\CS_{a+\gamma_1} \,%
	\CS_{b} \, \CS_{b+\gamma_1} \,\CS_{b-\gamma_2}\, %
	\CS_{c} \, \CS_{c-\gamma_2} \,\CS_{c+\gamma_3} \,.
\ee
It is understood here and below that $a,b,c$ solitons refer to the parallel transport of the original ones in the W region, crossing curves of marginal stability as described.
There are other such `trivial' walls of marginal stability along the way to the next region S, which we have omitted for simplicity. On the curve of marginal stability between SW and S the following central charges align
\be
	\arg(Z_{-a}) = \arg(Z_{b-\gamma_2})\,,\qquad%
	\arg(Z_{-b}) = \arg(Z_{c+\gamma_3})\,,\qquad%
	\arg(Z_{c}) = \arg(Z_{a+\gamma_1})\,,
\ee
by direct inspection of the $\CS$-walls, the spectrum generator into region S reads
\be
	\mathbb{S_{\text{S}}} = %
	\CS_{c}\,\CS_{a+\gamma_1} \,%
	\CS_{-a} \, \CS_{b-\gamma_2}\, %
	\CS_{-b} \, \CS_{c+\gamma_3} \,.
\ee
There are now six soliton states, three of them decayed by the Cecotti-Vafa wall crossing mechanism. To see this, we will make use of the wall-crossing identities \cite{Gaiotto:2011tf}
\be
\begin{split}
	\CS_{ij}\CS_{ik} = & \CS_{ik} \CS_{ij} \,,\qquad \CS_{ij}\CS_{kj} = \CS_{kj} \CS_{ij}  \\
	& \CS_{jk} \CS_{ij} = \CS_{ij}\CS_{ik}\CS_{jk}
\end{split}
\ee
where $\CS_{ij}$ denotes the presence of a soliton whose charge belongs to the $(ij)$ topological sector.
Recall that the spectrum generator is defined up to conjugation, i.e. its expression depends on a choice of half-plane in the complex plane of central charges. 
We shall also adopt clockwise/counterclockwise rotations of this half-plane, which correspond to conjugating respectively by
\be
	\mathbb{S} \ \to \ \CS_{-X}\, \mathbb{S}\, \CS_{X} \qquad \text{or} \qquad %
	\mathbb{S} \ \to \ \CS_{X}\, \mathbb{S}\, \CS_{-X}\,.
\ee
We now show the equivalence (up to wall crossing) of the spectrum in the two regions 
\be
\begin{split}
	\mathbb{S_{\text{W}}} %
	& = %
	\CS_{a-\gamma_3}\,\CS_{a}\,\CS_{a+\gamma_1} \,%
	\CS_{b+\gamma_1} \, \CS_{b} \, \CS_{b-\gamma_2}\, %
	\CS_{c-\gamma_2} \, \CS_{c} \, \CS_{c+\gamma_3} \\
	& = %
	 \CS_{b}\,\CS_{a+\gamma_1} \,%
	\CS_{b+\gamma_1}  \, \CS_{b-\gamma_2}\, %
	\CS_{c-\gamma_2} \, \CS_{c} \, \CS_{c+\gamma_3} \,%
	\CS_{-a+\gamma_3}\,\CS_{-a} \\
	& = %
	\CS_{a+\gamma_1} \,\CS_{b+\gamma_1}\, \CS_{c}  \, %
	\CS_{b-\gamma_2}\, \CS_{c-\gamma_2}\,\CS_{-a} \,%
	\CS_{c+\gamma_3} \, \CS_{-a+\gamma_3}  \, \CS_{-b} \\
	& = %
	\CS_{c} \, \CS_{a+\gamma_1}   \, %
	\CS_{-a} \, \CS_{b-\gamma_2} \,%
	\CS_{-b} \, \CS_{c+\gamma_3} \, \\
	& = \mathbb{S_{\text{S}}} \,.
\end{split}
\ee
Moving further into region SE, we find six infinite towers of 2d soliton states, as well as solitons interpolating between a vacuum and itself but with nontrivial topological charges ($(ii)$-type solitons). Their charges and phase ordering are encoded into the spectrum generator
\be
\begin{split}
	\mathbb{S}_{\text{SE}} %
	& = %
	\CK^{-1}_{-\gamma_3} \,%
	\left[  \prod^{\searrow}_{n=0} \CS_{(-c-\gamma_3)-n\gamma_3}  \right] \, %
	\left[  \prod^{\nearrow}_{n=0} \CS_{-a+n\gamma_1}  \right] \, \\
	& \qquad \qquad \qquad \times \CK^{-1}_{\gamma_1} %
	\left[  \prod^{\searrow}_{n=0} \CS_{(a+\gamma_1)+n\gamma_1}  \right] \, %
	\left[  \prod^{\nearrow}_{n=0} \CS_{-b-n\gamma_2}  \right] \, \\
	& \qquad \qquad \qquad \qquad \qquad \qquad \times \CK^{-1}_{-\gamma_2} %
	\left[  \prod^{\searrow}_{n=0} \CS_{(b-\gamma_2)-n\gamma_2}  \right] \, %
	\left[  \prod^{\nearrow}_{n=0} \CS_{-c+n\gamma_3}  \right] \,. %
\end{split}
\ee

\begin{figure}[ht]
	\centering
	\includegraphics[width=.4\textwidth]{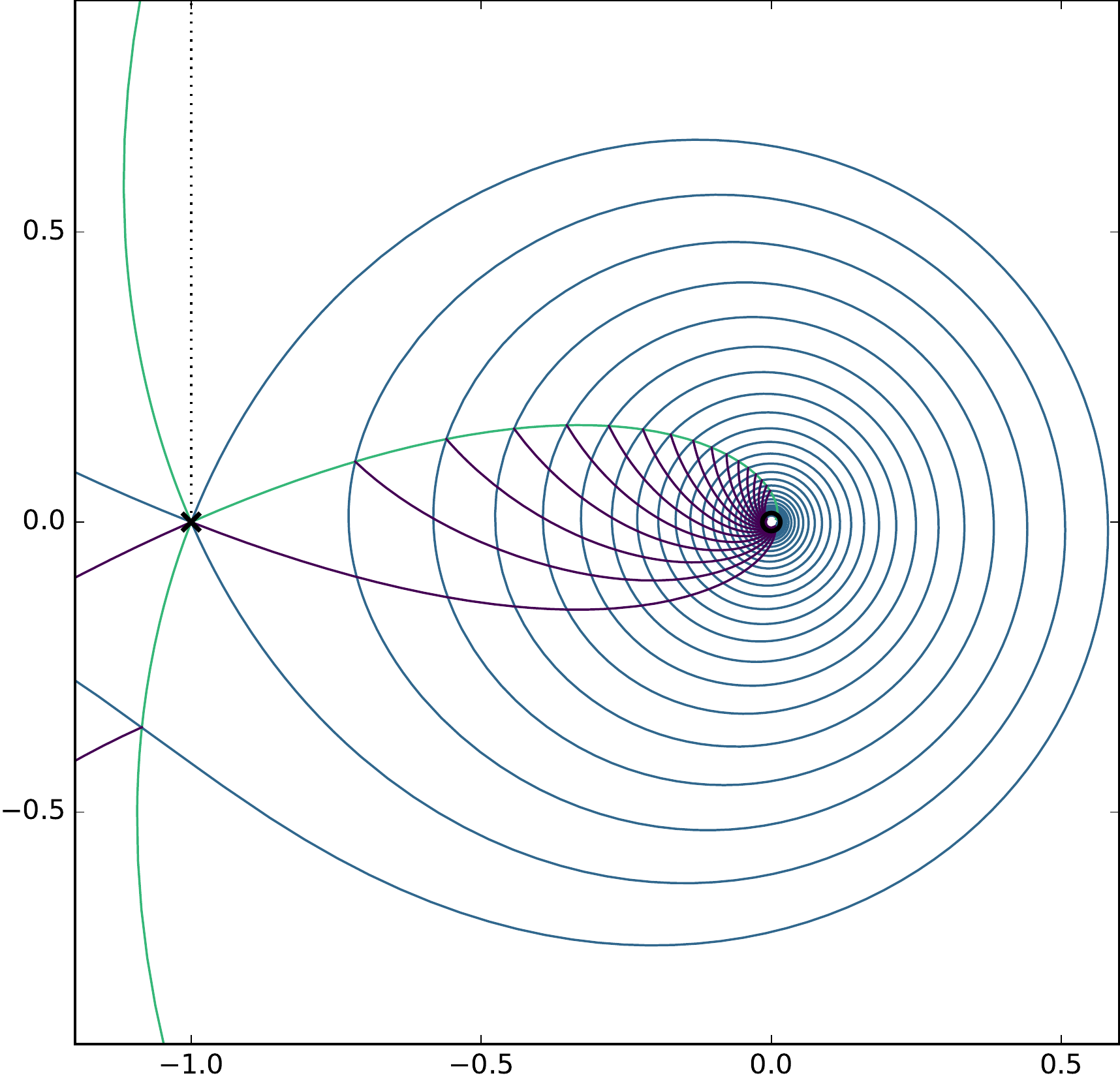}
	\caption{The new towers of solitons appearing in the E chamber.}
	\label{fig:CP_2_E_chamber_towers}
\end{figure}

We finally consider one more chamber, denoted as E in figure \ref{fig:CP2-chambers-merged}. There are infinitely many curves of marginal stability encountered in passing from SE to E, nevertheless we can again read off the spectrum directly from the network.
In addition to the states encountered in region SE, there are new towers of states, their charges are supported on $\CS$-walls like the ones depicted in figure \ref{fig:CP_2_E_chamber_towers}, with charges
\be
\begin{split}
	& a+n\gamma_3\,, \qquad  a-n\gamma_2\,,\qquad -c-n\gamma_2\,,\qquad -c+n\gamma_1\,,\qquad -b+n\gamma_1\,,\qquad -b-n\gamma_3\,,
\end{split}
\ee
with integer $n\geq 1$. If we keep moving closer to the puncture at $\tilde z=0$ the spectrum undergoes further transitions, due to infinitely many more curves of marginal stability as shown in Figure \ref{fig:CP2SN_2d_BPS_Wall_02}.

To conclude let us comment on the relation between our results and those obtained via field theoretic techniques. In particular we consider the work of \cite{Bolokhov:2012dv}, where a detailed study of the $\ICP^2$ sigma model with $\IZ_3$ symmetric twisted masses was performed.
The perspective in \cite{Bolokhov:2012dv} is slightly different from ours: while we fix a 4d vacuum and study variations of the 2d complexified FI-$\theta$ coupling, in \cite{Bolokhov:2012dv} the coupling is fixed while the values of twisted masses are varied. But the dictionary between the two perspectives is  simple. First of all, note that our choice of 4d vacuum coincides with having $\IZ_3$-symmetric twisted masses: taking the 4d vector multiplet scalar to have a vev $\Phi = M_0\, \text{diag}(1, e^{2\pi i/3}, e^{4\pi i/3})$ gives $u_2 = 0, u_3 \sim (M_0)^3$, therefore the $u_{h^\vee}$-plane parametrizes symmetric twisted masses for the defect theory. 
In \cite[eq. (3.3)]{Bolokhov:2012dv} the vacuum manifold of the 2d theory is given as 
\begin{align}
	\sigma_p^N = 1+m_0^N, 
\end{align}
which is a special case of (\ref{eq:weak-coupling-curves}) with $\sigma=\sigma_p$, $2 u_N / \Lambda^N = m_0^N$ and specializing to $z=1$.\footnote{The  mismatch in dimensionality of $2 u_N / \Lambda^N = m_0^N$ is due to fixing the 2d dynamical scale to 1 in \cite{Bolokhov:2012dv}, and restoring it resolves this issue.}
The $m_0$-plane is therefore a $3$-fold cover of the $u_3$-plane at fixed $\tilde z$.
The region near $m_0 = 0$ then maps to the region near $\tilde z\to \infty$, while the regime $m_0\to\infty$ maps to $\tilde z\to 0$. The three AD points identified in \cite{Bolokhov:2012dv} at the roots of $-1$ in the $m_0$-plane all map to the branch point we found at $\tilde z=-1$.

At strong coupling (for $m_0\to 0$) we find perfect agreement of 2d BPS spectra, with 9 soliton states arranged in triplets of $\SU(3)$, cf. \cite[eq. (5.10)]{Bolokhov:2012dv}.
In the weak coupling regime we recover the three ``elementary quanta'' and the six towers of solitons that were found in \cite[eq. (5.19)]{Bolokhov:2012dv}. 
We can also match the curves of marginal stability with those in \cite[Fig. 10]{Bolokhov:2012dv} (also see \cite{Dorey:2012dq} for related results). Noting that they are drawn on the $m_0^3$ plane, they should map by inversion $x\to 1/x$ to the curves in figure \ref{fig:CP2-chambers-merged}. To a good approximation they appear to agree.

\subsection{Semiclassical limit of the canonical defect of  \texorpdfstring{$\SO(8)$}{SO(8)} pure gauge theory}\label{sec:semiclassical_D4}
\begin{figure}[ht]
\begin{center}
\includegraphics[width=0.48\textwidth]{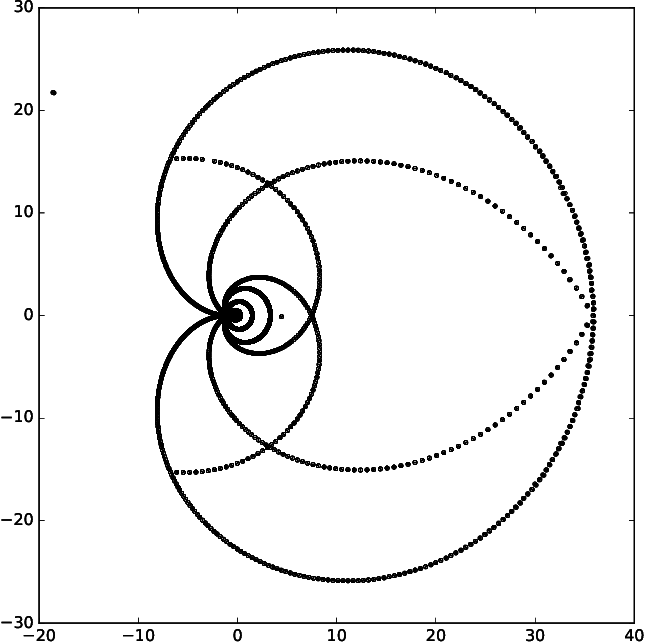}\hfill
\includegraphics[width=0.48\textwidth]{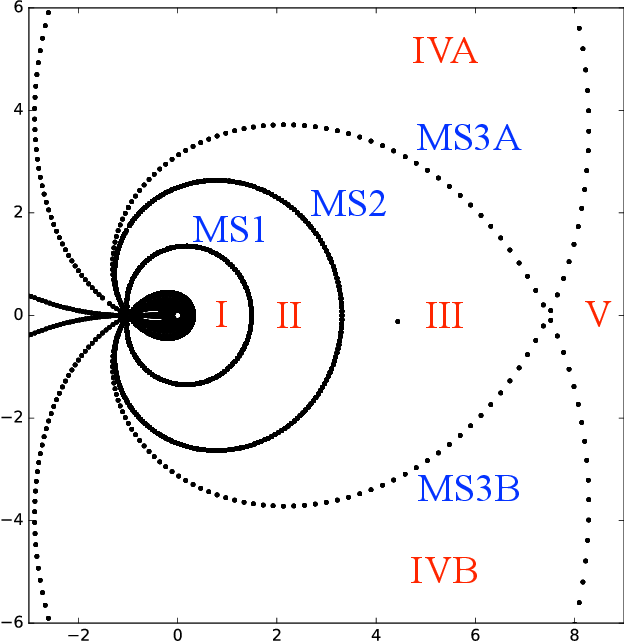}
\caption{
Left, the 2d walls of marginal stability for the 2d theory obtained from the decoupling limit of the SO(8) SYM surface defect. Right: a detail of the walls near the puncture. The branch point sourcing all $\CS$-walls is located at $z=-1$ while the puncture is located at $z=0$.
}
\label{fig:so8-2d-ms-walls}
\end{center}
\end{figure}

%
%
The semiclassical limit of the $\SO(8)$ canonical defect is considerably more complex than the $\ICP^2$ model studied in the previous section. The 2d walls of marginal stability are depicted in Figure \ref{fig:so8-2d-ms-walls}. We will analyze the BPS spectrum in chambers $I$ through $V$ and its jumps across the walls of marginal stability that divide the chambers.

2d vacua are in 1-1 correspondence with weights of the vector representation of SO(8), we will label vacua according to the following convention
\be\label{eq:so8-weights}
\begin{array}{c|c}
	\hline
	\mu_{1} & (1,0,0,0) \\
	\mu_{2} & (0,1,0,0) \\
	\mu_{3} & (0,0,1,0) \\
	\mu_{4} & (0,0,0,1) \\
	\mu_{5} & (-1,0,0,0) \\
	\mu_{6} & (0,-1,0,0) \\
	\mu_{7} & (0,0,-1,0) \\
	\mu_{8} & (0,0,0,-1) 
\end{array}
\ee
Two of the vacua are entirely degenerate, because two sheets of $\Sigma$ are located at $\tilde\sigma=0$ throughout $C$. 
With our choice of trivialization these correspond to $\mu_4$ and $\mu_8$. 
Nevertheless there are no massless degrees of freedom in the 2d theory, because there is no root of $\gD_4$ corresponding to $\mu_4 - \mu_8$, and therefore no soliton can connect the two vacua \cite{Longhi:2016rjt}.
Another potential issue could be due to marginal stability: one may wonder if there may be two solitons that could form a bound state, whose central charges acquire the same phase for all $z\in C$. 
This issue does not arise, because two solitons have central charges of identical phase for all $z$ only if their \emph{soliton trees} are identical \cite{Longhi:2016rjt}. 
In turn, if the trees were identical, they would propagate on overlapping $\CS$-walls, but roots of overlapping $\CS$-walls are always orthogonal, so the solitons supported by those walls cannot form bound states. We observe this directly in the spectral networks of the $SO(8)$ SYM theory.

$\CS$-walls are labeled by roots of SO(8),  for example a wall of type $\CS_\alpha$ with $\alpha = (1,-1,0,0) = \mu_{1}-\mu_2 = \mu_6-\mu_5$ carries solitons which interpolate between the two pairs vacua $(21)$ and $(56)$. Any $\CS$-wall carries $k_\rho = 2$ types of 2d solitons (see \cite{Longhi:2016rjt} for the definition of $k_\rho$).
The counter-clockwise sheet monodromy around the branch point at $z=-1$ permutes sheets as
\be
	1\to 2\to 3\to 5\to 6\to 7\to 1\qquad 4\leftrightarrow 8\,.
\ee
It is therefore possible to choose a set of closed cycles in $H_1(\Sigma, \IZ)$ based on the contours shown in Figure \ref{fig:CP2-cycles}, with the solid and dashed lines running on the following pairs of sheets
\be\label{eq:flavor-charges}
\begin{array}{c|cccccccc}
	& \gamma^f_{1}  & \gamma^f_{2} & \gamma^f_{3} & \gamma^f_{4} & \gamma^f_{5} & \gamma^f_{6} & \gamma^f_{7} & \gamma^f_{8} \\
	\hline
	\text{solid} & 1 & 2 & 3 & 5 & 6 & 7 & 4 & 8\\
	\text{dashed} & 2 & 3 & 5 & 6 & 7 & 1 & 8 & 4 
\end{array}
\ee
As anticipated there are \emph{overlapping} $\CS$-walls with different root types. This does not pose an issue of 2d marginal stability, because solitons carried by two overlapping walls cannot form boundstates, this is ensured by the fact that the corresponding roots are orthogonal. 
We will refer to a group of overlapping $\CS$-walls as a \emph{degenerate $\CS$-wall}.
In the presence of degenerate $\CS$-walls, new types of joints can appear in a spectral network. 
These correspond to intersections of a degenerate $\CS$-wall with a regular one, or to intersections of two degenerate $\CS$-walls. We will call these \emph{degenerate joints}.

Three distinct types of degenerate joints will play a role in the forthcoming analysis. 
A detailed study of 2d wall crossing across each of these joints can be found in Appendix \ref{sec:degenerate-joints}, here we simply state the results and employ them to compute the spectrum.
We will use the fact that every incoming $\CS$-walls for each joint of interest carries a {single soliton}, and describe the soliton content of outgoing $\CS$-walls by simply counting them. 
More detailed data, such as the precise charges of the outgoing solitons (i.e. the relative homology classes of the corresponding paths), can be recovered from the full analysis given in Appendix \ref{sec:degenerate-joints}.
%
%


\begin{figure}[h!]
	\centering
	\begin{subfigure}[b]{.32\textwidth}
	\includegraphics[width=0.99\textwidth]{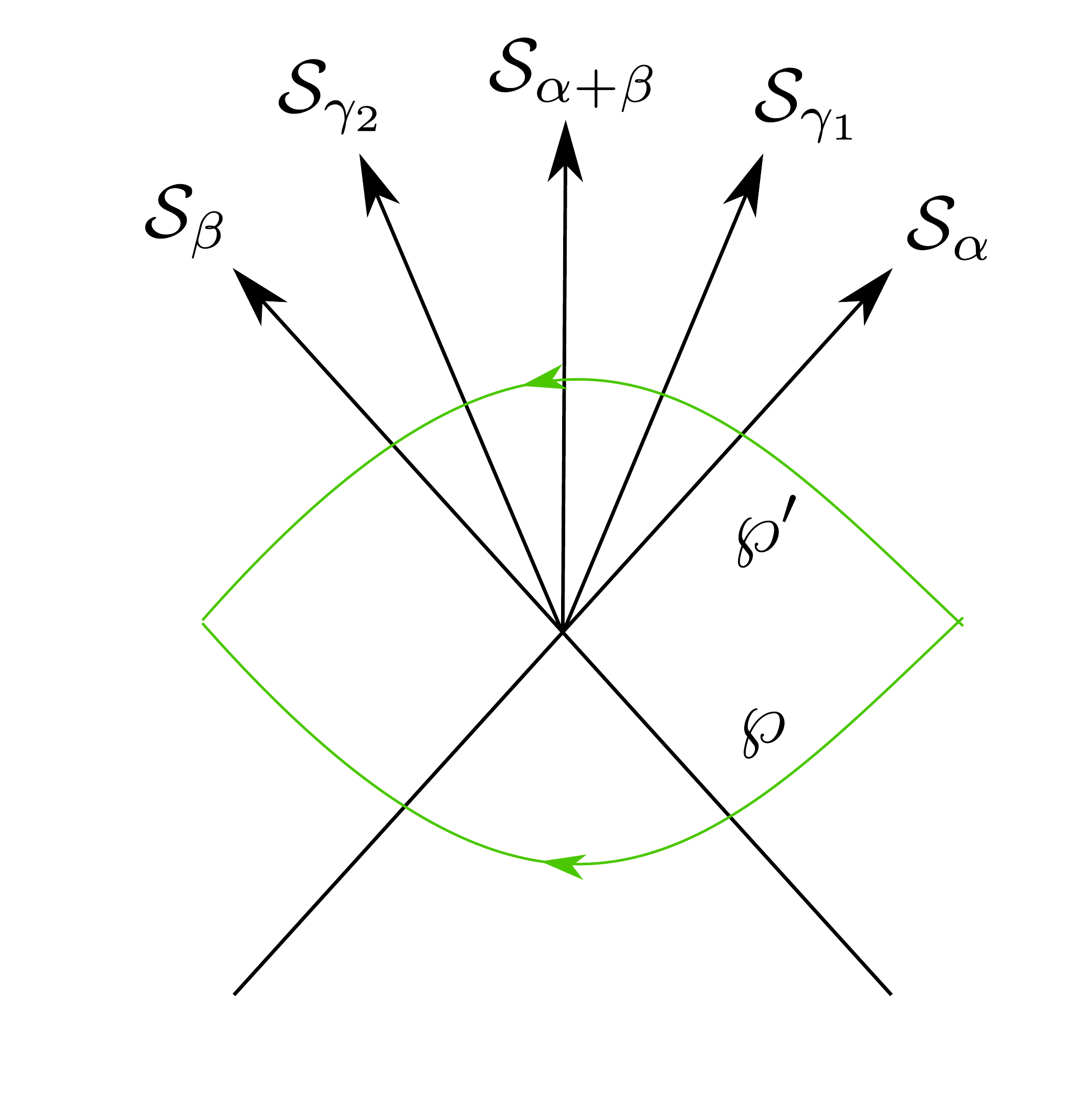}
	\caption{Type I}
	\label{fig:type-I-joint}
	\end{subfigure}
	\begin{subfigure}[b]{.32\textwidth}
	\includegraphics[width=0.99\textwidth]{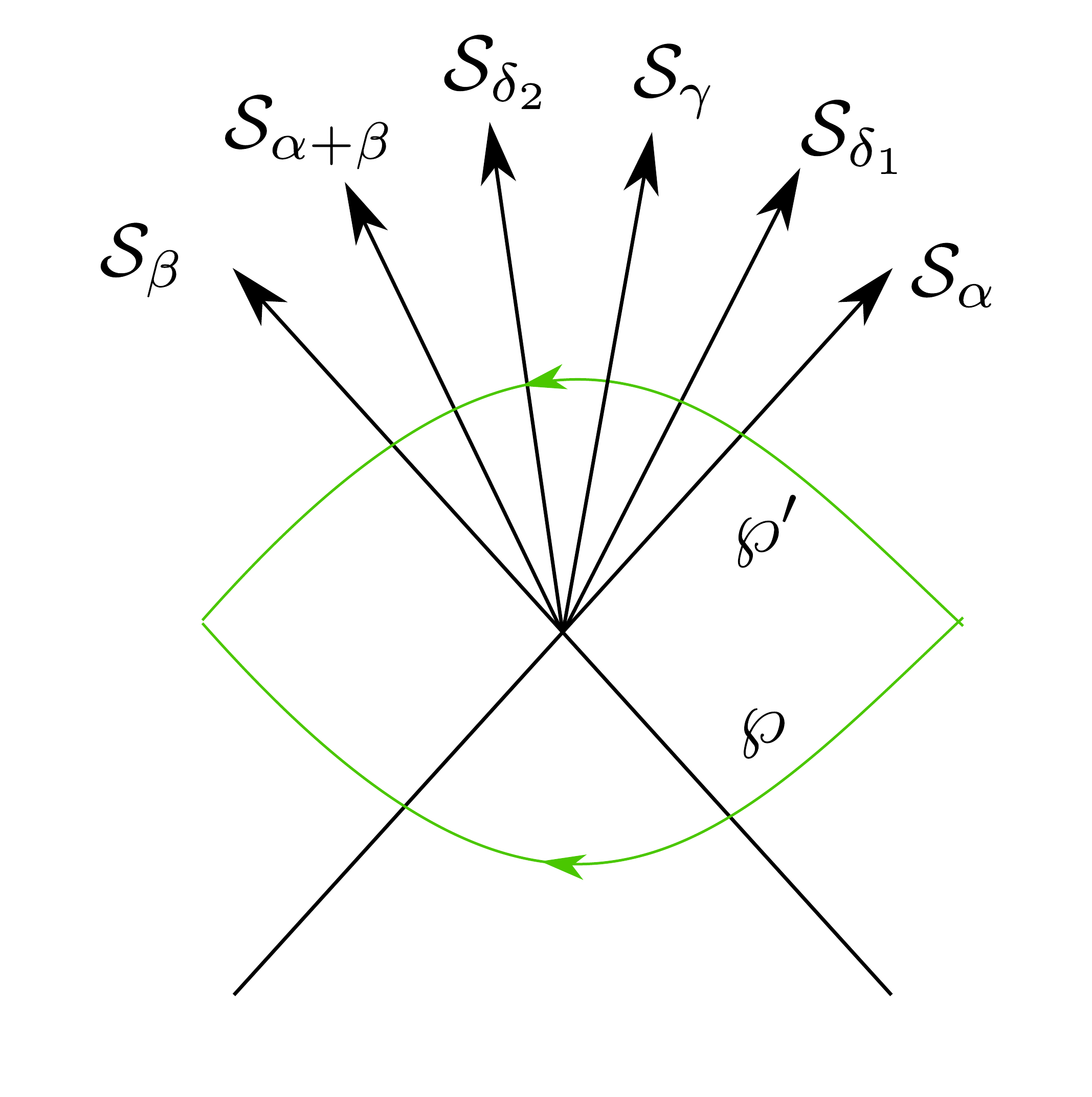}
	\caption{Type II}
	\label{fig:type-II-joint}
	\end{subfigure}
	\begin{subfigure}[b]{.32\textwidth}
	\includegraphics[width=0.99\textwidth]{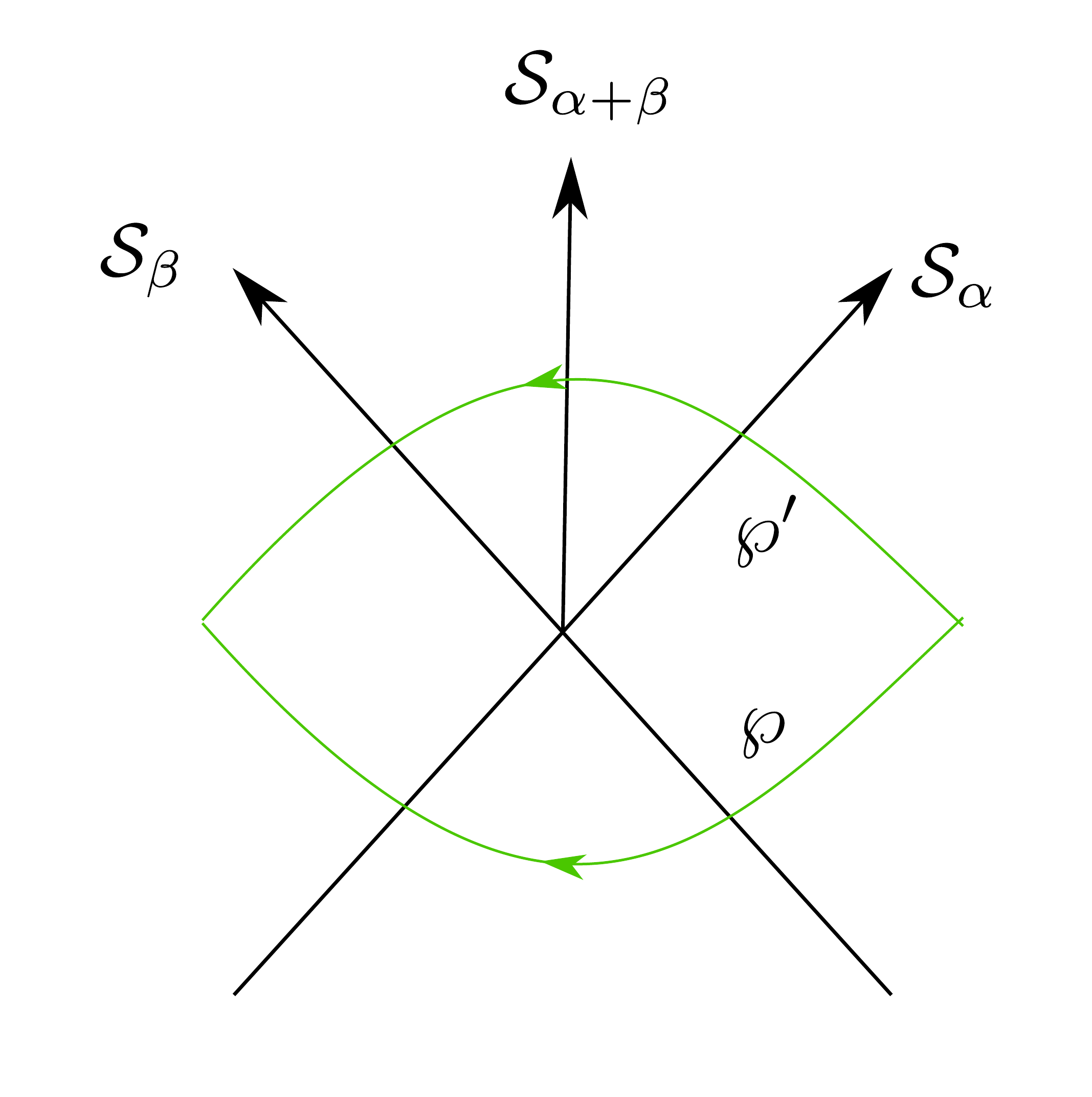}
	\caption{Type III}
	\label{fig:type-III-joint}
	\end{subfigure}
	\caption{Three types of degenerate joints appearing in the decoupling limit of SO(8) SYM. Green curves denote paths $\wp,\wp'$ for computing joint equations through invariance of the formal parallel transport.}
  \label{fig:degenerate-joints-I-II-III}
\end{figure}

The \emph{type-I} degenerate joint is shown in Figure \ref{fig:type-I-joint}. 
Both $\CS_\alpha$ and $\CS_\beta$ are degenerate $\CS$-walls, in the sense that each corresponds to three overlaping $\CS$-walls, each of a different root type. 
The newborn wall $\CS_{\alpha+\beta}$ is also degenerate, carrying three root types, while $\CS_{\gamma_i}$ are regular $\CS$-walls.
Let $\alpha_i,\, i=1,2,3$ be the root types supported on the degenerate wall $\CS_\alpha$ and  $\beta_i,\, i=1,2,3$ be the root types supported on $\CS_\beta$. 
This type of joint occurs if the set $\{\alpha_i+\beta_j\}_{i,j=1,2,3}$ contains more than one root. 
The following table gives an explicit example of possible root types involved.
\be\label{eq:type-I-roots}
\begin{array}{c|c|c}
	\text{$\CS$-wall} & \text{roots} \\
	\hline
	\CS_\alpha & \ \{%
	\alpha_1 = (0,0,1,1) \,, %
	\alpha_2 = (-1,1,0,0) \,, %
	\alpha_3 = (0,0,1,-1) %
	\}  & %
	\\
	\CS_\beta & \ \{%
	\beta_1 = (1,0,0,-1) \,, %
	\beta_2 = (1,0,0,1) \,, %
	\beta_3 = (0,1,-1,0) %
	\}  &  %
	\\
	& \ \{%
	(1,0,1,0) = \alpha_1 + \beta_1 = \alpha_3 +\beta_2 \,,\\ %
	\CS_{\alpha+\beta} & (0,1,0,1) = \alpha_2 + \beta_2  = \alpha_1 + \beta_3\,,\\ %
	& (0,1,0,-1)  = \alpha_3 + \beta_3 =  \alpha_2 + \beta_1 %
	\}  & %
	\\
	\CS_{\gamma_1} & \ \{%
	\gamma_1 = (0,1,1,0) = \alpha_1 + (\alpha_3+\beta_3) = \alpha_2+(\alpha_1+\beta_1) \equiv \alpha_3 + (\alpha_2+\beta_2) %
	\}  & %
	\\
	\CS_{\gamma_2} & \ \{%
	\gamma_2 = (1,1,0,0)= \beta_1 + (\alpha_2+\beta_2)= \beta_2 + (\alpha_3+\beta_3) = \beta_3 + (\alpha_1+\beta_1)%
	\}  & %
\end{array}
\ee
The soliton content of outgoing walls follows from the equations obtained in Appendix \ref{sec:degenerate-joints}.
In applications to the theory we are studying, the incoming walls $\CS_{\alpha_i}$ and $\CS_{\beta_i}$ always carry a single soliton for each {soliton type} $(ij)$ associated to the corresponding root. 
For example, $\CS_{\alpha_1}$ carries solitons of types $(74)$ and $(83)$, corresponding to $\mu_4-\mu_7 = \mu_3-\mu_8 = \alpha_1$.
Let us denote these soliton charges by $a_i, \bar a_i, b_i, \bar b_i$ respectively for $\CS_{\alpha_i}, \CS_{\beta_i}$ (see table (\ref{eq:soliton-types-type-I}) where the soliton types are listed, the corresponding soliton charges we are using here match those types by ordering them accordingly as $\{a_1, \bar a_1, a_2, \bar a_2, a_3, \bar a_3\}$ for $\alpha_i$, and so on). 
Then each of the $\CS$-walls $\CS_{\alpha_i+\beta_i}$ carries {two solitons of each soliton type}, whose charges correspond to homology classes obtained from concatenations $a_i \bar b_j$. The pairs of labels $i,j$ appearing in each concatenation are simply determined by the relations among roots reported in table (\ref{eq:type-I-roots}). 
For example, the outgoing wall $\CS_{(1,0,1,0)}$ carries solitons with charges $a_1 \bar b_1$ and $a_3 \bar b_2$, both of the same soliton type; it also carries solitons $b_1 \bar a_1, b_2 \bar a_3$ of a second soliton type.
Likewise, walls $\CS_{\gamma_1}$ and $\CS_{\gamma_2}$ carry six solitons each: for example $\CS_{\gamma_1}$ carries solitons of charge $a_2 b_1 \bar a_1, a_2 b_2 \bar a_3$, $b_3 a_1 \bar a_3$ all of the same soliton type, as well as three more solitons of another soliton type.\footnote{The soliton charge $b_3 a_1 \bar a_3$ appears in the equation, but choosing a different relative ordering of the $\CS$-wall factors $\CS_{\alpha_i}$, may exchange it for $b_3 a_3 \bar a_1$. This may at first seem strange, since the relative ordering of $\CS$-wall factors with orthogonal roots should not matter.
Equivalence of the two charges was in fact proved in \cite{Longhi:2016rjt}, in the study of joints of $\CS$-wall with orthogonal roots.}

The \emph{type-II} degenerate joint is shown in Figure \ref{fig:type-II-joint}. Now $\CS_\alpha$ is degenerate while $\CS_\beta$ is a regular $\CS$-wall, as a result the joint is asymmetric. 
The newborn walls $\CS_{\alpha+\beta}, \CS_{\gamma}$ are also degenerate, carrying three roots each, while $\CS_{\delta_i}$ are regular $\CS$-walls.
The following table gives an explicit example of possible root types involved.
\be\label{eq:type-II-roots}
\begin{array}{c|c|c}
	\text{$\CS$-wall} & \text{roots} \\
	\hline
	\CS_\alpha & \ \{%
	\alpha_1 = (-1,0,0,-1) \,, %
	\alpha_2 = (0,-1,1,0) \,, %
	\alpha_3 = (-1,0,0,1) %
	\}  & %
	\\
	\CS_\beta & \ \{%
	\beta = (1,1,0,0) \, %
	\}  &  %
	\\
	\CS_{\alpha+\beta} & \ \{%
	\alpha_1 + \beta = (0,1,0,-1) \,, %
	\alpha_2 + \beta = (0,-1,1,0) \,, %
	\alpha_3 + \beta = (-1,0,0,1)  %
	\}  & %
	\\
	& \ \{%
	\gamma_1 = (0,0,1,1) = \alpha_2 + \alpha_3+\beta\,, \\
	\CS_{\gamma}  &\ \ \ %
	\gamma_2 = (-1,1,0,0) = \alpha_1 + \alpha_3+\beta\,, \\
	&\ \ \ %
	\gamma_3 = (0,0,1,-1) = \alpha_1 + \alpha_2+\beta %
	\}  & %
	\\
	\CS_{\delta_1} & \ \{%
	\delta_1= (-1,0,1,0)= \alpha_n+\gamma_n%
	\}  & \\%
	\CS_{\delta_2} & \ \{%
	\delta_2= (0,1,1,0)= \alpha_n+\gamma_n+\beta%
	\}  & %
\end{array}
\ee
The outgoing soliton content for this joint is relatively simple: each of the newborn $\CS$-walls carries {one soliton of each soliton type}.  
This simplicity in the  combinatorics of soliton concatenation is due to the absence of extra relations among roots of incoming walls, unlike in the case of type I degenerate joints.
For concreteness, let $a_i, \bar a_i$ the solitons carried by $\CS_{\alpha_i}$ and by $b,\bar b$ the solitons carried by $\CS_\beta$, their soliton types are reported in table (\ref{eq:soliton-types-type-II}). The outgoing $\CS$-wall $\CS_{\alpha_i+\beta_i}$ carries a single soliton of charge $\bar a_1 \bar b_1$ of one soliton type, and a single soliton with charge $b_1 a_1$ of the other soliton type. Similarly the charges of solitons carried by other outgoing walls can be reconstructed easily from the relations in table (\ref{eq:type-II-roots}), together with the soliton types reported in table (\ref{eq:soliton-types-type-II}).

The \emph{type-III} degenerate joint is shown in Figure \ref{fig:type-III-joint}. Both $\CS_\alpha$ and $\CS_\beta$ are degenerate, each carries three root types. The newborn wall $\CS_{\alpha+\beta}$ is however of regular type.
Let $\alpha_i,\, i=1,2,3$ be the root types supported on the degenerate wall $\CS_\alpha$ and  $\beta_i,\, i=1,2,3$ be the root types supported on $\CS_\beta$. 
This type of joint occurs if the set $\{\alpha_i+\beta_j\}_{i,j=1,2,3}$ contains exactly one root. 
The following table gives an explicit example of possible involved root types.
\be\label{eq:type-III-roots}
\begin{array}{c|c|c}
	\text{$\CS$-wall} & \text{roots} \\
	\hline
	\CS_\alpha & \ \{%
	\alpha_1 = (0,0,1,1) \,, %
	\alpha_2 = (-1,1,0,0) \,, %
	\alpha_3 = (0,0,1,-1) %
	\}  & %
	\\
	\CS_\beta & \ \{%
	\beta_1 = (0,1,0,-1) \,, %
	\beta_2 = (1,0,1,0) \,, %
	\beta_3 = (0,1,0,1) \,, %
	\}  &  %
	\\
	\CS_{\alpha+\beta} & \ \{%
	\alpha_n + \beta_n = (0,1,1,0)  %
	\}  & %
\end{array}
\ee
The soliton content of $\CS_{\alpha+\beta}$ is once again encoded by the relation among roots in table (\ref{eq:type-III-roots}). There are \emph{three solitons} for each of the two soliton types carried by $\CS_{\alpha+\beta}$. Let $a_i, \bar a_i$ be the soliton charges carried by $\CS_{\alpha_i}$ and $b_i, \bar b_i$ those on $\CS_{\beta_i}$
Then the solitons on $\CS_{\alpha+\beta}$ have charges corresponding to the concatenations $a_1 \bar b_1, b_2 \bar a_2, a_3 b_3$ for one soliton type, and $b_1 \bar a_1, a_2 \bar b_2, \bar b_3\bar a_3$ for the other one (see table (\ref{eq:soliton-types-type-II}), where $a_i, \bar a_i$ are reported in the order $a_1, \bar a_1, a_2, \bar a_2,a_3, \bar a_3$).

With the tools of degenerate joints at hand, we are now ready to discuss the wall crossing of the 2d BPS spectrum across $C$. The simplest spectrum occurs in region I shown in Figure \ref{fig:so8-2d-ms-walls}. 
In this case there are precisely two solitons of each $(ij)$ type, for each root of SO(8), for a total of $2\times 2\times 12 = 48$ solitons (one factor of 2 is $k_\rho$, while 12 is the number of positive roots of SO(8)), plus their CPT conjugates. 
One of the two solitons of each type is supported on an $\CS$-wall traveling from the branch point and around the puncture from above, the other is supported on a wall traveling below the puncture. The two soliton charges thus differ by a flavor charge, which can be computed with the conventions established in table (\ref{eq:flavor-charges}). For example, the two solitons of types $(12)$ supported on walls $\CS_{(-1,1,0,0)}$ would differ by $\gamma^f_1$. This is so because  the concatenation of one soliton with the inverse of the other would result in a closed cycle running around the puncture on sheets $1$ and $2$.

The first wall of marginal stability we cross is denoted  $MS1$ in the figure. Here several walls intersect: we find six Type-II degenerate joints, with the following root types
\be
\begin{array}{c|c|c}
	\text{joint} & \CS_\alpha & \CS_\beta \\
	\hline
	1 & (1,0,0,1),(0,1,-1,0),(1,0,0,-1) & (-1,0,1,0) \\
	2 & (0,-1,1,0),(-1,0,0,1),(-1,0,0,-1) & (1,1,0,0) \\
	3 & (1,0,1,0),(0,10,1),(0,1,0,-1) & (-1,-1,0,0) \\
	4 & (-1,0,-1,0),(0,-1,0,-1),(0,-1,0,1)& (0,1,1,0) \\
	5 & (0,0,1,1),(-1,1,0,0),(0,0,1,-1)& (0,-1,-1,0) \\
	6 & (0,0,-1,-1),(1,-1,0,0),(0,0,-1,1)& (-1,0,1,0)
\end{array}
\ee
plus their CPT conjugates (obtained by reversing the signs of all root vectors).
As a result the spectrum in chamber II gets enhanced by $6 \times 8 \times 2  = 96$ new solitons (six joints, each creating 8 new $\CS$-walls, each carrying two solitons because $k_\rho=2$), plus their CPT conjugates. 
There are $96/(2\times 12)=4$  new solitons of each type $(ij)$, thus the spectrum in region II consists of 6 BPS solitons of each allowed type.

Moving into region III, we cross the marginal stability wall MS2, where six distinct joints occur (for phases of the network contained betweem $0$ and $\pi$, plus another six CPT conjugate joints for phases between $\pi$ and $2\pi$).
Half of these are Type-I joints, while the other half are regular joints, involved root types are 
\be
\begin{array}{c|c|c}
	\text{joint} & \CS_\alpha & \CS_\beta \\
	\hline
	7 & (0,0,1,1),(-1,1,0,0),(0,0,1,-1) & (1,0,0,1),(0,1,-1,0),(1,0,0,-1) \\
	8 & (-1,0,1,0) & (1,1,0,0) \\
	9 & (0,-1,1,0),(-1,0,0,1),(-1,0,0,-1) & (1,0,1,0),(0,1,0,1),(0,1,0,-1) \\
	10 & (-1,-1,0,0) & (0,1,1,0) \\
	11 & (0,-1,0,1),(-1,0,-1,0),(0,-1,0,-1)& (0,0,1,1),(-1,1,0,0),(0,0,1,-1) \\
	12 & (0,-1,-1,0)&(-1,0,1,0)
\end{array}
\ee
The wall crossings corresponding to these joints generate $3\times 12\times 2 + 3\times 1 \times 2 = 78$ new solitons, plus their CPT conjugates (a Type I joint generates $24=12\times 2$ new solitons overall). 
This time however the 2d solitons are created in different amounts for different soliton types. There are 7 new solitons for each $(ij)$ type in $\CP_\alpha$ for $\alpha$ equal to $(0,1,1,0), (1,0,-1,0), (1,1,0,0)$ (plus the CPT conjugates). There are instead 2 new solitons for each $(ij)$ corresponding to the remaining 9 positive roots.

Proceeding into region IVA, we cross the wall MS3A, where 3 distinct joints occur (together with their CPT conjugates, as usual). These are all of type III, their root types are
\be
\begin{array}{c|c|c}
	\text{joint} & \CS_\alpha & \CS_\beta \\
	\hline
	13 & (0,0,1,1),(-1,1,0,0),(0,0,1,-1) & (1,0,1,0),(0,1,0,1),(0,1,0,-1) \\
	14 & (0,-1,1,0),(-1,0,0,1),(-1,0,0,-1) & (0,0,1,1),(-1,1,0,0),(0,0,1,-1)  \\
	15 &  (-1,0,-1,0),(0,-1,0,-1),(0,-1,0,1) & (-1,0,0,-1) , (0,-1,1,0), (-1,0,0,1)
\end{array}
\ee
The wall crossings corresponding to these joints generate $3\times 3\times 2  = 18$ new solitons, plus their CPT conjugates. 
There are 3 new solitons for each $(ij)$ type in $\CP_\alpha$ for $\alpha$ equal to $(0,1,1,0), (-1,0,1,0), (1,1,0,0)$ (plus the CPT conjugates).\footnote{
The $\CS$-walls responsible for these joints are not all primary $\CS$-walls. 
Each joint involves both a primary $\CS$-wall (i.e. one sourced directly by the branch point) and an $\CS$-wall generated by a joint from MS1. 
Since MS1 has only Type-II degenerate joints, and since each $\CS$-wall generated at these joints carries precisely one soliton of each type, the condition that incoming walls for MS3A carry exactly one soliton of each type is nontrivially verified. 
Similar considerations apply to the walls generating MS3B.
}

If instead we moved into region IVB from region III, we would have crossed the wall MS3B. Here as well 3  distinct joints occur (together with their CPT conjugates, as usual). Again these are all of type III, their root types are
\be
\begin{array}{c|c|c}
	\text{joint} & \CS_\alpha & \CS_\beta \\
	\hline
	16 & (1,0,1,0),(0,1,0,1),(0,1,0,-1) & (0,0,1,1),(-1,1,0,0),(0,0,1,-1)  \\
	17 &  (0,0,1,1),(-1,1,0,0),(0,0,1,-1) &   (-1,0,0,-1) , (0,-1,1,0), (-1,0,0,1) \\
	18 &  (-1,0,0,-1) , (0,-1,1,0), (-1,0,0,1) & (-1,0,-1,0),(0,-1,0,-1),(0,-1,0,1)
\end{array}
\ee
The wall crossings corresponding to these joints generate $3\times 3\times 2  = 18$ new solitons, plus their CPT conjugates. 
There are again 3 new solitons for each $(ij)$ type in $\CP_\alpha$ for $\alpha$ equal to $(0,1,1,0), (-1,0,1,0), (1,1,0,0)$ (plus the CPT conjugates). 

Finally we can easily get the spectrum into region V as well. We simply combine the spectrum of region IVA with the jump across MS3B. In region V the spectrum is thus that of region III augmented by  6 new solitons for each $(ij)$ type in $\CP_\alpha$ for $\alpha$ equal to $(0,1,1,0), (-1,0,1,0), (1,1,0,0)$ (plus the CPT conjugates). 

\section{Superconformal limit of canonical defects}\label{sec:superconformal_limit} 

The class $\CS$ spectral curves describing vacua of the superconformal limit of canonical surface defects are given above in (\ref{eq:curves-scft-limit}), which we reproduce here using $\lambda = x\, \dd z$.\footnote{%
We are dropping $\,{\tilde{}}\,$ in this section to lighten notation. Here $z$ and $x$ correspond to $\tilde z$ and $\tilde{\sigma}$ in (\ref{eq:curves-scft-limit}), respectively.
}
\begin{align}
	\fg = \gA_{N-1} &: \lambda^{N} - z\, \dd z^{N}= 0, \label{eq:A_type_local_model}\\
	\fg = \gD_{N} &: \lambda^{2N} - z\, \dd z^{N-2}\, \lambda^2 = 0, \label{eq:D_type_local_model}\\
	\fg = \gE_6 &: \lambda^{27} - 5z\, \dd z^{12} \lambda^{15} -\frac{1}{108} z^2\, \dd z^{22}\, \lambda^3 = 0, \label{eq:E_6_type_local_model}\\
	\fg = \gE_7 &: \lambda^{56} + \frac{2458}{3} z\, \dd z^{18}\, \lambda^{38} + \frac{8371}{27} z^2\, \dd z^{36}\,\lambda^{20} + \frac{1}{729} z^3\, \dd z^{54} \, \lambda^{2} = 0. \label{eq:E_7_type_local_model}
\end{align} 

In each case the UV curve is a sphere with an irregular singularity at $z = \infty$, whose ramification structure can be deduced from the spectral curve.
By a mild abuse of terminology we shall say that each of these spectral curves is taken in the \emph{vector} representation
$\rho_\text{v}$ of $\fg$ in the sense of (\ref{eq:spectral_curve_definition}). 
For $\fg=\gA_N$ and for $\fg = \gD_{N\geq3}$ the vector representation is understood to be the first fundamental representation, while it corresponds to the \textbf{27} for $\gE_6$ and the \textbf{56} for $\gE_7$.
As described in \cite{Longhi:2016rjt}, using the same class $\CS$ data we can define analogous curves in other minuscule representations:
$\rho = \omega_k$, $1 < k < N$ for $\fg = \gA_{N-1}$ and $\rho = \omega_{N}$ for $\fg = \gD_{N}$\footnote{We can choose either one of the two spinor representation of $\fg = \gD_{N}$, $\rho = \omega_{N-1}$ and $\omega_{N}$, because of the $\mathbb{Z}_2$ outer automorphism of $\fg = \gD_{N}$.
}, where $\omega_k$ is the $k$-th fundamental weight and we denote a representation $\rho$ by its highest weight if there is no confusion. Changing the choice of a representation should not modify the 4d theory but only the choice of defect coupled to it. 
Also note that by taking a superconformal limit the 2d-4d BPS spectrum of the defect is decoupled from the 4d physics as discussed in Section \ref{sec:superconformal_limit_definition}.
We will denote a 2d defect described by such a spectral curve by 
$\SD{\mathfrak{g}}{\rho}{z}$.

When the defect ends at $z=0$, the effective 2d theory on the defect is expected to be a superconfonformal field theory (SCFT). 
In \cite{Hori:2013ewa} it was argued that the IR limit of $\mathcal{S}(\gA_{N-1},\omega_k)$ is described by a Kazama-Suzuki coset model \cite{Kazama:1988qp} associated with $G = \SU(N)$ and $\rho = \omega_k$,
\begin{align}
  \frac{\SU(N)_1}{\mathrm{S}[\UU(k)\times \UU(N-k)]}. \label{eq:KS coset A-type level 1}
\end{align}
When the defect is instead located at $z_*\neq 0$, the 2d theory corresponds to a specific deformation of the coset model. It is described by a Landau-Ginzburg (LG) model with the most relevant deformation parametrized by $z_*$.

There are additional coset models described in \cite{Kazama:1988qp} based on $G = \SO(2N)$, $\gE_6$ and $\gE_7$, see Table \ref{tab:KS models}. 
\begin{table}[ht] 
	\centering
	\begin{tabular}{| c | c | c | c | c |}
		\hline
		$G/H$ & $\mathfrak{g}$ & $\rho$ & $w_i$ & $h^{\vee}$\\ 
		\hline\hline
		$\SU(N)/\mathrm{S}[\UU(k)\times \UU(N-k)]$ & $\gA_{N-1}$ & $\omega_k$,\ $1 \leq k < N$ & $2,3,\ldots,N$ & $N$ \\
		\hline
		$\SO(2N)/\SO(2N-2) \times \UU(1)$ & \multirow{2}{*}{$\gD_{N}$} & $\rho_\text{v}$ & \multirow{2}{*}{$2,4,\ldots,2(N-1);N$} & \multirow{2}{*}{$2(N-1)$} \\
    \cline{1-1}\cline{3-3}
		$\SO(2N)/\SU(N) \times \UU(1)$ & & $\omega_{N}$ & & \\
		\hline
		$\UE_6/\SO(10) \times \UU(1)$ & $\gE_6$ & $\rho_\text{v}$ & $2,5,6,8,9,12$ & 12 \\
		\hline		
		$\UE_7/\UE_6 \times \UU(1)$ & $\gE_7$ & $\rho_\text{v}$ & $2, 6, 8, 10, 12, 14, 18$ & 18 \\
		\hline
	\end{tabular}
	\caption{Kazama-Suzuki coset models associated with simply-laced Lie algebras}
	\label{tab:KS models}
\end{table}
Each coset model is associated with a Lie algebra $\fg$ and a minuscule representation $\rho$ of $\mathfrak{g}$ whose dimension is the same as $W(G)/W(H)$, where $W(G)$, $W(H)$ are Weyl groups of $G$ and $H$, respectively. Each of these coset models admits an LG description \cite{Lerche:1989uy}. In particuar $\frac{\SU(N)}{[\SU(N-1) \times \UU(1)]}$ and $\frac{\SO(2N)}{[\SO(2N-2) \times \UU(1)]}$ correspond to $\mathcal{N} = 2$  minimal models of types $\gA_N$ and $\gD_{2N}$, respectively. 

Here we will argue that the surface defect $\SD{\fg}{\rho}{z_*}$ is described by an LG model with the most relevant deformation $z_*$, which flows in the IR to a superconformal coset model associated with $(\mathfrak{g}, \rho)$. 
We support this claim by matching the vacua and soliton spectra. 

We will first review the LG description of coset models, casting them into a form that is suitable for comparing them with superconformal limits of surface defects. 
We will later study the vacua and the BPS spectra of the surface defects, and show that they coincide with those of the LG models.

\subsection{LG descriptions of Kazama-Suzuki coset models} 

\paragraph{Superpotentials of the LG models with $\rho = \rho_\text{v}$} 
The coset models associated with the vector representation $\rho_\text{v}$ are the IR fixed points of the LG models with the following superpotentials \cite{Lerche:1989uy,Lerche:1991re,Eguchi:2001fm}
\begin{align}
	\fg = \gA_{N-1}&: W_0(X) = \frac{1}{N+1}X^{N+1}, \label{eq:W_0_A}\\
	\fg = \gD_{N}&: W_0(X, Y) = \frac{1}{2} X Y^2 + \frac{1}{2(2N-1)}X^{2N-1},\ Y = X_{N-1}, \label{eq:W_0_D}\\
	\fg = \gE_6&: W_0(X, Y) = X^{13} + X^{9} Y + a X Y^3,\ a = \left( \frac{5}{13} \right)^2, Y = X_{4} \label{eq:W_0_E_6}\\
	\fg = \gE_7&: W_0(X, Y, Z) = X^{19} + Y^2 Z + X Z^2 + a\, X^{14} Y + b\, X^{10} Z, \nonumber \\
		&\quad a = 37 \left( \frac{19}{2791} \right)^{3/4}, \quad %
		b = -21 \left( \frac{19}{2791} \right)^{1/2}, \quad%
		Y = X_{5} ,\quad %
		Z = X_{9}. \label{eq:W_0_E_7}
\end{align}
Here a degree-$w_i$ variable $X_{w_i}$ is a chiral field of scaling dimension $\Delta_i = \frac{w'_i}{h^{\vee} + 1}$ defined by $W(\lambda^{\Delta_i} \cdot X_{w_i}) = \lambda \cdot W(X_{w_i})$,
where $w_i$ is the degree of the $i$-th Casimir invariants of $H$\footnote{We assume, following \cite{Lerche:1989uy}, that the $\UU(1)$ factor of $H$ has a Casimir invariant of degree 1.}, $h^{\vee}$ is the dual Coxeter number of $\mathfrak{g}$, and $X = X_1$. For simplicity the above superpotentials have all deformations set to zero, see \cite{Eguchi:2001fm} for the full expressions.\footnote{More specifically, (3.18) and (A.16) of \cite{Eguchi:2001fm} for $\gE_{6}$ and  $\gE_{7}$, respectively.}

\paragraph{Superpotentials of the LG models with other minuscule representations} 

For $(\fg, \omega_k)$  with 
\begin{itemize}
	\item $1 < k \leq N$ for $\fg = \gA_{N-1}$ and 
	\item $k = N$ for $\fg = \gD_{N}$
\end{itemize}	
a procedure to obtain $W(X_i)$ from that of $\rho = \rho_\text{v}$ is described in \cite{Eguchi:2001fm}. 

For $(\fg = \gA_{N-1}, \omega_k)$, the LG superpotential, with all deformations set to zero, is
\begin{align}
	 W_0(X_1,\ldots, X_k)=  \sum_{i=1}^{k} \xi_i^{N+1},
\end{align}
where $\xi_i$ are auxiliary variables such that $X_j$ are their elementary symmetric polynomials,
\begin{align}
	X_i = \sum_{1 \leq l_1 < l_2 < \cdots < l_i \leq k} \xi_{l_1} \xi_{l_2} \cdots \xi_{l_i},\ i = 1, \ldots, k.
\end{align}

There is no closed-form expression for the LG superpotential of $(\fg = \gD_N, \omega_N)$ with general $N$, but we will describe a few examples in Section \ref{sec:D_N_spinor} and illustrate how to obtain the superpotentials in Section \ref{sec:D_N_omega_N_class_S_defect} using the procedure described in \cite{Eguchi:2001fm}.

\paragraph{Ground states and BPS solitons of the LG models} 
When we introduce the most relevant deformation to the LG description of a Kazama-Suzuki coset model, its ground states and BPS solitons are encoded in the minuscule representation $\rho$: the ground states of the LG model correspond to the weights of $\rho$, and the solitons correspond to the roots connecting two different weights of $\rho$ \cite{Lerche:1991re}. 

To obtain the central charges of the solitons, we project the weights and the roots onto a Coxeter plane that is the eigenspace of the Coxeter element\footnote{The Coxeter element can be represented as a product of all simple Weyl reflections and the order of the product does not matter as far as the projection is concerned.} of the Weyl group of $\mathfrak{g}$ with eigenvalue $\exp(2\pi i/h^{\vee})$, where $h^{\vee}$ is the dual Coxeter number of $\mathfrak{g}$. Then we identify the Coxeter plane with the superpotential plane $W$ of the 2d theory, and read out the central charge of a soliton by measuring the projected lengths of the root corresponding to the soliton on the plane \cite{Lerche:1991re}. Because $W(X_i) = W_0(X_i) + z\, X_1$ is a quasi-homogeneous potential with $\Delta(z) = 1 - i/(h^\vee + 1)$, the $W$-plane can also be identified with the $X_1$-plane when only the most relevant deformation $z$ is turned on.

In the following we will illustrate these structures with explicit examples.

\subsubsection{\texorpdfstring{$(\gA_{N-1}, \omega_{1 \leq k < N-1})$}{A N-1, omega 1 leq k lt N-1} LG models}

Consider introducing the most relevant deformation to the LG model by changing the superpotential to
\begin{align}
	W(X_i) = W_0(X_i) + u_N \, X_1.
\end{align}
Then the ground states of the LG model correspond to the weights of $\rho_k$, the $k$-th fundamental representation \cite{Lerche:1991re}. The vector space associated with this representation is $V_k = \wedge^k \mathbb{C}^{N}$ and its dimension is 
\begin{align}
	|V_k| = \binom{N}{k}, 
\end{align}	
which is the number of the ground states of the coset model. 
BPS solitons of the LG model correspond to the roots connecting two different weights of $\rho_k$. The number of solitons is
\begin{align}
	2 \cdot \binom{N}{2} \cdot \binom{N-2}{k-1} = \frac{N!}{(N-k-1)!(k-1)!}.
\end{align}

\begin{figure}[t]
	\centering
	\begin{subfigure}[b]{.25\textwidth}
		\vspace{2em}
		\includegraphics[width=\textwidth]{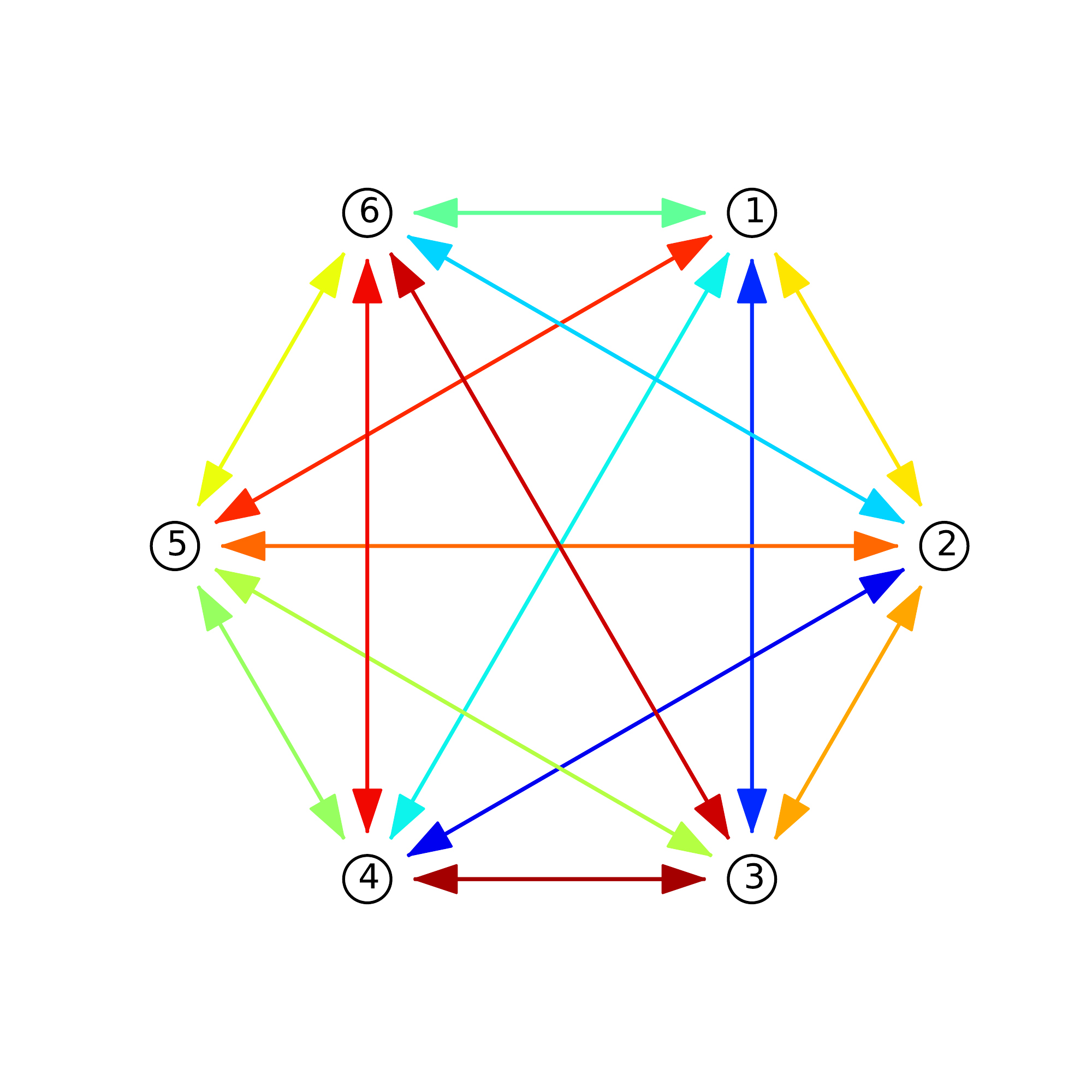}
		\vspace{1em}
		\caption{$\rho_1$}
		\label{fig:coxeter_projection_A_5_1}
	\end{subfigure}
	\begin{subfigure}[b]{.35\textwidth}
    	\includegraphics[width=\textwidth]{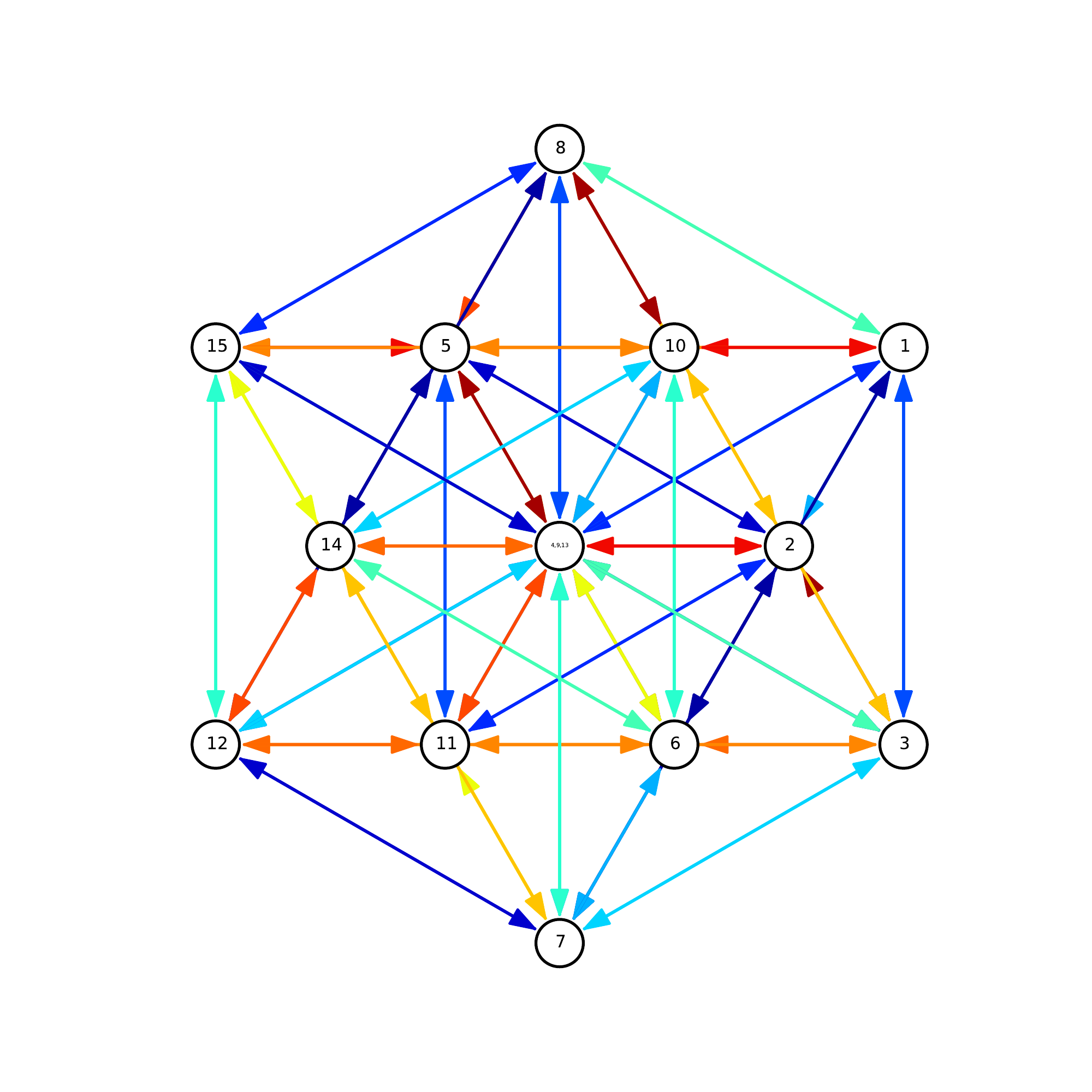}
    	\caption{$\rho_2$}
    	\label{fig:coxeter_projection_A_5_2}
	\end{subfigure}	
	\begin{subfigure}[b]{.35\textwidth}
    	\includegraphics[width=\textwidth]{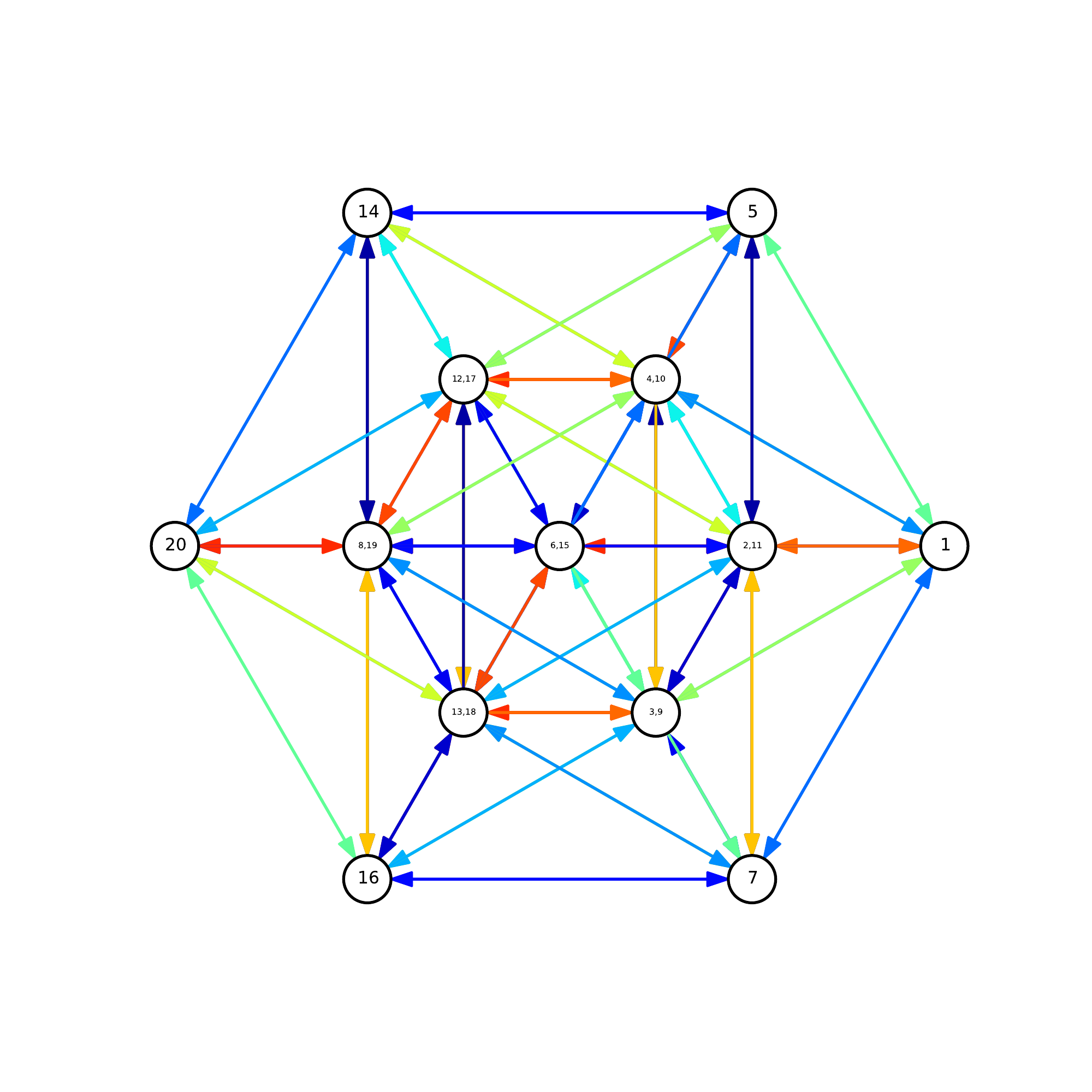}
    	\caption{$\rho_3$}
    	\label{fig:coxeter_projection_A_5_3}
	\end{subfigure}		
	\caption{Coxeter projections for fundamental representations of $A_5$. Each node is a weight and corresponds to a ground state, while each edge connecting a pair of weights is a root and corresponds to a BPS soliton. A root $\alpha$ and its negative $-\alpha$ are both drawn with the same color. These diagrams are generated using \cite{cproj}.
}
\label{fig:coxeter_projection_A_5}
\end{figure}
As an example consider $\mathfrak{g} = A_5$, Coxeter projections for the first three fundamental representations are shown in Figure \ref{fig:coxeter_projection_A_5}. As explained before, the central charge of a soliton can be read out from the Coxeter projection.

\begin{itemize}

\item $\rho = \rho_1$: the Coxeter projection is shown in Figure \ref{fig:coxeter_projection_A_5_1}, with more details available at \href{http://het-math2.physics.rutgers.edu/cproj_A_5_1}{this link} \cite{cproj_A_5_1}. There are $\binom{N}{2} \times 2 = 30$ solitons interpolating every pair of the weights.

\item $\rho = \rho_2$: the Coxeter projection is shown in Figure \ref{fig:coxeter_projection_A_5_2}, with more details available at \href{http://het-math2.physics.rutgers.edu/cproj_A_5_2}{this link} \cite{cproj_A_5_2}. A given ground state is connected to 8 other ground states by solitons, as shown at
\href{http://het-math2.physics.rutgers.edu/cproj_A_5_2_1}{this link} \cite{cproj_A_5_2_1} for solitons from a ground state labeled by $\weight_1$, and at \href{http://het-math2.physics.rutgers.edu/cproj_A_5_2_4}{this link} \cite{cproj_A_5_2_4} for solitons from a ground state labeled by $\weight_4$. These show that there are $8 \times 15 = 120$ solitons, which are $\binom{N-2}{k-1} = \binom{4}{1} = 4$ copies of $\rho_1$ solitons.

\item $\rho = \rho_3$: the Coxeter projection is shown in Figure \ref{fig:coxeter_projection_A_5_3}, with more details available at \href{http://het-math2.physics.rutgers.edu/cproj_A_5_3}{this link} \cite{cproj_A_5_3}. A given ground state is connected to 9 other ground states by solitons, as shown at
\href{http://het-math2.physics.rutgers.edu/cproj_A_5_3_2}{this link} \cite{cproj_A_5_3_2} for solitons from a ground state labeled by $\weight_2$, and at \href{http://het-math2.physics.rutgers.edu/cproj_A_5_3_6}{this link} \cite{cproj_A_5_3_6} for solitons from a ground state labeled by $\weight_6$. Therefore there are $9 \times 20 = 180$ solitons, which are $\binom{N-2}{k-1} = \binom{4}{2} = 6$ copies of $\rho_1$ solitons.

\end{itemize}

\subsubsection{\texorpdfstring{$(\gD_N, \rho_\text{v})$}{D N, rho v} LG models}
For the coset model $\frac{\SO(2N)_1}{\SO(2N-2) \times \UU(1)}$, the minuscule representation of $\gD_N$ corresponding to its ground states is the vector representation of $\SO(2N)$, $\rho_\textrm{v}$, whose dimension is $2N$, the same as the number of ground states. 
When $N \geq 3$, $\rho_\textrm{v}$ has the first fundamental weight $\omega_1 = (1,0,\ldots,0)$ as the highest weight.

The LG superpotential with all deformations turned on is \cite{Eguchi:2001fm}
\begin{align}
	W(X, Y) = \frac{1}{2} X Y^2 + \frac{1}{2(2N-1)}X^{2N-1} + \sum_{i=1}^{N-1} \frac{u_{2i}}{2(2N-1-2i)}X^{2N-1-2i} - \tilde{u}_N Y, \label{eq:LG-DN-superpotential}
\end{align}
from which we get the following vanishing relations
\begin{align}
	\partial_X W &= \frac{1}{2} \left( Y^2 + X^{2N-2} + \sum_{i=1}^{N-1} u_{2i}X^{2N-2-2i} \right) = 0, \label{eq:D_chiral_ring_1}\\
	\partial_Y W &= XY - \tilde{u}_N = 0. \label{eq:D_chiral_ring_2}
\end{align}

When all deformations are set to zero except for $u_{2(N-1)} = z$, the ground states and BPS solitons are identified with weights of $\rho_1$ and the roots connecting them, respectively.  Overall the BPS spectrum contains $2N(2N-2)$ solitons connecting $2N$ ground states.

\begin{figure}[t]
	\centering
	\begin{subfigure}{.45\textwidth}
		\includegraphics[width=\textwidth]{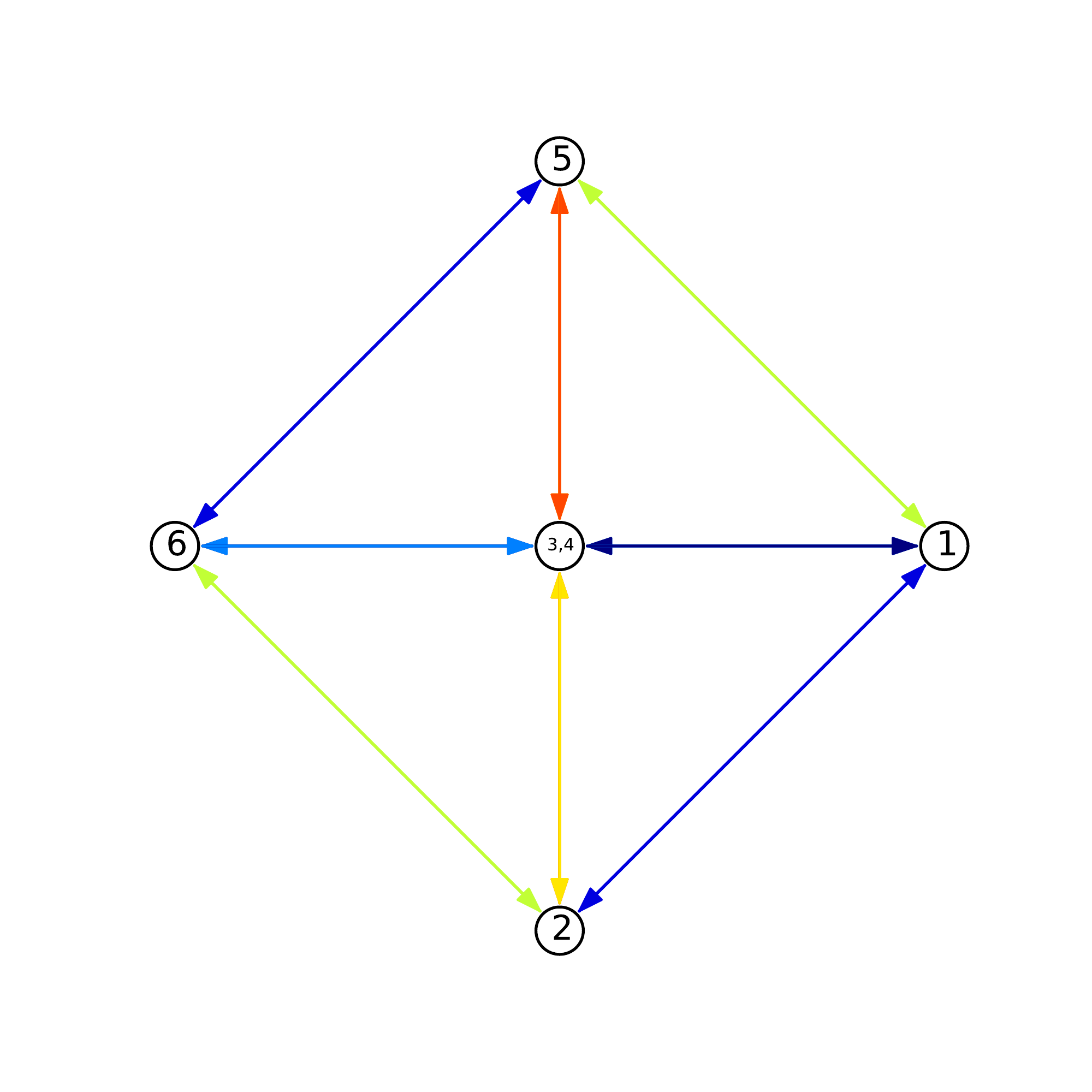}
		\caption{$N=3$}
		\label{fig:coxeter_projection_D_3_1}
	\end{subfigure}
	\begin{subfigure}{.45\textwidth}
		\includegraphics[width=\textwidth]{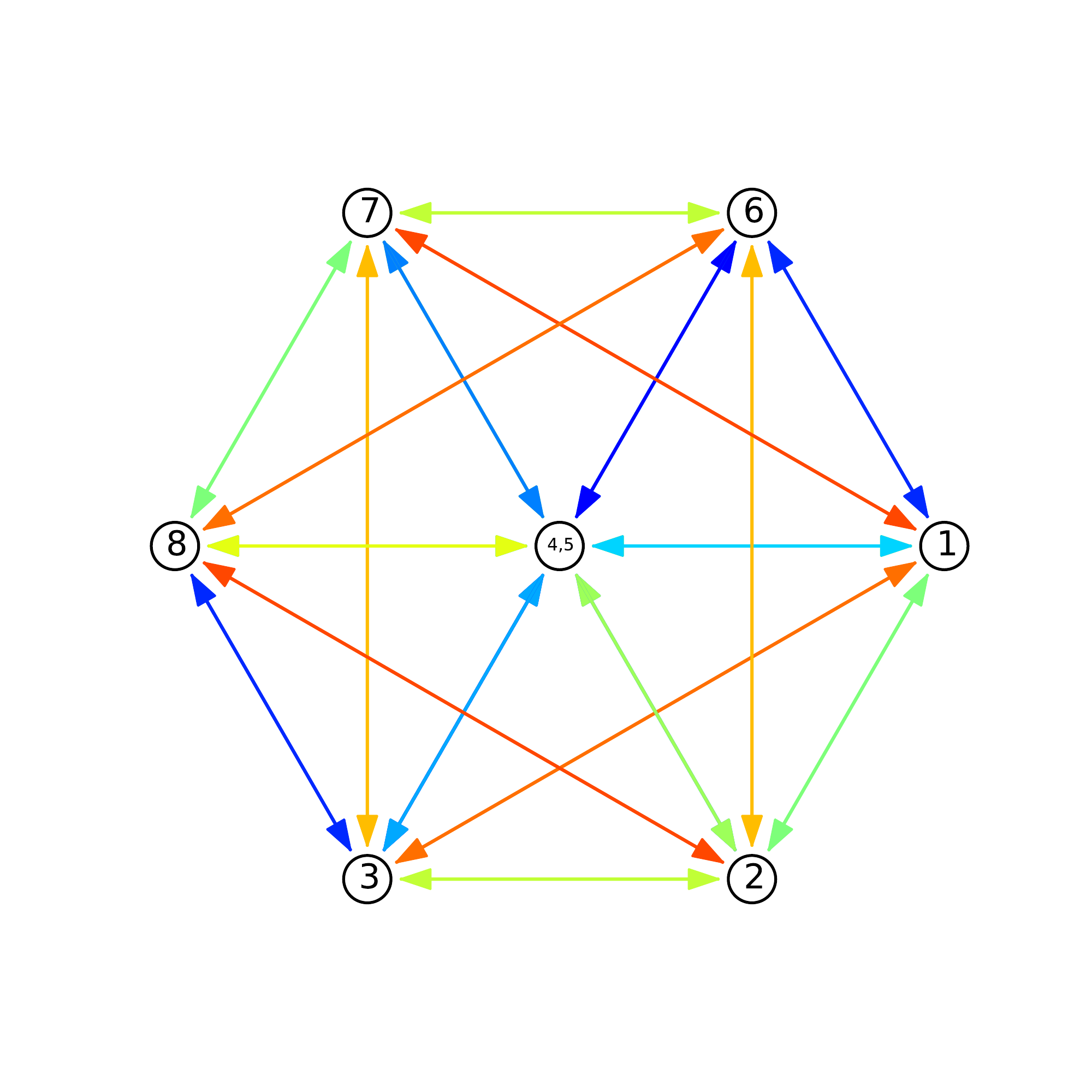}
		\caption{$N=4$}
		\label{fig:coxeter_projection_D_4_1}
	\end{subfigure}
	\caption{Coxeter projection of the vector representation of $\mathfrak{g} = D_N$}
	\label{fig:coxeter_projection_D_N_1}
\end{figure}

Here we consider $\fg = \gD_3$, $\gD_4$ as examples. Figure \ref{fig:coxeter_projection_D_N_1} show the Coxeter projections of their vector representations, with more details available at \href{http://het-math2.physics.rutgers.edu/cproj_D_3_1}{this link} \cite{cproj_D_3_1} for $\gD_3$ and at \href{http://het-math2.physics.rutgers.edu/cproj_D_4_1}{this link} \cite{cproj_D_4_1} for $\gD_4$. There are two things that are obscured by the projection:
\begin{itemize}
	\item There are two weights $\pm e_N$ that are projected to the origin of the plane.%
    \footnote{To see that the weights are projected onto the origin, first note that for the Coxeter element $c = w_N w_{N-1} \cdots w_1$, it acts on $e_N$ as $c(e_N) = (w_N w_{N-1}) (e_N) = w_N(e_{N-1}) = -e_N$. For the Coxeter vector $v_c$ we use for the projection, by definition $c(v_c) = \exp(2\pi i/h^\vee)v_c$. Then $v_c \cdot e_N = c(v_c) \cdot c(e_N) = -\exp(2\pi i/h^\vee) v_c \cdot e_N$, where the first equality comes from the fact that any Weyl group element is an orthogonal transformation. Therefore $v_c \cdot e_N = 0$.} They correspond to $(X, Y) = (0, \pm i \sqrt{z})$. There is no root connecting the pair of weights.
	\item The other $2N-2$ weights are on a circle around the origin, and two weights $\pm e_i$ are located oppositely from the origin, therefore there is no root connecting the pairs.
\end{itemize}

\subsubsection{\texorpdfstring{$(\gD_N, \omega_N)$ LG models}{D N, omega N}}\label{sec:D_N_spinor}

\begin{figure}[t]
	\centering
	\begin{subfigure}{.45\textwidth}
		\includegraphics[width=\textwidth]{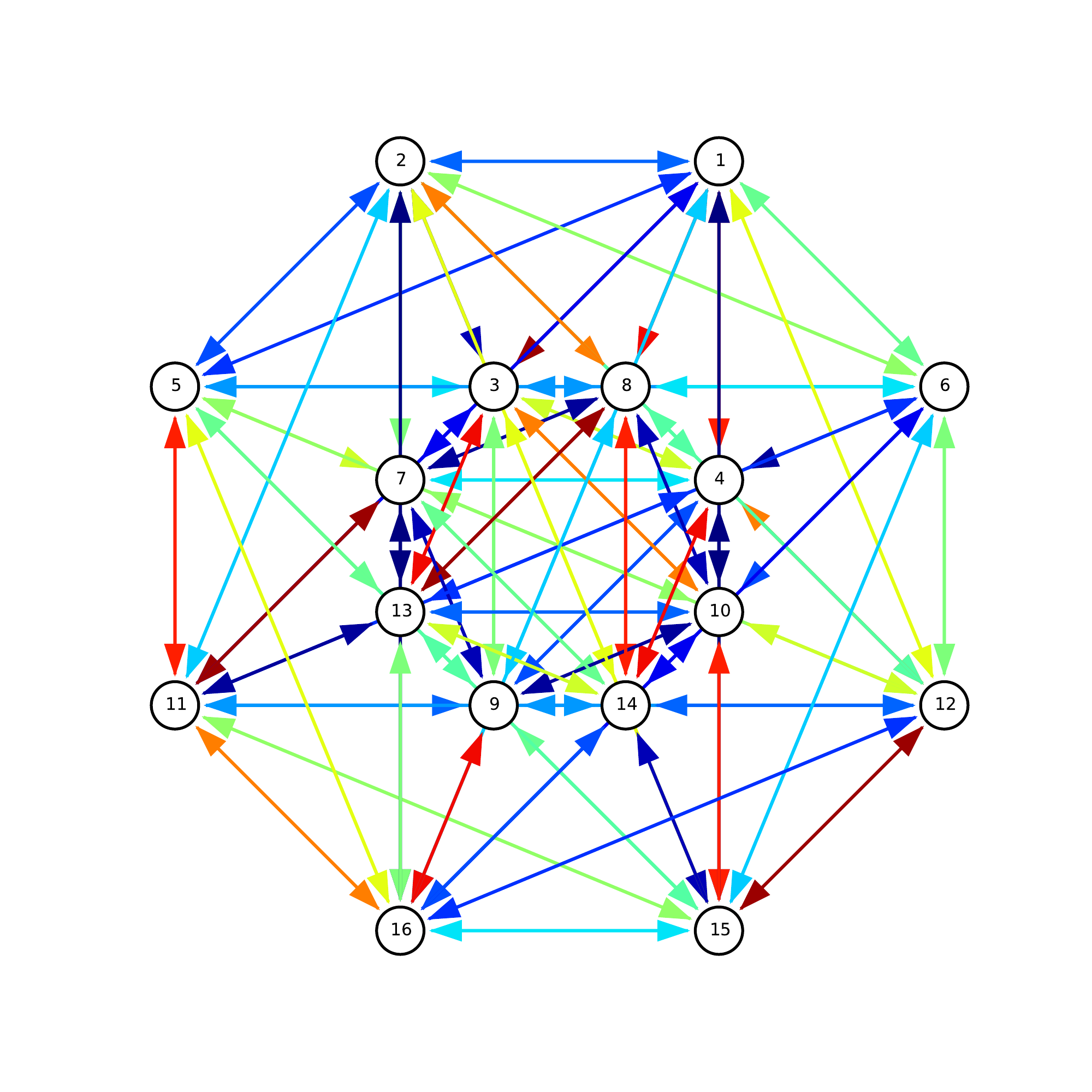}
		\caption{$N=5$}
		\label{fig:coxeter_projection_D_5_5}
	\end{subfigure}
	\begin{subfigure}{.45\textwidth}
		\includegraphics[width=\textwidth]{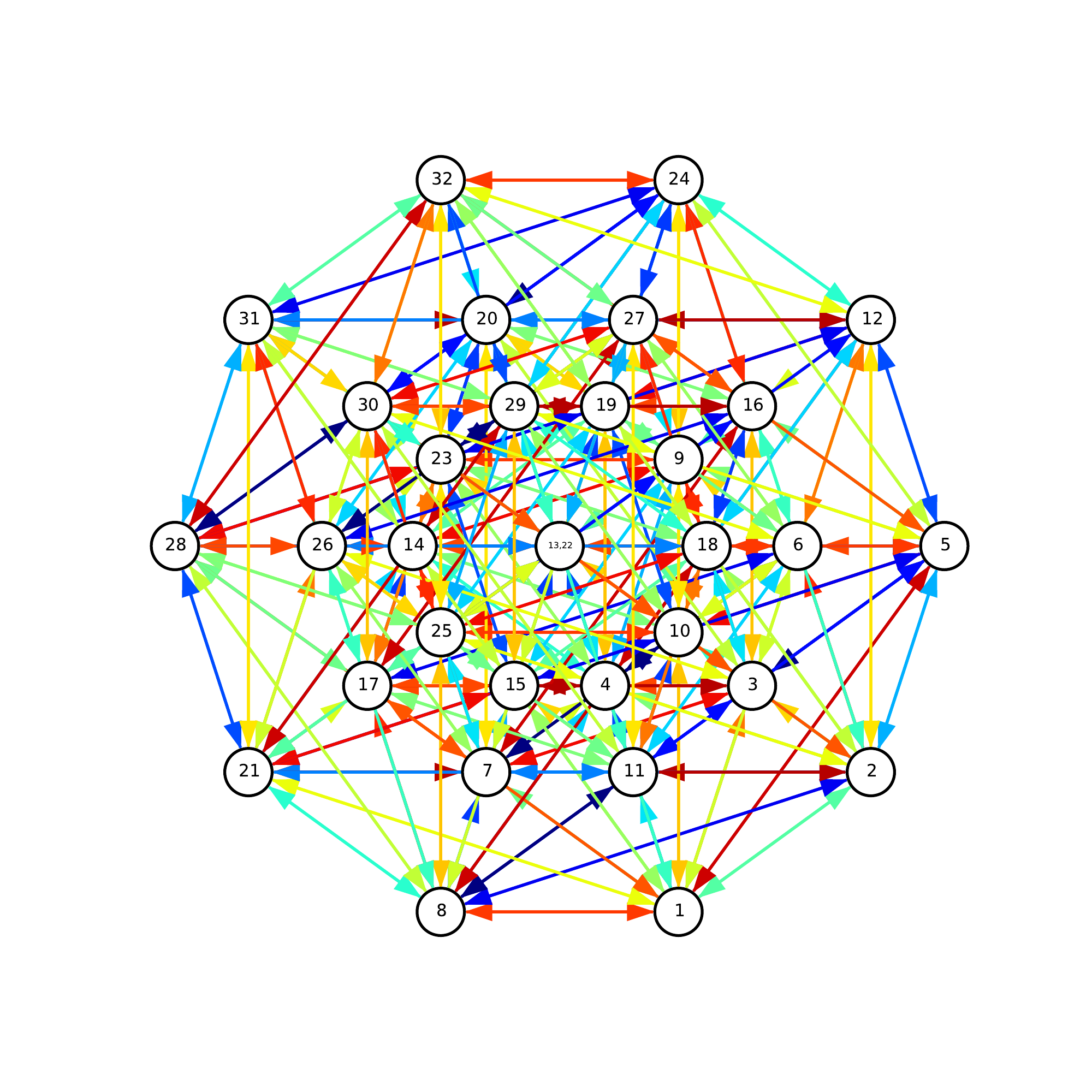}
		\caption{$N=6$}
		\label{fig:coxeter_projection_D_6_6}
	\end{subfigure}
	\caption{Coxeter projection, $\mathfrak{g} = \gD_N$, $\rho = \omega_N$.}
	\label{fig:coxeter_projection_D_N_N}
\end{figure}

The other coset model associated with $\mathfrak{g} = \gD_N$, $\frac{\SO(2N)_1}{\SU(N) \times \UU(1)}$, corresponds to a spinor representation of $\SO(2N)$. Here we focus on its LG model with only the most relevant deformation being turned on. Let's consider an example of $\mathfrak{g} = \gD_5$, $\frac{\SO(10)_1}{\SU(5) \times \UU(1)}$. The superpotential is \cite{Eguchi:2001fm}
\begin{align}
	W(X, Y) = \frac{2372}{25^3 \cdot 9} X^9 + \frac{92}{25^2} X^6\,Y + \frac{36}{25} X^3\, Y^2 + \frac{1}{6} Y^3 - u_8 X,
\end{align}
where $Y = X_3$. 
The vanishing relations are
\begin{align}
  \partial_X W &= \frac{2372}{25^3} X^8 + \frac{552}{25^2} X^5\, Y + \frac{108}{25} X^2\, Y^2 - u_8 = 0,\\
	\partial_Y W &= \frac{92}{25^2} X^6 + \frac{72}{25}X^3\,Y + \frac{1}{2} Y^2 = 0.
\end{align}
There are 16 ground states, corresponding to the 16 weights of a spinor representation of $\SO(10)$. 
Their locations in the $W$-plane($\sim X$-plane) can be found by using the chiral ring equations to eliminate $Y$. 
This coincides with a Coxeter projection shown in Figure \ref{fig:coxeter_projection_D_5_5}, with more details available at \href{http://het-math2.physics.rutgers.edu/cproj_D_5_5}{this link} \cite{cproj_D_5_5}.

\subsubsection{\texorpdfstring{$(E_6, \rho_\text{v})$}{E6, rho v} LG model}

\begin{figure}[t]
  \centering
  \begin{subfigure}{.45\textwidth}
    \includegraphics[width=\textwidth]{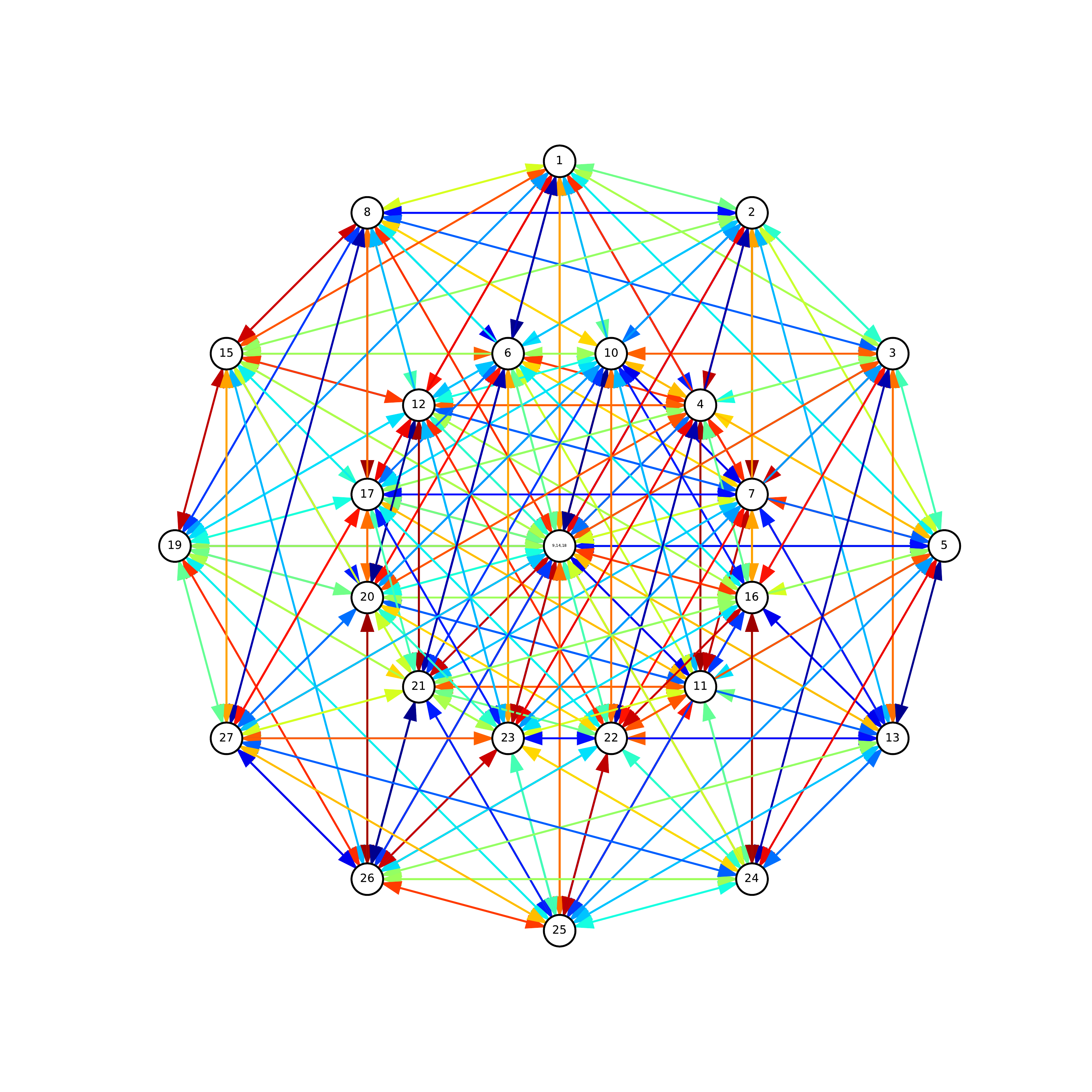}
    \caption{All solitons}
    \label{fig:coxeter_projection_E_6_1}
  \end{subfigure}
  \begin{subfigure}{.45\textwidth}
    \includegraphics[width=\textwidth]{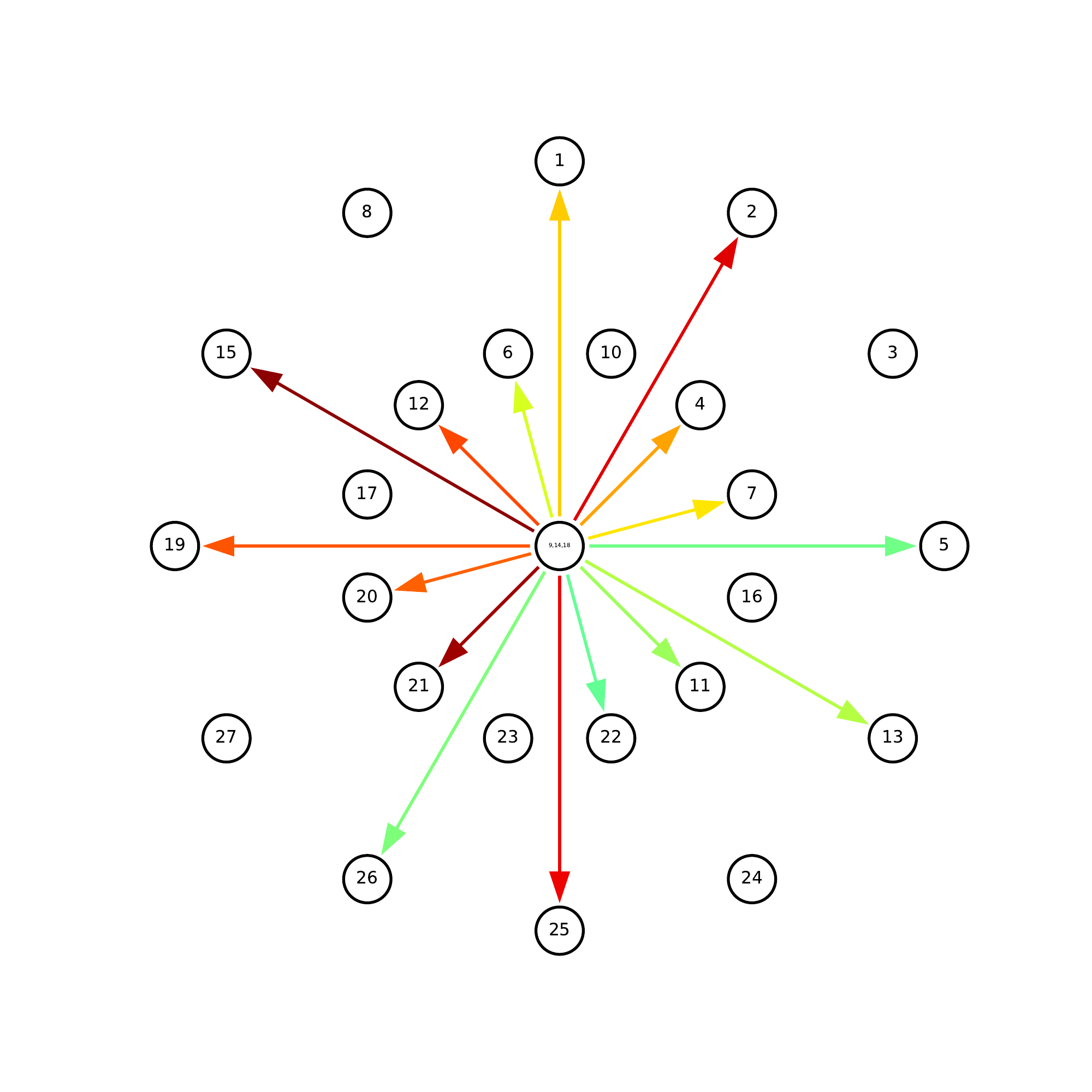}
    \caption{Solitons from $\nu_9$}
    \label{fig:coxeter_projection_E_6_1_9}
  \end{subfigure}
  \caption{Coxeter projection for the ${\bf 27}$ of $\gE_6$.}
  \label{fig:coxeter_projection_E_6}
\end{figure}

Deforming the superpotential of the LG model as $W(X, Y) = W_0(X, Y) + u_{12} \,X$ we obtain the following vanishing relations
\begin{align}
  \partial_X W &= 13 X^{12} + 9 X^8 Y + a Y^3 + u_{12} = 0,\\
  \partial_Y W &= X^9 + 3a X Y^2 = 0.
\end{align}
Solving the above equations yields 3 ground states located at
\begin{align}
  (X, Y) = (0, \omega^k \sqrt{-u_{12}/a}),\, \omega^3 = 1,
\end{align}
with $W = 0$. There are also 24 ground states arranged into two sets of 12, located at
\begin{align}
  X^{12} + c_\pm z = 0,\, Y = \pm\sqrt{3a} X^4,\, c_\pm = 13 \pm 9\sqrt{3a} \pm a(3a)^{3/2},
\end{align}
whose values of $W$ arrange into two concentric circles around $W = 0$. The locations of the ground states on the $W$-plane is the same as the Coxeter projection of \textbf{27} of $\gE_6$, which is shown in Figure \ref{fig:coxeter_projection_E_6_1}, with more details available at \href{http://het-math2.physics.rutgers.edu/cproj_E_6_1}{this link} \cite{cproj_E_6_1}. From each ground state we have 16 solitons connecting the ground state to other ground states as shown in Figure \ref{fig:coxeter_projection_E_6_1_9} and at \href{http://het-math2.physics.rutgers.edu/cproj_E_6_1_9}{this link} \cite{cproj_E_6_1_9}, therefore there are $16 \times 27 = 432$ solitons in total.


\subsubsection{\texorpdfstring{$(E_7, \rho_\text{v})$}{E7, rho v} LG model}
\begin{figure}[t]
  \centering
  \begin{subfigure}{.45\textwidth}
    \includegraphics[width=\textwidth]{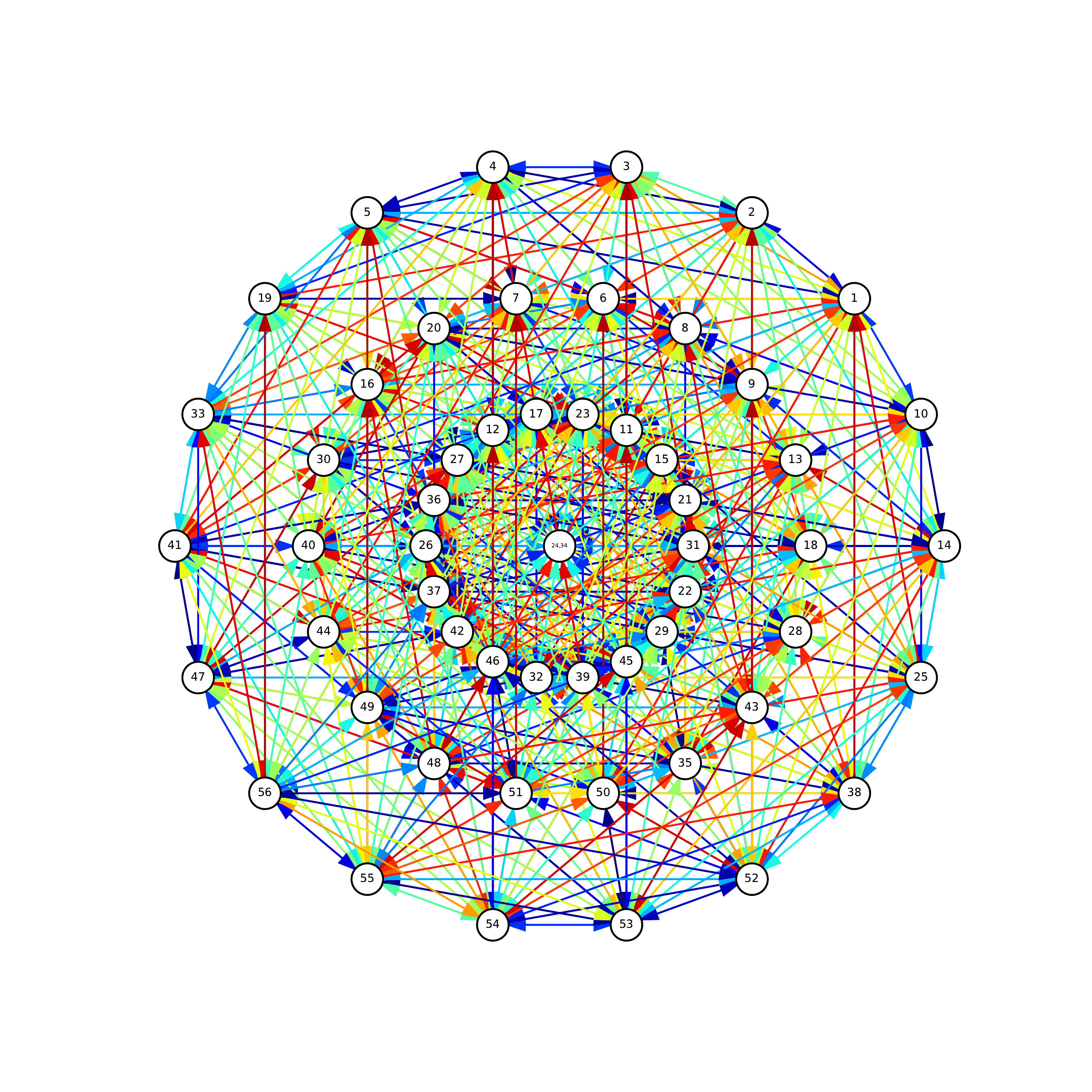}
    \caption{All solitons}
    \label{fig:coxeter_projection_E_7_7}
  \end{subfigure}
  \begin{subfigure}{.45\textwidth}
    \includegraphics[width=\textwidth]{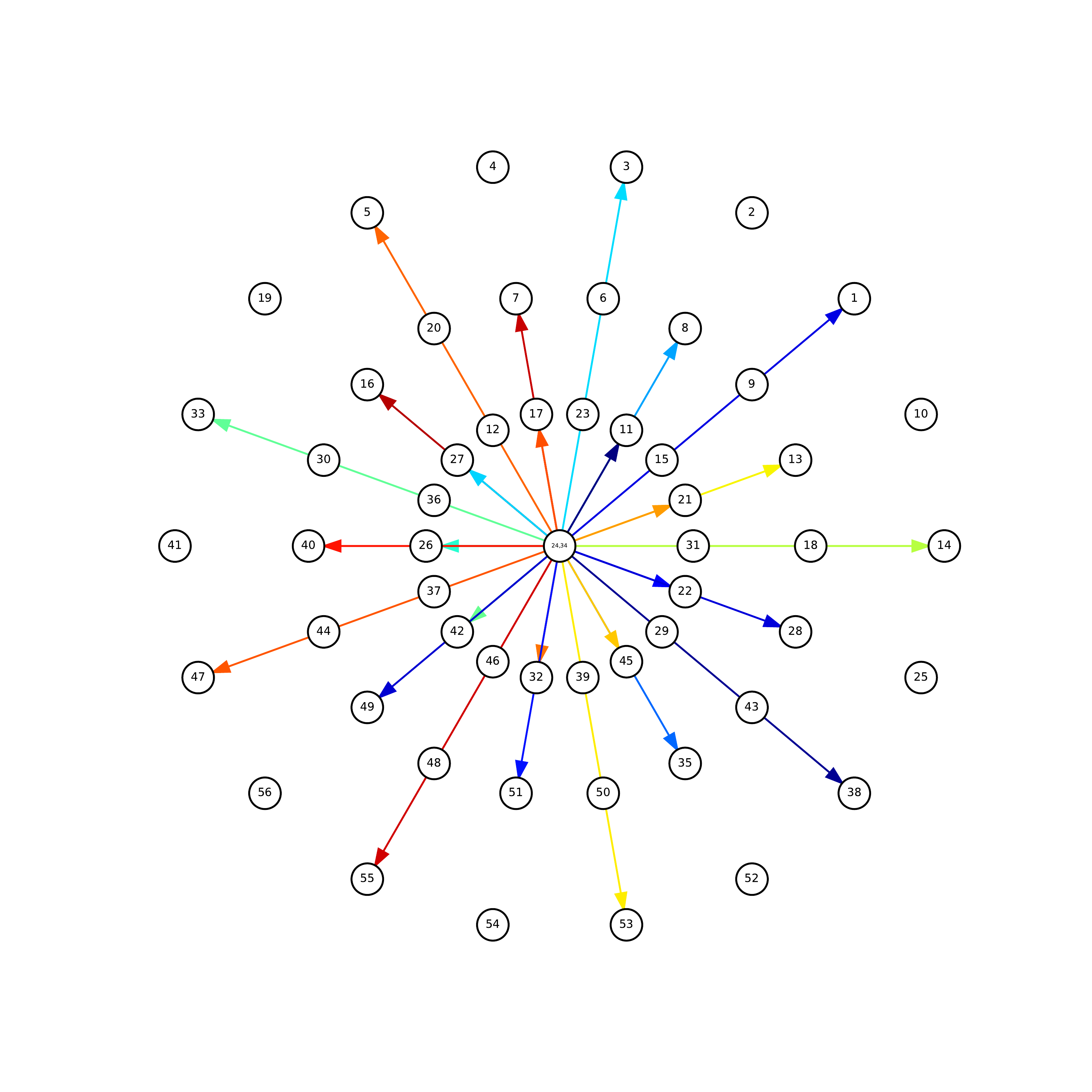}
    \caption{Solitons from $\nu_{24}$}
    \label{fig:coxeter_projection_E_7_7_24}
  \end{subfigure}
  \caption{Coxeter projection, $\mathfrak{g} = \gE_7$.}
  \label{fig:coxeter_projection_E_7}
\end{figure}

Deforming the superpotential of the LG model as  $W(X, Y) = W_0(X, Y) + u_{18} \,X$ we get the following vanishing relations
\begin{align}
	\partial_X W &= 19 X^{18} + Z^2 + 14a X^{13} Y + 10b X^9 Z + u_{18}=0,\\
	\partial_Y W &= 2YZ + a^{14}=0,\\
	\partial_Z W &= Y^2 + 2X Z + bZ^{10}=0.
\end{align}
Solving the above equations gives 2 ground states 
\begin{align}
  (X, Y, Z) = (0, 0, \pm \sqrt{-u_{18}})
\end{align}
with $W = 0$. There are 54 additional ground states, arranged into three sets of 18 ground states, located at $(X, Y, Z)$ that satisfy
\begin{align}
  X^{18}  &= -\frac{u_{18}}{19 + {c_i}^2 + 7a^2/c_i + 10\,b\, c_i},\\
  Y &= \frac{a}{2\, c_i} X^5,\\
  Z &= c_i X^9,
\end{align}
where $c_i$ $(i = 1, 2, 3)$ are the three solutions of
\begin{align}
	a^2 + 4\,b\, {c_i}^2 + 8 {c_i}^3 = 0.
\end{align}
On the $W$-plane ground states from each set arrange into a concentric circle around $W = 0$, as shown in Figure \ref{fig:coxeter_projection_E_7}, with more details available at \href{http://het-math2.physics.rutgers.edu/cproj_E_7_7}{this link} \cite{cproj_E_7_7}. From each ground state we have 27 solitons connecting the ground state to other ground states as shown in Figure \ref{fig:coxeter_projection_E_7_7_24} and at \href{http://het-math2.physics.rutgers.edu/cproj_E_7_7_24}{this link} \cite{cproj_E_7_7_24}, therefore there are $27 \times 56 = 1512$ solitons in total. 

\subsection{LG models and superconformal limits of canonical defects}
We first observe that we can relate the LG superpotential of a coset model associated with a minuscule representation $\rho$ of $\fg$ to the spectral curve $F(z, x)$ associated with  a superconformal limit of a canonical defect $\SD{\fg}{\rho}{z}$. 
For the sake of generality we shall temporarily work with a slightly generalized spectral curve, and turn on all relevant deformations for both the LG model and the defect. 
Later we will  specialize to the superconformal limit of the defect.
The starting point is to reformulate the description of \cite{Brandhuber:1995zp, Lerche:1996an, Eguchi:2001fm} into the framework of class $\cS$ theories.%
\footnote{All the following algebraic manipulations are already described in \cite{Eguchi:2001fm}, for example see Section 3.4 and Appendix A.1 in the paper. The only new point here is we identify the characteristic polynomial $P^{\mathcal{R}}_{G}$ of the paper with a spectral cover associated with $\rho = \mathcal{R}$ and $\mathfrak{g} = \mathrm{Lie}(G)$.
} 
Consider a class $\CS$ theory of type $\fg$ on $C=\IC$ parametrized by the following Casimir invariants. For every degree-$w_i$ Casimir invariant of $\fg$ choose 
\begin{align}
	\phi_{w_i} = u_{w_i} (\dd z)^{w_i},
\end{align}
where $u_{w_i}$ are complex parameters that are independent of $z$ for all $w_i$ except the highest degree $w_i = h^\vee$, for which we set
\begin{align}
	\phi_{h^\vee} = (u_{h^\vee} - z)\, (\dd z)^{h^\vee}. 
\end{align}
Out of this data we can construct a spectral curve $F(z, x)$ associated with a minuscule representation $\rho$ of $\fg$ as explained in \cite{Longhi:2016rjt},
\begin{align}
    F(z, x)\, (\dd z)^{d}= \det \left[ \lambda\, \mathbb{I}_d - \rho(\varphi(z)) \right] = \det \left[ (x \,\dd z) \mathbb{I}_d - \rho(\varphi(z)) \right], \label{eq:spectral_curve_def}
\end{align}
where $\varphi$ is a $\fg$-valued differential whose Casimir invariants are $\phi_{w_i}$, and $d = \dim(\rho)$. The explicit form of $F(z, x)$ for $(\fg, \rho_\text{v})$ can be found in Appendix F of \cite{Longhi:2016rjt} in terms of Casimir invariants of $\fg$. The resulting curve is the spectral curve of $\SD{\fg}{\rho}{z}$, a superconformal limit of a canonical defect, whose explicit forms with all $u_{w_i}$ set to zero are given in (\ref{eq:A_type_local_model}--\ref{eq:E_7_type_local_model}) for the vector representation $\rho = \rho_\text{v}$.
Solving $F(z, x)$ for $z$ gives a function $z = \tau(x; u_{w_i})$ called the \emph{prepotential of the single-variable potential} $\mathcal{W}(X)$ in \cite{Eguchi:2001fm}, defined as
\begin{align}
	\tau(X; u_{w_i}) = \partial_X \mathcal{W}(X; u_{w_i})\,. 
\end{align}
The single-variable potential $\mathcal{W}(X = X_1; u_{w_i})$ can be obtained from the superpotential $W(X_i; u_{w_i})$ of the LG model by eliminating chiral fields $X_{i > 1}$ except the lowest degree one $X = X_1$ using vanishing relations $\{\partial_{X_{i>1}} W(X_i; u_{w_i}) = 0\}$.
Therefore we see that the superpotential $W(X_i)$ is related via the single-variable potential $\mathcal{W}(X)$ to the class $\CS$ spectral curve associated with $\SD{\fg}{\rho}{z}$.

$\mathcal{W}(X)$ encodes part of the holomorphic data of the original superpotential $W(X_i)$ of the LG model. 
This includes the locations of the critical points $\{W_a\}$ in the $W$-plane, and the central charges $Z_{ab} = W_b - W_a$ of solitons  interpolating between two critical points $(W_a, W_b)$. Therefore we can conclude that, even with all the relevant deformations turned on, the ground states of the LG model and the defect match with each other.

However, it is not obvious that we can recover the degeneracies of BPS solitons just from $\widetilde W$, which is encoded in the original superpotential $W$ in a non-holomorphic way.\footnote{We thank Greg Moore for clarifying this difference between $\mathcal{W}(X)$ and $W(X_i)$.} Therefore it is a nontrivial fact that the BPS spectrum of a surface defect, computed via spectral networks, coincides with the BPS spectrum of the LG model. 
For $\fg = \gA_n$ this correspondence was first observed in \cite{Hori:2013ewa}. Here we use ADE spectral networks to extend the correspondence between LG models of types $\gD$ and $\gE$ and superconformal limits of corresponding canonical defects.

Before we delve into detailed studies of each defect, let us briefly describe how ADE spectral covers and networks look like around a branch point of order $h^\vee$. 
For $z$ close to the branch point the $x$-coordinates of the fiber of the spectral curve $\Sigma_\rho$ coincide with the Coxeter projection of $\rho$ up to an overall complex rescaling. When the defect is at $z$, these locations on the $x$-plane correspond to ground state of the defect, therefore they coincide with the ground states of the corresponding LG model. 
From the branch point emanate $\CS$-walls, and there is an $\CS$-wall for every root of $\fg$.\footnote{There are more $\CS$-walls than roots of $\fg$, which is required by the consistency of a spectral network.} 
The BPS solitons of the surface defect are counted by the $\mathcal{S}$-walls that sweep through $z$ as the phase $\vartheta$ of the spectral network varies over $2 \pi$, and exactly one $\CS$-wall passes through $z$ for every root of $\fg$, resulting in a 2d BPS soliton for every root of $\fg$. 
This BPS spectrum coincides exactly with that of the corresponding LG model. 

The remainder of this section is devoted to illustrating the correspondence with explicit examples.

\subsubsection{\texorpdfstring{$(\gA_{N-1}, \omega_1)$}{A N-1, omega 1} surface defect}
The spectral curve that describes $\SD{\gA_{N-1}}{\omega_1}{z}$ is \cite{Hori:2013ewa}
\begin{align}
	F(z, x) = x^{N} + u_2 x^{N-2} + \cdots + u_{N-1} x + (u_N - z) = 0\,, \label{eq:curve_A_omega_1}
\end{align}
which corresponds to the choice of degree-$w_i$ differentials $\phi_{w_i}$ as
\begin{align}
\begin{split}
	\phi_{j} &= u_{j}(dz)^{j},\ j = 2, \ldots, N-1, \\
    \phi_{N} &= (u_N - z)\, (dz)^N. \label{eq:A_type_class_S_data}
\end{split}
\end{align}
Solving $F(z, x) = 0$ for $z$ gives the prepotential of $\mathcal{W}(X)$, from which we get
\begin{align}
    \mathcal{W}(X) = \frac{1}{N+1} X^{N+1} + \frac{u_2}{N-1} X^{N-1} + \cdots + \frac{u_{N-1}}{2} X^2 + u_N\, X.
\end{align}
Introducing the same deformations to the LG model associated with $(\gA_{N-1}, \omega_1)$ gives the same single variable potential we can get from the spectral curve, which implies that the LG model and the defect have the same ground states even when the deformations are turned on. 

When setting all $u_j=0$, we obtain the superconformal limit of the canonical surface defect described in Section \ref{sec:superconformal_limit_definition}. As claimed in \cite{Hori:2013ewa}, in this limit the spectral network reproduces the BPS spectrum of the LG model \cite{Lerche:1991re}.

\subsubsection{\texorpdfstring{$(\gA_{N-1}, \omega_{k > 1})$}{A N-1, omega k gt 1} surface defect}

The spectral curve that describes $\SD{\gA_{N-1}}{\omega_{k > 1}}{z}$ can be obtained from the curve of $\rho = \omega_1$, we refer the reader to \cite{Longhi:2016rjt} for more details. The class $\CS$ data is the same for all minuscule representation, and the spectral curve $F(z, x)$ is obtained via (\ref{eq:spectral_curve_def}). Because the single variable potential $\mathcal{W}(X)$ can be obtained via the same characteristic polynomial \cite{Eguchi:2001fm}, we can conclude that the LG model associated with $(\gA_{N-1}, \omega_k)$ is related via $\mathcal{W}(X)$ to the defect, with all the relevant deformations being turned on. This implies that they have the same ground states.

Spectral network remains essentially identical for all representations $\rho = \omega_k$: what changes is the soliton spectrum carried by $\CS$-walls. Thanks to the \emph{soliton symmetry} found in \cite{Longhi:2016rjt}, we can infer that the 2d solitons of the defect in a higher fundamental representation  
have the same \emph{soliton trees} as the 2d solitons of the defect in the first fundamental representation. In practice, this means that the central charges coincide, and that all 2d BPS degeneracies are enhanced by a factor of $\binom{N-2}{k-1}$. This result is consistent both with the 2d BPS spectra of the LG theories \cite{Lerche:1991re} and with the BPS spectrum computed from the analysis of brane configurations \cite{Hori:2013ewa}.

\subsubsection{\texorpdfstring{$(\gD_N, \rho_{\text{v}})$}{D N, rho v} surface defect}
The spectral curve that describes $\SD{\gD_{N}}{\rho_{\text{v}}}{z}$ is
\begin{align}
    F(z, x) = x^{2N} + \sum_{j=m_i+1}^{2N-2} u_j\, x^{2N-j} + \left( \tilde{u}_N \right)^2,
\end{align}
corresponding to the choice of degree-$w_i$ differentials $\phi_{w_i}$ as
\begin{align}
\begin{split}
	\phi_{j} &= u_{j}(dz)^{j},\ j = 2, 4, \ldots, 2N-4;\\
	\phi_{2N-2} &= \left(u_{2N-2} - z \right)\, (dz)^{2N-2},\\
    \phi_{N} &= \tilde{u}_N (dz)^N. \label{eq:D_N_class_S_data}
\end{split}
\end{align}
By solving $F(z, x) = 0$ for $z$ we get the corresponding single-variable potential as
\begin{align}
    \mathcal{W}(X) = \frac{1}{2N-1}X^{2N-1} + \sum_{i=1}^{N-1} \frac{u_{2i}}{2N-1-2i}X^{2N-1-2i} - \frac{\left(\tilde{u}_N\right)^2}{X},
\end{align}
which agrees with what we get from the superpotential (\ref{eq:LG-DN-superpotential}) by eliminating $Y$ using the vanishing relation (\ref{eq:D_chiral_ring_2}) up to a numerical factor of $\frac{1}{2}$. This again provides the relations between the ground states of the LG model and the defect with all the relevant deformations being nonzero.

\begin{figure}[t]
  \centering
  \begin{subfigure}{.45\textwidth}
    \includegraphics[width=\textwidth]{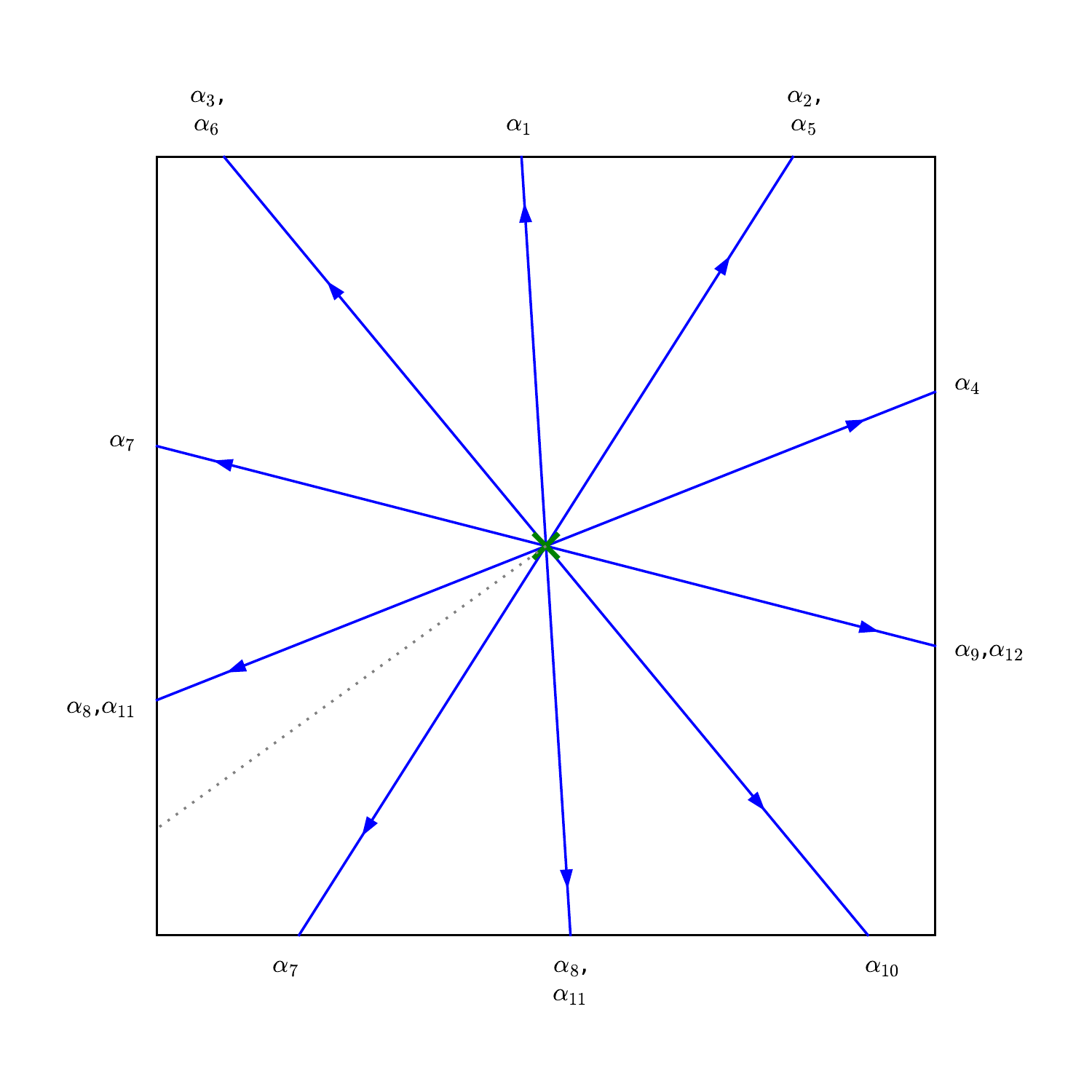}
    \caption{$N=3$}
    \label{fig:D_3_SN}
  \end{subfigure}
  \begin{subfigure}{.45\textwidth}
    \includegraphics[width=\textwidth]{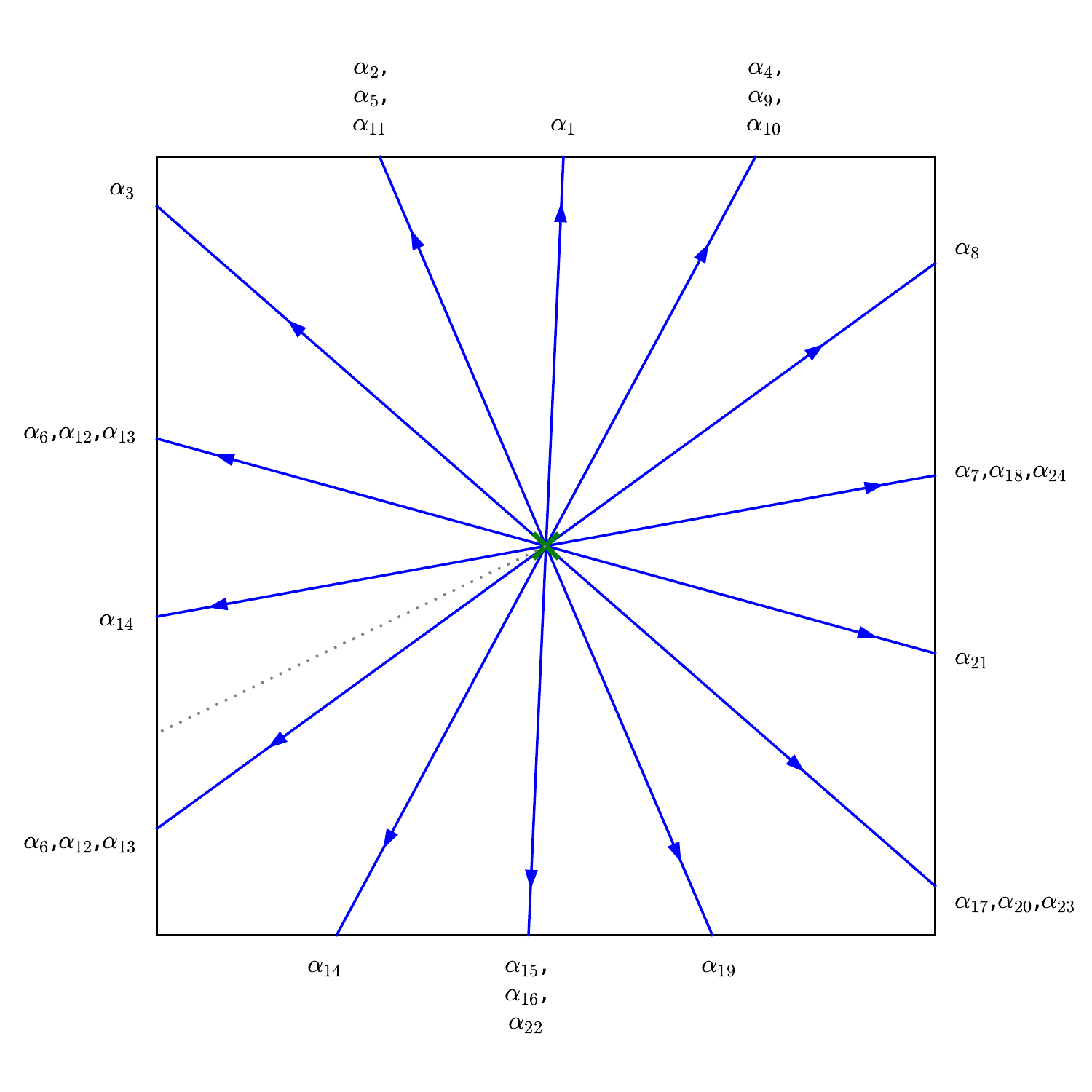}
    \caption{$N=4$}
    \label{fig:D_4_SN}
  \end{subfigure}
  \caption{Spectral networks for the superconformal limits of canonical surface defects of $\SO(2N)$ pure gauge theories, corresponding to the vector representation. 
  Each $\CS$-wall is labeled by associated roots, see \href{http://het-math2.physics.rutgers.edu/cproj_D_3_1}{this link} \cite{cproj_D_3_1} for $\gD_3$ and \href{http://het-math2.physics.rutgers.edu/cproj_D_4_1}{this link} \cite{cproj_D_4_1} for $\gD_4$ for labeling conventions of the roots.} 
  \label{fig:D_N_SN}
\end{figure}

The superconformal limit of the defect corresponds to taking all $u_j = 0$. One way to visualize such a limit in terms of class $\CS$ theory corresponds to ``zooming in'' near one of the two branch points of the spectral curve of pure $\SO(2N)$ gauge theory that is described in Section 5.3 of \cite{Longhi:2016rjt}, in Sections 5.3.1 and 5.3.3 of which describe examples of $\fg = \gD_3$ and $\gD_4$ in detail. The resulting network is shown in Figure \ref{fig:D_N_SN}. The analysis carried out in \cite{Longhi:2016rjt} shows that the BPS spectrum of the defect obtained using spectral networks coincides with that of the LG model with the most relevant deformation $u_{2N-2} \neq 0$. In the following we illustrate this with examples of $\fg = \gD_3$ and $\gD_4$.

Figure \ref{fig:D_3_SN} shows the spectral network associated with the defect $\SD{\gD_3}{\omega_1}{z}$.
There are 15  $\CS$-walls sourced by a branch point, and the fan of walls rotates by $2\pi \left( \frac{2N-2}{2N-1} \right)= 2\pi \left(\frac{4}{5}\right)$ when we change the phase of the spectral network by $2\pi$, resulting in $24$ BPS solitons. 
To see this we fix $z$ near the branch point, and count the $\CS$-walls sweeping through $z$. 
There are 4 pairs of overlapping $\CS$-walls that move across $z$, corresponding to 4 quartets of BPS solitons with the same central charges because each wall carries 2 solitons. 
In addition there are 4 regular $\CS$-walls, carrying 2 solitons each. Overall this gives 12 BPS solitons plus 12 conjugates. This result is consistent with the BPS spectrum of the LG model obtained from the Coxeter projection of Figure \ref{fig:coxeter_projection_D_3_1}. 

A similar analysis can be carried out for $\fg = \gD_4$, with corresponding spectral network shown in Figure \ref{fig:D_4_SN}. There are now 28 $\CS$-walls, and the spectral network rotates by $2\pi \left( \frac{2N-2}{2N-1} \right)= 2\pi \left(\frac{6}{7}\right)$ when we change the phase of the spectral network by $2\pi$. Druing the rotation there are 6 triplets of overlapping $\CS$-walls moving across fixed $z$, corresponding to $6$ sextets of BPS solitons with the same value of central charges. In addition there are 6 regular $\CS$-walls moving across the defect, each carrying 2 BPS states. This gives us $24$ BPS solitons plus $24$ conjugates, which is consistent with the BPS spectrum of the LG model obtained from the Coxeter projection of Figure \ref{fig:coxeter_projection_D_4_1}.

\subsubsection{\texorpdfstring{$(\gD_N, \omega_N)$}{D N, omega N} surface defect}\label{sec:D_N_omega_N_class_S_defect}
As is the case for the defects in higher fundamental representations of $\gA_N$, we can obtain the spectral curve for the spinor representation from that of the vector representation. The spectral curve of $\SD{\gD_N}{ \omega_N}{z}$ is
\begin{align}
	F(z, x) = 
	\prod^{2^{N-1}} \left[x + \left(\sum_{i=1}^{N} \pm \frac{1}{2}a_i\right) \right], \label{eq:spin_D-type_characteristic_polynomial}
\end{align}
where
\begin{align}
	(-1)^{N+j}\sum_{1 \leq i_1 < i_2 < \cdots < i_j \leq N} a_{i_1}^2 a_{i_2}^2 \cdots a_{i_j}^2 &=
    \begin{cases}
        u_{j},\quad j = 2, 4, \ldots, 2(N-2),\\
        u_{2N-2} - z,\quad j = 2N-2,
    \end{cases}\nonumber \\
	a_{1} a_{2} \cdots a_{N} &= \tilde{u}_N. \nonumber
\end{align}
This curve is obtained from (\ref{eq:spectral_curve_def}) using the same class $\CS$ data (\ref{eq:D_N_class_S_data}) of the vector representation but in the spinor representation, therefore it leads to the same single variable potential $\mathcal{W}(X)$ that can be obtained from the superpotential of the LG model associated with $(\gD_N, \omega_N)$ with all the nonzero relevant deformations, hence the same ground states for both the LG model and the defect. 

To consider the superconformal limit of such a defect, we set $u_{w_i} = 0$. Then (\ref{eq:spin_D-type_characteristic_polynomial}) is an even function of $x$, and is a homogeneous polynomial of degree $2^{N-1}$, where $x$ has degree 1 and $z$ has degree $2(N-1)$. Here are a few explicit examples:
\begin{itemize}
	\item $N=4$: $F(z, x) = x^8 - z\, x^2$ up to a rescaling of $z$. This gives us the same spectral curve as that of $\SD{\gD_4}{ \omega_1}{z}$ because of the triality of $\gD_4$.
	
	\item $N=5$: $F(z, x) = x^{16} + c\, z\, x^8 + z^2$, where $c$ can be determined by considering the Coxeter projection of the 16-dimensional spin weights shown in Figure \ref{fig:coxeter_projection_D_5_5}, with more details available at \href{http://het-math2.physics.rutgers.edu/cproj_D_5_5}{this link} \cite{cproj_D_5_5}.
	
	\item $N=6$: $(z, x) = x^{32} + c_1\, z\, x^{22} + c_2\, z^2 x^{12} + z^3 x^2 = x^2 \left( x^{30} + c_1\, z\, x^{20} + c_2\, z^2 x^{10} + z^3 \right)$, where $c_i$ can be determined by considering the Coxeter projection of the 32-dimensional spin weights shown in Figure \ref{fig:coxeter_projection_D_6_6}, with more details available at \href{http://het-math2.physics.rutgers.edu/cproj_D_6_6}{this link} \cite{cproj_D_6_6}.
	Note that the projection results in 2 degenerate images at the origin and three groups of 10 weights on concentric circles around the origin.

\end{itemize}

As in the case of the vector representation, we can use the analysis of Section 5.3 of \cite{Longhi:2016rjt} to show that in the superconformal limit the BPS spectrum of the surface defect matches that of the LG model. See in particular Section 5.3.2 of that reference for an explicit example with $\fg = \gD_3$.

\subsubsection{\texorpdfstring{$(\gE_N, \rho_\text{v})$}{E N, rho v} surface defect}
The spectral curve associated with $\SD{\gE_N}{\omega_N}{z}$, with all the deformation turned on, is obtained from the class $\CS$ data of $\phi_{w_i} = u_{w_i}\left(\mathrm{d}z\right)^{w_i}$, $w_i < h^\vee$ and $\phi_{h^\vee} = \left(u_{h^\vee} - z\right) \left(\mathrm{d}z\right)^{h^\vee}$. For the explicit form of the curve, see Appendix F of \cite{Longhi:2016rjt}. By construction the spectral curve is related via the single variable potential to the superpotential of the LG model associated with $(\gE_N, \omega_N)$, resulting in the same ground state stuctures in both 2d models.

\begin{figure}[t]
  \centering
  \begin{subfigure}{.45\textwidth}
    \includegraphics[width=\textwidth]{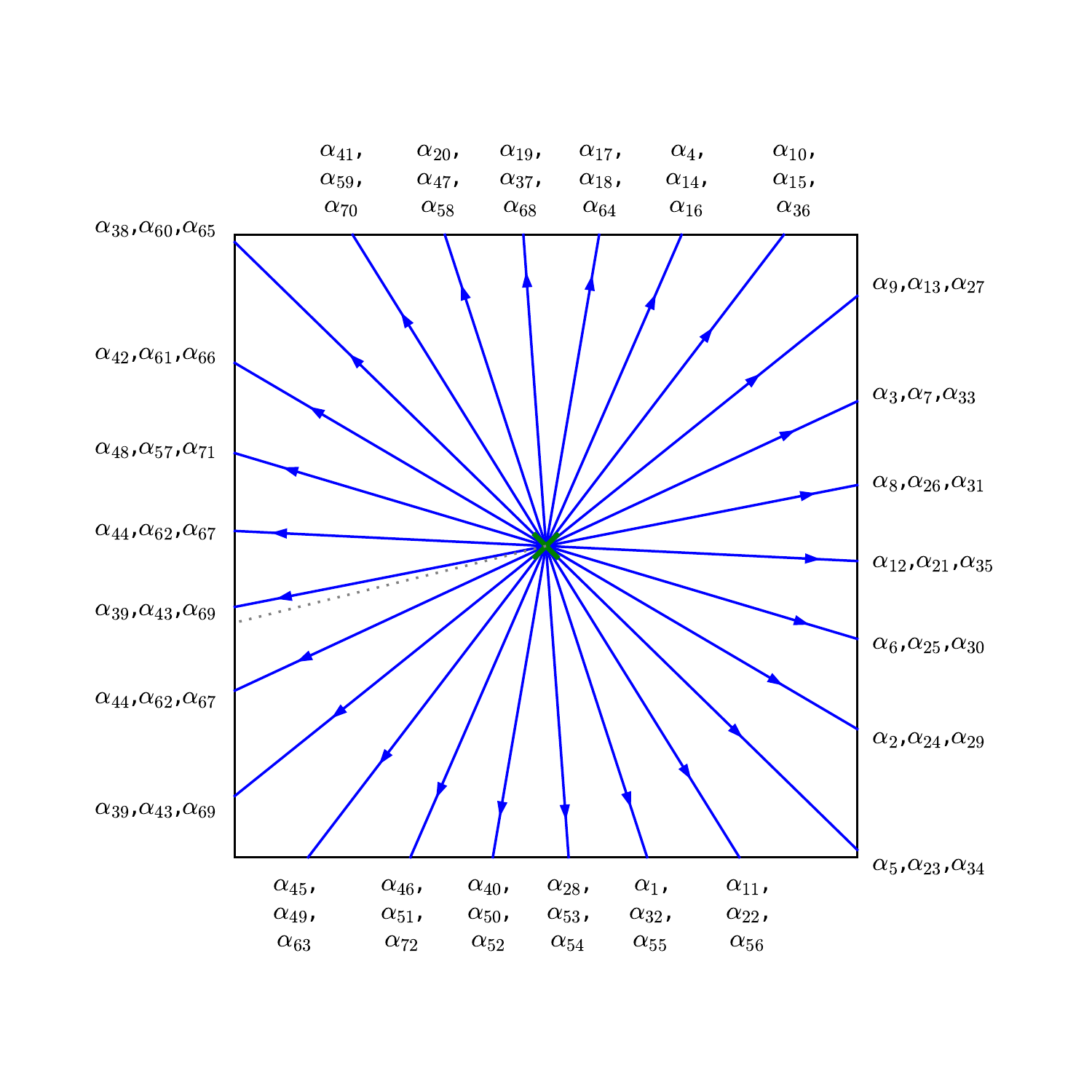}
    \caption{$N=6$}
    \label{fig:E_6_SN}
  \end{subfigure}
  \begin{subfigure}{.45\textwidth}
    \includegraphics[width=\textwidth]{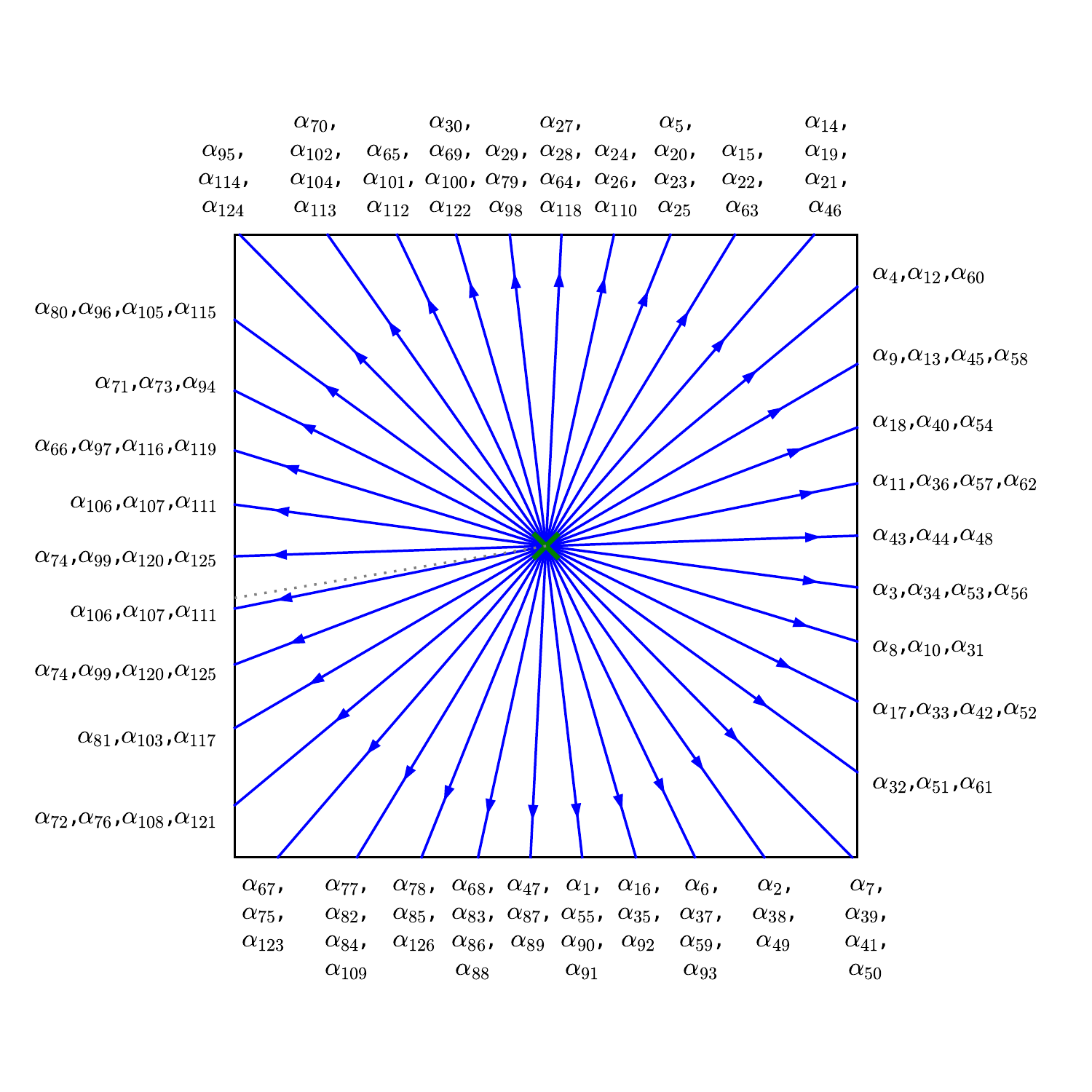}
    \caption{$N=7$}
    \label{fig:E_7_SN}
  \end{subfigure}
  \caption{Spectral networks for the superconformal limits of canonical surface defects of $\gE_N$ pure gauge theories, corresponding to the vector representation. Each $\CS$-wall is labeled by associated roots, see \href{http://het-math2.physics.rutgers.edu/cproj_E_6_1}{this link} \cite{cproj_E_6_1} for $\gE_6$ and \href{http://het-math2.physics.rutgers.edu/cproj_E_7_7}{this link} \cite{cproj_E_7_7} for $\gE_7$ for labeling conventions of roots.} 
  \label{fig:E_N_SN}
\end{figure}

\paragraph{$\boldsymbol{N}\mathbf{=6}$} 

When $u_{w_i} = 0$, the spectral curve of $\SD{\gE_6}{\omega_6}{z}$ simplifies to
\begin{align}
	F(z, x) = x^{27} - 5\, z\, x^{15} - \frac{1}{108} z^2 \, x^3=0.
\end{align}
For a fixed value of $z$ there are 27 sheets of $x$, three of which is at $x = 0$. 
This curve has a single branch point at $z=0$, where all 27 sheets collide and three branch cuts of orders 3, 12, and 12 come out. Its spectral network is shown in Figure \ref{fig:E_6_SN}.
The branch point emanates $78$ walls, each carrying $6$ solitons ($k_\rho=6$ in the conventions of \cite{Longhi:2016rjt}). 
As discussed in Section 5.3.4 of \cite{Longhi:2016rjt}, in the spectral network there are 3 $\CS$-walls on top of each other, thereby showing only 26 flow lines emanating from the branch point. Note that the three $\CS$-walls are labeled by different roots, therefore they are genuinely different $\CS$-walls. Fixing $z$ near the branch point, we observe 36 $\CS$-walls sweeping through $z$ when we change the phase of the spectral network by $\pi$. We therefore find $36 \times 6 = 216$ BPS solitons, plus $216$ CPT conjugates. 
This coincides with the number of solitons predicted by the Coxeter projection shown in Figure \ref{fig:coxeter_projection_E_6}, and when we calculate their central charges we can show that the BPS spectrum of the defect coincides with that of the LG model \cite{Lerche:1991re}.

\paragraph{$\boldsymbol{N}\mathbf{=7}$} 
When $u_{w_i} = 0$, the spectral curve of $\SD{\gE_7}{\omega_7}{z}$  simplifies to
\begin{align}
	F (z, z) = x^{56} + \frac{2458}{3} \, z\, x^{38} + \frac{8371}{27} \, z^{2}\, x^{20} + \frac{1}{729} \, z^{3}\, x^{2}.
\end{align}
For a fixed value of $z$ there are 56 sheets of $x$, two of which is at $x = 0$. This curve has a single branch point at $z = 0$, where all the 56 sheets meet and three branch cuts of orders 2, 18, and 18 come out. Section 5.3 of \cite{Longhi:2016rjt} provides a general description of a spectral network around the branch point at $z = 0$, and combining this with the analysis of spectral networks around a branch point given in Section 2.2.1 of \cite{Maruyoshi:2013fwa} we can understand the spectral network without explicit data, which is shown in Figure \ref{fig:E_7_SN}. 
$\gE_7$ has 126 roots and therefore there are at least 126 $\CS$-walls emanating from the branch point. The order of an orbit of each root under the action of the Coxeter element corresponding to the branch point is 18, therefore the differential equation that an $\CS$-wall satisfies around the branch point is
\begin{align}
    \partial_t \langle \alpha, \varphi(z) \rangle \propto z^{1/18} \frac{\dd z}{\dd t} = \e^{i \vartheta}.
\end{align}
Then each $\CS$-wall rotates by $2\pi \left(\frac{18}{19}\right)$ when the phase of the spectral network $\vartheta$ changes by $2\pi$, and the consistency of the spectral network requires that there should be $126 \times \frac{19}{18} = 133$ $\CS$-walls coming out from the branch point. There are $\frac{133}{3 + 4} \times 2 = 38$ flow lines from the branch point, each flow line corresponding to either 3 or 4 overlapping $\CS$-walls, the two groups alternating one after the other. Each $\CS$-wall carries 12 solitons, which is the number of pair of weights in \textbf{56} of $\gE_7$ differing by a single root of $\gE_7$ (i.e. $k_\rho=12$ in conventions of \cite{Longhi:2016rjt}). 63 $\CS$-walls move across a fixed $z$ when the phase of the spectral network varies by $\pi$, which is the same as the number of positive roots of $\gE_7$. Therefore the defect has $63 \times 12 = 756$ BPS solitons and 756 conjugates. These BPS solitons fit into the picture of the Coxeter projection of the \textbf{56} as shown in Figure \ref{fig:coxeter_projection_E_7}, and this is the same as that of the LG model \cite{Lerche:1991re}.

\section*{Acknowledgements}
We thank Joe Minahan and Greg Moore for discussions.
PL thanks the Carl Tryggers foundation for support during completion of this work. 
The research of PL is supported by the grants ``Geometry and Physics'' and  ``Exact Results in Gauge and String Theories'' from the Knut and Alice Wallenberg foundation. The work of CYP is supported in part by the National Science Foundation under Grant No.\ NSF PHY-140405.

\appendix

\section{Derivation of degenerate joint equations}\label{sec:degenerate-joints}
In this section we give the details of soliton propagation across joints of degenerate type. 
A degenerate joint is one which involves $\CS$-walls that overlap entirely, they appear at special loci of Coulomb branches in higher-rank $\mathrm{ADE}$ theories of class $\CS$.
In this section we will focus on the three types of degenerate joints that appear in the case of SO(8) SYM, studied in Section \ref{sec:semiclassical_D4}.
The main tool in deriving the equations will be the principle of homotopy invariance of the formal parallel transport \cite{Gaiotto:2012rg} (also see \cite{Longhi:2016rjt} for the notation used here).

The \emph{type-I} degenerate joint is shown in Figure \ref{fig:type-I-joint}. Both $\CS_\alpha$ and $\CS_\beta$ are degenerate, in the sense that each consists of the overlap of three distinct $\CS$-walls, with different root types. 
The newborn wall $\CS_{\alpha+\beta}$ is also degenerate, carrying three root types as well, while $\CS_{\gamma_i}$ are regular $\CS$-walls.
Let $\alpha_i,\, i=1,2,3$ be the root types supported on the degenerate wall $\CS_\alpha$ and  $\beta_i,\, i=1,2,3$ be the root types supported on $\CS_\beta$. 
This type of joint occurs if the set $\{\alpha_i+\beta_j\}_{i,j=1,2,3}$ contains more than one root. 
An explicit realization of the root types involved in this type of joint is given above in table (\ref{eq:type-I-roots}), conventions in the following shall refer to that.

To compute the soliton content of the outgoing walls in terms of the incoming ones, we use flatness of the formal parallel transport across the joint, using paths $\wp,\wp'$ shown in Figure \ref{fig:type-I-joint}. For the definition of formal parallel transport in the conventions adopted here we refer the reader to our previous paper \cite{Longhi:2016rjt}.
The incoming $\CS$-wall factors are
\be
\begin{split}
	\CS_\alpha & =\prod_{n=1}^{3} \CS_{\alpha_n}  = \prod_{n=1}^{3} \left(1 + \Xi_{\alpha_n}\right) = \prod_{n=1}^{3} \left(1+\sum_{(ij)\in\CP_{\alpha_n}}\nu_{ij}\right) \\
	\CS_\beta & =\prod_{n=1}^{3} \CS_{\beta_n}  = \prod_{n=1}^{3} \left(1 + \Xi_{\beta_n}\right) = \prod_{n=1}^{3} \left(1+\sum_{(ij)\in\CP_{\beta_n}}\nu_{ij}\right) \\
\end{split}
\ee
The outgoing ones
\be
\begin{split}
	\CS'_\alpha & =\prod_{n=1}^{3} \CS'_{\alpha_n}  = \prod_{n=1}^{3} \left(1 + \Xi'_{\alpha_n}\right) = \prod_{n=1}^{3} \left(1+\sum_{(ij)\in\CP_{\alpha_n}}\tau_{ij}\right) \\
	\CS'_\beta & =\prod_{n=1}^{3} \CS'_{\beta_n}  = \prod_{n=1}^{3} \left(1 + \Xi'_{\beta_n}\right) = \prod_{n=1}^{3} \left(1+\sum_{(ij)\in\CP_{\beta_n}}\tau_{ij}\right) \\
	\CS'_{\alpha+\beta} & = \prod_{n=1}^{3} \CS'_{\alpha_n+\beta_n}  = \prod_{n=1}^{3} \left(1 + \Xi'_{\alpha_n+\beta_n}\right) = \prod_{n=1}^{3} \left(1+\sum_{(ij)\in\CP_{\alpha_n+\beta_n}}\tau_{ij}\right) \\
	\CS'_{\gamma_1} & = 1 + \Xi'_{\gamma_1} = 1+\sum_{(ij)\in\CP_{\gamma_1}}\tau_{ij} \\
	\CS'_{\gamma_2} & = 1 + \Xi'_{\gamma_2} = 1+\sum_{(ij)\in\CP_{\gamma_2}}\tau_{ij} 
\end{split}
\ee
Recall that each root corresponds to two soliton types $(ij), (i'j')$. The types of solitons carried by each wall are obtained by comparing (\ref{eq:type-I-roots}) with (\ref{eq:so8-weights}), they read
\be\label{eq:soliton-types-type-I}
\begin{array}{c|c|c}
	\text{$\CS$-wall} 
	& \text{weight pairs} &  \\
	\hline
	\CS_\alpha & 
	\{(74),(83),(65),(12),(78),(43)\} %
	\\
	\CS_{\gamma_1} & 
	\{(63),(72)\} %
	\\
	\CS_{\alpha+\beta} & 
	\{(71),(53),(64),(82),(68),(42)\} %
	\\
	\CS_{\gamma_2} &
	\{(61),(52)\} %
	\\
	\CS_\beta & \ %
	\{(58),(41),(54),(81),(67),(32)\} %
	\\
\end{array}
\ee
The formal parallel transport along $\wp, \wp'$ takes schematically the following form\footnote{For notational simplicity we are omitting the ``diagonal'' pieces of the parallel transport, denoted as $D(\wp_\pm)$ in \cite{Longhi:2016rjt}.}
\be
\begin{split}
	F(\wp) & = \CS_\beta\CS_\alpha \\
	& = 1+\sum_{n=1}^{3}\Xi_{\beta_n}+\sum_{n=1}^{3}\Xi_{\alpha_n} \\
	& + \nu _{{41}}\nu _{{12}} +\nu _{{54}} \nu _{{41}}+ \nu _{{54}} \nu _{{43}}+ \nu _{{74}}\nu _{{43}}+\nu _{{67}} \nu _{{74}}+\nu _{{67}} \nu _{{78}}+ \nu _{{81}}\nu _{{12}} +\nu _{{58}} \nu _{{83}} \\
	& + \nu _{{54}} \nu _{{41}} \nu _{{12}} + \nu _{{67}}  \nu _{{74}}\nu _{{43}} 
\end{split}
\ee
\be
\begin{split}
	F(\wp') & = \CS'_\alpha\CS'_{\gamma_{1}}\CS'_{\alpha+\beta}\CS'_{\gamma_{2}}\CS'_\beta \\
	& = 1+\sum_{n=1}^{3}\Xi'_{\beta_n}+\sum_{n=1}^{3}\Xi'_{\alpha_n}+\sum_{n=1}^{3}\Xi'_{\alpha_n+\beta_n}+\Xi'_{\gamma_1}+\Xi'_{\gamma_2} \\
	& + \tau _{{43}}\tau _{{32}} +\tau _{{74}} \tau _{{43}}+\tau _{{53}}\tau _{{32}} +\tau _{{54}}\tau _{{41}} + \tau _{{63}}\tau _{{32}}+\tau _{{64}}\tau _{{41}} +\tau _{{64}}\tau _{{42}} +\tau _{{65}}\tau _{{52}}  \\
	& + \tau _{{65}}\tau _{{53}}+\tau _{{65}}\tau _{{54}} +\tau _{{65}}\tau _{{58}} + \tau _{{74}}\tau _{{41}}+ \tau _{{74}}\tau _{{42}} +\tau _{{68}} \tau _{{81}}+\tau _{{78}} \tau _{{81}}+\tau _{{78}} \tau _{{82}}+\tau _{{83}}\tau _{{32}} \\
	& +   \tau _{{65}}\tau _{{53}}\tau _{{32}}+ \tau _{{65}}\tau _{{54}}\tau _{{41}} + \tau _{{74}} \tau _{{43}}\tau _{{32}} \,.
\end{split}
\ee
It is convenient to separate the equations according to soliton types. 
First of all there are 12 easy equations $\Xi'_{\alpha_n}=\Xi_{\alpha_n}$ and $\Xi'_{\beta_n}=\Xi_{\beta_n}$, whose solutions are simply 
\be\label{eq:trivial-walls}
	\tau_{ij} = \nu_{ij} \,,\qquad \forall (ij) \in \CP_{\alpha_n}\cup \CP_{\beta_n} \,.
\ee
These equations simply state that the solitons carried by $\CS_\alpha,\CS_\beta$ do not change across the joint.
The soliton data on $\CS_{\alpha_n+\beta_n}$ is instead determined by the following equations
\be\label{eq:diagonal-eqs}
\begin{split}
	& \tau_{71} + \tau_{74} \tau_{41} + \tau_{78}\tau_{81} = 0 \\
	& \tau_{42} + \tau_{43} \tau_{32} = \nu_{41}\nu_{12}  \\
	& \tau_{82} + \tau_{83}\tau_{32} = \nu_{81}\nu_{12} \\
	& \tau_{53} = \nu_{58}\nu_{83}+\nu_{54}\nu_{43}\\
	& \tau_{68}+\tau_{65}\tau_{58} = \nu_{67}\nu_{78}\\
	& \tau_{64}+\tau_{65}\tau_{54} = \nu_{67}\nu_{74} \,.
\end{split}
\ee
Using (\ref{eq:trivial-walls}) it is straightforward to solve these to obtain the soliton content on $\CS_{\alpha+\beta}$.
There is one slightly tricky point however, one must take care of accounting for extra signs, due to the twisting of the formal $X_\gamma$ associated with winding of paths around the circle fiber of the circle bundle over $\Sigma$ \cite{Gaiotto:2012rg}. 
For concreteness, let us assume that $\nu_{ij}$ all carry a single soliton (this will be the case in actual applications below), although the formulae are valid in full generality. 
Then, for example the second equation becomes $\tau_{42} = - \nu_{43} \nu_{32}+ \nu_{41}\nu_{12}$ and the $(42)$-soliton contained in $ \nu_{43} \nu_{32}$ carries precisely an extra unit of winding with respect to the one containe din $\nu_{41}\nu_{12}$, therefore their contributions add up.
Taking the signs into account in this way, we find that the three overlapping $\CS$-walls of types $\alpha_n+\beta_n$ each carry \emph{two solitons} of type $ij$ for each $(ij)\in\CP_{\alpha_n+\beta_n}$.

Moving on to the soliton content of $\CS_{\gamma_1}$ and $\CS_{\gamma_2}$, this is determined by 
\be\label{eq:other-eqs}
\begin{split}
	& \tau_{63} + \tau_{65} \tau_{53} = \nu_{67}  \nu_{74}  \nu_{43}\\
	& \tau_{52}+\tau_{53}\tau_{32} = \nu_{54}\nu_{41}\nu_{12} \\
	& \tau_{72}+\tau_{74}\tau_{42}+\tau_{78}\tau_{82} +\tau_{74}\tau_{43}\tau_{32}= 0\\
	& \tau_{61} +\tau_{64}\tau_{41} +\tau_{68}\tau_{81} +\tau_{65}\tau_{54}\tau_{41}= 0
\end{split}
\ee
The first two equations readily imply that there are \emph{three} solitons of types $(63)$ and $(52)$ carried by $\CS_{\gamma_1},\CS_{\gamma_2}$ respectively (Again taking signs correctly into account).
By virtue of the \emph{soliton symmetry} found in \cite{Longhi:2016rjt}, one should find the same number of solitons of types $(72), (61)$.
Let us focus on the third equation: here $\tau_{42}$ and $\tau_{82}$ both carry two solitons, while $\tau_{72}=\nu_{72}$ and $\tau_{78}=\nu_{78}$ carry one each. On the other hand, the term $\tau_{74}\tau_{43}\tau_{32}= \nu_{74}\nu_{43}\nu_{32}$ contributes one soliton, and it comes with an extra loop of winding around the $S^1$ fiber above $\Sigma$. As a result, the contribution of the latter cancels one of the four solitons carried by the former two terms, giving a net number of three. 
The same applies to the last equation: $\tau_{63}$ carries three solitons after cancellations. 
As a result we recover a soliton content for outgoing streets that is compatible with the soliton symmetry of \cite{Longhi:2016rjt}.

The soliton content on all streets has been determined, however flatness of the formal parallel transport imposes three extra equations (the overall system is overcomplete, consisting of 25 equations in 22 variables). These have the role of consistency conditions, they read
\be
\begin{split}
	& \tau_{54} \tau_{41} = \nu_{54}\nu_{41} \\
	& \tau_{74} \tau_{43} = \nu_{74}\nu_{43} \\
	& \tau_{63}\tau_{32} + \tau_{65}\tau_{52}+ \tau_{64}\tau_{42}+\tau_{65}\tau_{53}\tau_{32} = 0\,.
\end{split}
\ee
The first two are trivially satisfied, the third equation requires more care.
The first two terms involve $\tau_{63}$ and $\tau_{52}$, therefore each carries three solitons ($\tau_{32}=\nu_{32}$ and $\tau_{65}=\nu_{65}$ each carry one soliton). The concatenations involved contribute a unit of winding around the circle fiber over $\Sigma$, thus each of the first two terms contributes with a negative sign.
The third term carries four solitons since $\tau_{64}$ and $\tau_{42}$ carry two each, moreover the overall winding number due to the concatenation is zero in this case, because the walls carrying these solitons are overlapping. 
The last term contributes two solitons because $\tau_{53}$ carries two, while $\tau_{65}=\nu_{65}$ and $\tau_{32}=\nu_{32}$. 
The sign of the last term is positive because the concatenations add two loops of winding around the circle fiber, contributing $(-1)^2$. Overall the last equation is thus automatically satisfied as well.

\medskip

The \emph{type-II} degenerate joint is shown in Figure \ref{fig:type-II-joint}. Now $\CS_\alpha$ is degenerate while $\CS_\beta$ is a regular $\CS$-wall, as a result the joint is asymmetric. 
Both newborn walls $\CS_{\alpha+\beta}$ and $\CS_{\gamma}$ are also degenerate, carrying three roots each, while $\CS_{\delta_i}$ are regular $\CS$-walls.
Table (\ref{eq:type-II-roots}) contains an explicit example of involved root types.

The factors of incoming $\CS$-wall are
\be
\begin{split}
	\CS_\alpha & =\prod_{n=1}^{3} \CS_{\alpha_n}  = \prod_{n=1}^{3} \left(1 + \Xi_{\alpha_n}\right) = \prod_{n=1}^{3} \left(1+\sum_{(ij)\in\CP_{\alpha_n}}\nu_{ij}\right) \\
	\CS_\beta & =1 + \Xi_{\beta} = 1+\sum_{(ij)\in\CP_{\beta}}\nu_{ij}
\end{split}
\ee
The outgoing ones
\be
\begin{split}
	\CS'_\alpha & =\prod_{n=1}^{3} \CS'_{\alpha_n}  = \prod_{n=1}^{3} \left(1 + \Xi'_{\alpha_n}\right) = \prod_{n=1}^{3} \left(1+\sum_{(ij)\in\CP_{\alpha_n}}\tau_{ij}\right) \\
	\CS'_\beta & =1 + \Xi'_{\beta} = 1+\sum_{(ij)\in\CP_{\beta}}\tau_{ij}\\
	\CS'_{\alpha+\beta} & = \prod_{n=1}^{3} \CS'_{\alpha_n+\beta}  = \prod_{n=1}^{3} \left(1 + \Xi'_{\alpha_n+\beta}\right) = \prod_{n=1}^{3} \left(1+\sum_{(ij)\in\CP_{\alpha_n+\beta}}\tau_{ij}\right) \\
	\CS'_\gamma & =\prod_{n=1}^{3} \CS'_{\gamma_n}  = \prod_{n=1}^{3} \left(1 + \Xi'_{\gamma_n}\right) = \prod_{n=1}^{3} \left(1+\sum_{(ij)\in\CP_{\gamma_n}}\tau_{ij}\right) \\
	\CS'_{\delta_1} & = 1 + \Xi'_{\delta_1} = 1+\sum_{(ij)\in\CP_{\delta_1}}\tau_{ij} \\
	\CS'_{\delta_2} & = 1 + \Xi'_{\delta_2} = 1+\sum_{(ij)\in\CP_{\delta_2}}\tau_{ij} 
\end{split}
\ee
The types of 2d solitons carried by each wall are now
\be\label{eq:soliton-types-type-II}
\begin{array}{c|c|c}
	\text{$\CS$-wall} 
	& \text{weight pairs} &  \\
	\hline
	\CS_\alpha & 
	\{(18),(45),(23),(76),(14),(85)\} %
	\\
	\CS_{\delta_1} & 
	\{(13),(75)\} %
	\\
	\CS_{\gamma} & 
	\{(83),(74),(65),(12),(43),(78)\} %
	\\
	\CS_{\delta_2} & 
	\{(63),(72)\} %
	\\
	\CS_{\alpha+\beta} & 
	\{(42),(68),(71),(53),(82),(64)\} %
	\\
	\CS_\beta & \ %
	\{(61),(52)\} %
	\\
\end{array}
\ee
Again we use flatness of the formal parallel transport to derive the outgoing soliton data. Choosing paths $\wp,\wp'$ as in figure this reads
\be
\begin{split}
	F(\wp) & = \CS_\beta\CS_\alpha \\
	& = 1+\Xi_{\beta}+\sum_{n=1}^{3}\Xi_{\alpha_n} \\
	& + \nu _{{23}} \nu _{{52}}+\nu _{{14}} \nu _{{61}}+\nu _{{18}} \nu _{{61}}+\nu _{{18}} \nu _{{85}}  +\nu _{{18}} \nu _{{61}} \nu _{{85}}
\end{split}
\ee
\be
\begin{split}
	F(\wp') & = \CS'_\alpha\CS'_{\gamma_{1}}\CS'_{\alpha+\beta}\CS'_{\gamma_{2}}\CS'_\beta \\
	& = 1+\Xi'_{\beta}+\sum_{n=1}^{3}\Xi'_{\alpha_n}+\sum_{n=1}^{3}\Xi'_{\alpha_n+\beta}+ \sum_{n=1}^{3} \Xi'_{\gamma_n} +\Xi'_{\delta_1}+\Xi'_{\delta_2} \\
	& + \tau _{{14}} \tau _{{42}}+\tau _{{74}} \tau _{{42}}+\tau _{{14}} \tau _{{43}}+\tau _{{45}} \tau _{{52}}+\tau _{{45}} \tau _{{53}}+\tau _{{52}} \tau _{{65}}+\tau _{{53}} \tau _{{65}}+\tau _{{43}} \tau _{{74}}\\
	& +\tau _{{52}} \tau _{{75}}+\tau _{{53}} \tau _{{75}}+\tau _{{61}} \tau _{{76}}+\tau _{{63}} \tau _{{76}}+\tau _{{64}} \tau _{{76}}+\tau _{{65}} \tau _{{76}}+\tau _{{68}} \tau _{{76}}+\tau _{{18}} \tau _{{82}}\\
	& +\tau _{{68}} \tau _{{82}}+\tau _{{78}} \tau _{{82}}+\tau _{{18}} \tau _{{83}}+\tau _{{18}} \tau _{{85}}+\tau _{{52}} \tau _{{85}}+\tau _{{53}} \tau _{{85}} \\
	& + \tau _{{52}} \tau _{{65}} \tau _{{76}}+\tau _{{53}} \tau _{{65}} \tau _{{76}}+\tau _{{68}} \tau _{{82}} \tau _{{76}}+\tau _{{18}} \tau _{{52}} \tau _{{85}}+\tau _{{18}} \tau _{{53}} \tau _{{85}}
\end{split}
\ee
Separating equations according to soliton types, we immediately find 8 simple ones $\Xi'_{\alpha_n}=\Xi_{\alpha_n}$ and $\Xi'_{\beta}=\Xi_{\beta}$. Their solutions are simply 
\be\label{eq:simple-type-II}
	\tau_{ij} = \nu_{ij} \,,\qquad \forall (ij) \in \CP_{\alpha_n}\cup \CP_{\beta} ,
\ee
which imply that solitons carried by $\CS_\alpha$ and $\CS_\beta$ do not change across the joint.
In addition we find the following equations for outgoing solitons on $\CS_{\alpha_n+\beta}$
\be
\begin{split}
	& \tau_{42} + \tau_{45} \tau_{52} = 0 \\
	& \tau_{68} = \nu_{61}\nu_{18}  \\
	& \tau_{71} + \tau_{76} \tau_{61} = 0 \\
	& \tau_{53} = \nu_{52}\nu_{23}  \\
	& \tau_{82} + \tau_{85} \tau_{52} = 0 \\
	& \tau_{64} = \nu_{61}\nu_{14} \,.
\end{split}
\ee
Using (\ref{eq:simple-type-II}) it is simple to solve these to obtain the outgoing solitons on $\CS_{\alpha_n+\beta}$.
As usual, one mast take into account signs due to the circle bundle twisting of the formal variables $X_\gamma$. In the end the result is that $\CS$-walls of types $\alpha_n+\beta_n$ each carry \emph{one soliton} of type $ij$ for each $(ij)\in\CP_{\alpha_n+\beta}$.\footnote{We are again assuming for simplicity that there is one incoming soliton of each type in $\nu_{ij}$. This will be the case in actual applications, although the equations are valid in full generality.}
The equations corresponding to soliton pairs of roots $\gamma_n$ are
\be
\begin{split}
	& \tau_{83} + \tau_{85} \tau_{53} = 0 \\
	& \tau_{74} + \tau_{76} \tau_{64} = 0 \\
	& \tau_{65} = \nu_{61} \nu_{18} \nu_{85} \\
	& \tau_{12} + \tau_{14} \tau_{42} + \tau_{18}\tau_{82} + \tau_{18}\tau_{85}\tau_{52} = 0 \\
	& \tau_{43} + \tau_{45} \tau_{53} = 0 \\
	& \tau_{78} + \tau_{76} \tau_{68} = 0 \,.
\end{split}
\ee
The first two and last two equations readily imply that there is \emph{one} soliton of each type $(83), (74), (43), (78)$, which are carried by $\CS_{\gamma_1}$ and $\CS_{\gamma_3}$. 
The third equation likewise implies that one soliton of type $(65)$ propagates on $\CS_{\gamma_2}$. The fourth equation requires a bit of care due to canceling contributions, but it is not hard to show that this also results in one soliton of type $(12)$ on $\CS_{\gamma_2}$, in agreement with the soliton symetry of \cite{Longhi:2016rjt}. 

Soliton equations for walls $\CS_{\delta_1},\CS_{\delta_2}$ read
\be
\begin{split}
	& \tau_{13} + \tau_{14} \tau_{43} + \tau_{18}\tau_{83} + \tau_{18}\tau_{85}\tau_{53} = 0 \\
	& \tau_{75} + \tau_{76} \tau_{65} = 0 \\
	& \tau_{72} + \tau_{74} \tau_{42} + \tau_{75}\tau_{52} + \tau_{78}\tau_{82} + \tau_{76}\tau_{65}\tau_{52} + \tau_{76}\tau_{68}\tau_{82} = 0 \\
	& \tau_{63} + \tau_{65} \tau_{53} = 0 	\,.
\end{split}
\ee
Again, taking due cancellations into account, the result is that these walls carry \emph{one soliton} of each type $(13),(75),(72),(63)$.

Flatness of the parallel transport also involves three additional equations (for a total of 27, in 24 variables), whose role is of consistency conditions.
They are
\be
\begin{split}
	& \tau_{18} \tau_{85} = \nu_{18}\nu_{85} \\
	& \tau_{68} \tau_{82} = \nu_{68}\nu_{82} \\
	& \tau_{74}\tau_{43}  + \tau_{75}\tau_{53} + \tau_{76}\tau_{63} + \tau_{76}\tau_{65}\tau_{53} = 0\,.
\end{split}
\ee
The first two are trivially satisfied. 
The first term in the third equation involves $\tau_{74}\tau_{43}$, there is no contribution from winding around the circle fiber since these are supported on the same wall, so the contribution is one soliton, with positive sign. 
The second contribution on the other hand involves a unit of winding and picks up a negative sign, again the contribution is just one soliton. Likewise the third and fourth terms come with opposite signs due to a difference by one in the winding number in the circle fiber.
Overall the last equation is thus automatically satisfied as well.

\medskip

The \emph{type-III} degenerate joint is shown in Figure \ref{fig:type-III-joint}. Both $\CS_\alpha$ and $\CS_\beta$ are degenerate, each carries three root types. The newborn wall $\CS_{\alpha+\beta}$ is however of regular type.
Let $\alpha_i,\, i=1,2,3$ be the root types supported on the degenerate wall $\CS_\alpha$ and  $\beta_i,\, i=1,2,3$ be the root types supported on $\CS_\beta$. 
This type of joint occurs if the set $\{\alpha_i+\beta_j\}_{i,j=1,2,3}$ contains exactly one root. 
In table (\ref{eq:type-I-roots}) we give an explicit example of possible involved root types.
The incoming $\CS$-wall factors are
\be
\begin{split}
	\CS_\alpha & =\prod_{n=1}^{3} \CS_{\alpha_n}  = \prod_{n=1}^{3} \left(1 + \Xi_{\alpha_n}\right) = \prod_{n=1}^{3} \left(1+\sum_{(ij)\in\CP_{\alpha_n}}\nu_{ij}\right) \\
	\CS_\beta & =\prod_{n=1}^{3} \CS_{\beta_n}  = \prod_{n=1}^{3} \left(1 + \Xi_{\beta_n}\right) = \prod_{n=1}^{3} \left(1+\sum_{(ij)\in\CP_{\beta_n}}\nu_{ij}\right)\,.
\end{split}
\ee
The outgoing ones
\be
\begin{split}
	\CS'_\alpha & =\prod_{n=1}^{3} \CS'_{\alpha_n}  = \prod_{n=1}^{3} \left(1 + \Xi'_{\alpha_n}\right) = \prod_{n=1}^{3} \left(1+\sum_{(ij)\in\CP_{\alpha_n}}\tau_{ij}\right) \\
	\CS'_\beta & =\prod_{n=1}^{3} \CS'_{\beta_n}  = \prod_{n=1}^{3} \left(1 + \Xi'_{\beta_n}\right) = \prod_{n=1}^{3} \left(1+\sum_{(ij)\in\CP_{\beta_n}}\tau_{ij}\right) \\
	\CS'_{\alpha+\beta} & =  \CS'_{\alpha_n+\beta_n}  =  \left(1 + \Xi'_{\alpha_n+\beta_n}\right) =  \left(1+\sum_{(ij)\in\CP_{\alpha_n+\beta_n}}\tau_{ij}\right) 
\end{split}
\ee
where we used the fact that $\alpha_1+\beta_1 = \alpha_2+\beta_2=\alpha_3+\beta_3$.

The soliton pairs carried by each wall are
\be\label{eq:soliton-types-type-III}
\begin{array}{c|c|c}
	\text{$\CS$-wall} 
	& \text{weight pairs} &  \\
	\hline
	\CS_\alpha & 
	\{(74),(83),(65),(12),(78),(43)\} %
	\\
	\CS_{\alpha+\beta} & 
	\{(72),(63)\} \\ %
	\CS_{\beta} & 
	\{(68),(42),(71),(53),(82),(64)\} %
\end{array}
\ee

Again we use flatness of the formal parallel transport to derive the outgoing soliton data. Choosing paths $\wp,\wp'$ as in figure it is straightforward to write down $F(\wp), F(\wp')$ we omit their expicit expressions for this type of joint.
Separating the equations according to soliton types, we find 12 simple ones: $\Xi'_{\alpha_n}=\Xi_{\alpha_n}$ and $\Xi'_{\beta}=\Xi_{\beta}$. Their solutions are
\be\label{eq:simple-type-III}
	\tau_{ij} = \nu_{ij} \,,\qquad \forall (ij) \in \CP_{\alpha_n}\cup \CP_{\beta} .
\ee
In addition we find the following equations for outgoing soliton data on $\CS_{\alpha+\beta}$
\be
\begin{split}
	& \tau_{63} + \tau_{65} \tau_{53} = 2 \nu_{64}\nu_{43} \\
	& \tau_{72} +2 \tau_{78}\tau_{82}= \nu_{71}\nu_{12}  
\end{split}
\ee
Using (\ref{eq:simple-type-III}) it is streightforward to solve these.
Taking into account signs due to twisting of formal variables, we find tha the $\CS$-wall of type $\alpha+\beta$ carries \emph{three solitons} of type $ij$ for each $(ij)\in\CP_{\alpha_n+\beta_n}$. 

\bibliographystyle{jhep}

\bibliography{ade_2d}

\end{document}